\documentclass[12pt]{article}

\usepackage{amsmath}
\usepackage{amsfonts}
\usepackage{amsthm}
\usepackage[english]{babel} 
\usepackage{dsfont}
\usepackage{amssymb}
\usepackage{graphicx}        
\usepackage{cite}   
\usepackage{caption}
\usepackage{subcaption}
\usepackage{epstopdf}
\usepackage{epsfig}
\usepackage[table]{xcolor}
\usepackage{subfloat}
\usepackage{float}
\usepackage[thinlines]{easytable}
\usepackage{array}
\usepackage{mathtools}
\usepackage{hyperref}
\usepackage{multirow}
\allowdisplaybreaks
\numberwithin{equation}{section}

\newtheorem{remark}{Remark}[section]

\usepackage[linesnumbered,algoruled,boxed,lined]{algorithm2e}

\def\RR{\mathbb R}

\renewcommand{\d}{{\rm d}}			
\newcommand{\CFL}{\mathsf{CFL}}		

\newcommand{\U}{{\bf{U}}}
\newcommand{\V}{{\bf{V}}}
\newcommand{\F}{{\bf{F}}}
\newcommand{\G}{{\bf{G}}}
\renewcommand{\H}{{\bf{H}}}

\def\be{\begin{equation}}
\def\ee{\end{equation}}
\def\bea{\begin{eqnarray}}
\def\eea{\end{eqnarray}}

\title{Hyperbolic compartmental models for epidemic spread on networks with uncertain data: application to the emergence of Covid-19 in Italy}

\author{Giulia Bertaglia\footnote{Department of Mathematics and Computer Science, University of Ferrara, Via Machiavelli 30, 44121 Ferrara, Italy (giulia.bertaglia@unife.it)}\and
Lorenzo Pareschi\footnote{Department of Mathematics and Computer Science, University of Ferrara, Via Machiavelli 30, 44121 Ferrara, Italy (lorenzo.pareschi@unife.it)}}

\begin{document}
\maketitle

\begin{abstract}
The importance of spatial networks in the spread of an epidemic is an essential aspect in modeling the dynamics of an infectious disease. Additionally, any realistic data-driven model must take into account the large uncertainty in the values reported by official sources, such as the amount of infectious individuals. In this paper we address the above aspects through a hyperbolic compartmental model on networks, in which nodes identify locations of interest, such as cities or regions, and arcs represent the ensemble of main mobility paths. The model describes the spatial movement and interactions of a population partitioned, from an epidemiological point of view, on the basis of an extended compartmental structure and divided into commuters, moving on a suburban scale, and non-commuters, acting on an urban scale. Through a diffusive rescaling, the model allows us to recover classical diffusion equations related to commuting dynamics.
The numerical solution of the resulting multiscale hyperbolic system with uncertainty is then tackled using a stochastic collocation approach in combination with a finite-volume IMEX method. 
The ability of the model to correctly describe the spatial heterogeneity underlying the spread of an epidemic in a realistic city network is confirmed with a study of the outbreak of COVID-19 in Italy and its spread in the Lombardy Region. 
\end{abstract}


\tableofcontents

\section{Introduction}
The advent of the COVID-19 pandemic, which still afflicts the entire world, has prompted many researchers in the mathematical field and beyond to propose epidemic models increasingly suitable for the study of the evolution of this specific coronavirus, to support the definition of the best strategies for the control of its spread.
Most of the proposed models are rooted in the compartmental epidemiological modeling proposed by Kermack and McKendrick \cite{capasso1978,hethcote2000}, based on systems of ordinary differential equations (ODE), which describe only the temporal evolution of the spread of the epidemic, neglecting the spatial component in favor of an assumption of homogeneity of population and territory \cite{buonomo2020,franco2020,giordano2020,giordano2021,kantner2020,peirlinck2020, scarabel2021,tang2020,lolipiccolomini2020,parolini2021}. 
Recently, models have also been proposed in which we have moved away from the idea of a homogeneously mixed population to allow a better description in terms of social contact patterns, vulnerability and case fatality ratio, but still neglecting the spatial development of the epidemic spread \cite{albi2020a,bellomo2020}.

Generally, the concept of the average behavior of a population is sufficient to have a first reliable description of the development of an epidemic. However, the inclusion of the spatial component in epidemiological systems is crucial especially when there is a need to consider spatially heterogeneous interventions, as was and still is the case for the control of the spread of COVID-19 \cite{riley2015,pellis2015}.
Some recent works have attempted to fill these gaps by proposing epidemiological models based on partial differential equations (PDE), in which the spatial structure is taken into account. In \cite{colombo2020,viguerie2020,viguerie2021,boscheri2020} a two-dimensional (2D) space dependence is considered, obtaining models able to properly capture the complexity of the spatial transmission of the virus.  On the other hand, meta-population network approaches have also been proposed \cite{gatto2020,dellarossa2020}, which result in systems of ODE that provide no information about the spatial characteristics of the epidemic spread along the pathways connecting the different populations \cite{riley2015}. 

Another important aspect concerns the wide use of deterministic models, which, although more computationally efficient, are based on the assumption that initial conditions, boundary conditions and all the parameters involved are known. However, in practical applications, this assumption is rarely true. In the context of epidemic modeling, for example, initial conditions are certainly affected by uncertainty because data are limited and screening policy is always a matter of compromises. Also epidemic parameters, although normally estimated or calibrated, are candidates for being random variables. Therefore, when attempting
to solve the problem of interest numerically, one must take into account these limitations, recurring to more consistent stochastic models \cite{albi2020a,dellarossa2020,gatto2020,peirlinck2020,pareschi2020}.

In this work, a hyperbolic transport model on networks with uncertain data capable to describe the spatial spread of an epidemic phenomenon defined by an extended compartmental dynamics is proposed. The model takes inspiration by the multiscale transport model recently introduced in \cite{boscheri2020} by extending it to include more realistic compartmentalizations. In addition, the spatial structure of the system lays on the network modeling originally adopted in \cite{bertaglia2021}, in which nodes of the network identify locations of interests, depending on the spatial scale considered (villages, provinces, regions, nations), and arcs constitute the mobility paths connecting them (hence roads, railways, airline connections, etc.). Individuals moving on the network are subdivided into commuting and non-commuting, the latter acting only at the urban, nodal scale they belong to, not contributing in the transmission of the virus in the extra-urban domain. This particular feature prevents unrealistic mass migration effects where the entire population travels through the network \cite{boscheri2020}. Furthermore, thanks to its hyperbolic structure the model avoids the unphysical feature of infinite propagation speeds typical of reaction-diffusion systems, still recovering the parabolic behaviour in the diffusive limit \cite{bertaglia2021,barbera2013}.

The resulting model is solved numerically through an IMEX finite volume stochastic scheme that permits to preserve uniform accuracy with respect to the scaling parameters \cite{jin2015, bertaglia2021}. More precisely, the numerical scheme combines a stochastic collocation method, a \textit{non-intrusive} method which guarantees spectral convergence in the stochastic space and ease of implementation, avoiding the loss of important structural properties of the original system of governing equations \cite{poette2009,pareschi2020,xiu2005}, with an Implicit-Explicit (IMEX) Runge-Kutta finite-volume scheme that maintains the consistency in the asymptotic limit (i.e. the asymptotic-preserving (AP) property) \cite{boscarino2017}.

The rest of the paper is organized as follows. In Section \ref{sect:mathmodel} the mathematical model is introduced, first in its kinetic transport formulation and then deriving its macroscopic representation. Subsequently, the diffusion limit is formally computed and the details of the model extension to networks are presented. Moreover, a definition of the basic reproduction number of the epidemic is proposed, with the details of its derivation discussed in Appendix \ref{appendix:R0}.
In Section \ref{sect:numres} we present an application to the first outbreak of COVID-19 in Italy and its spread in the Lombardy Region. First, to validate the proposed numerical approach a convergence study is performed. Next, the capacity of the proposed model to capture the spatial heterogeneous characteristics underlying COVID-19 epidemics is assessed through several simulations based on data reported by official sources. The details of the stochastic collocation approach are summarized in Appendix \ref{appendix:SCM}, whereas the IMEX finite-volume scheme is described in Appendix \ref{appendix:IMEX}. 
Finally, conclusions and future perspectives are drawn in Section \ref{sect:conclusion}.

\section{Mathematical model}
\label{sect:mathmodel}
\subsection{A hyperbolic compartmental model for epidemic spread}
The epidemiological starting point of our model is given by a compartmental structure with a SEIAR partitioning \cite{tang2020,tang2020a,peirlinck2020}, in which the population is assumed to be divided in susceptible individuals $S$, exposed individuals in the latent period, which are not yet infectious $E$, infectious individuals manifesting severe symptoms $I$, infectious individuals with no/mild symptoms $A$ and removed (deceased or healed and immune) individuals $R$. We assume to have a population with subjects having no prior immunity. The vital dynamics represented by births and deaths is neglected because of the time scale considered. This model differs from the classic SEIR \cite{hethcote2000} for the presence of a subgroup of people who will never develop symptoms or very mild ones, which is an essential feature to take into account if one aims at analyzing the evolution of the COVID-19 pandemic. Indeed, individuals belonging to this group are very hard to be detected, deeply impacting the efficiency of the monitoring and making isolation, containment and tracing of individual cases very challenging \cite{peirlinck2020,gatto2020,giordano2020,buonomo2020}. They tend to behave like simple susceptible, but in fact highly contributing in augmenting the spread of the virus. 

To account for the spatial movement of the population, the model has its roots within discrete velocity models in kinetic theory \cite{Lions1997,bertaglia2021}, and follows the approach in \cite{boscheri2020}, where individuals of each compartment are subdivided in three classes, $S_{\pm,0}$, $E_{\pm,0}$, $I_{\pm,0}$, $A_{\pm,0}$, $R_{\pm,0}$, traveling in a one-dimensional space domain $\Omega\subseteq \RR$ with characteristic speeds $+\lambda_i,-\lambda_i$ and $0$ respectively, with $i \in \{S,E,I,A,R\}$. The total compartmental densities are defined as the sum of all the components of the subgroups
\begin{equation}
\begin{split}
S_T=S_++S_-+S_0,\qquad E_T=E_+&+E_-+E_0,\qquad I_T=I_++I_-+I_0,\\
A_T=A_++A_-+A_0,&\qquad R_T=R_++R_-+R_0\,.
\label{eq:seiart}
\end{split}
\end{equation}
The discrete-velocity system of the SEIAR epidemic transport model for commuters, associated to relaxation times $\tau_i$, $i \in \{S,E,I,A,R\}$, then reads
\begin{equation}
\begin{aligned}
\frac{\partial S_{\pm}}{\partial t} \pm \lambda_S \frac{\partial S_{\pm}}{\partial x} &= -f_I(S_{\pm}, I_T) -f_A(S_{\pm}, A_T) + \frac{1}{2\tau_S}\left(S_\mp - S_\pm\right)\\
\frac{\partial E_{\pm}}{\partial t} \pm \lambda_E \frac{\partial E_{\pm}}{\partial x} &= f_I(S_{\pm}, I_T) +f_A(S_{\pm},A_T) -a E_{\pm} + \frac{1}{2\tau_E}\left(E_\mp - E_\pm\right)\\
\frac{\partial I_{\pm}}{\partial t} \pm \lambda_I \frac{\partial I_{\pm}}{\partial x} &= a \sigma E_{\pm} -\gamma_I I_{\pm} + \frac1{2\tau_I}\left(I_\mp - I_\pm\right)\\
\frac{\partial A_{\pm}}{\partial t} \pm \lambda_A \frac{\partial A_{\pm}}{\partial x} &= a(1-\sigma) E_{\pm} -\gamma_A A_{\pm} + \frac{1}{2\tau_A}\left(A_\mp - A_\pm\right)\\
\frac{\partial R_{\pm}}{\partial t} \pm \lambda_R \frac{\partial R_{\pm}}{\partial x} &= \gamma_I I_{\pm} + \gamma_A A_{\pm} + \frac1{2\tau_R}\left(R_\mp - R_\pm\right)\, .
\end{aligned}
\label{eq.SEIARkinetic}
\end{equation}
This system is coupled with the following SEIAR model describing the evolution of a stationary population of non-commuters
\begin{equation}
\begin{aligned}
\frac{\d S_{0}}{\d t}  &= -f_I(S_{0}, I_T) +f_A(S_{0}, A_T) \\
\frac{\d E_{0}}{\d t}  &= f_I(S_{0}, I_T) +f_A(S_{0}, A_T) -a E_0 \\
\frac{\d I_{0}}{\d t}  &= a \sigma E_0 -\gamma_I I_{0}\\
\frac{\d A_{0}}{\d t}  &= a(1-\sigma) E_0 - \gamma_A A_0 \\
\frac{\d R_{0}}{\d t}  &= \gamma_I I_0 + \gamma_A A_0\, .
\end{aligned}
\label{eq.SEIARkinetic_noncommuters}
\end{equation}

\begin{figure}[t!]
\centering
\includegraphics[width=0.7\linewidth]{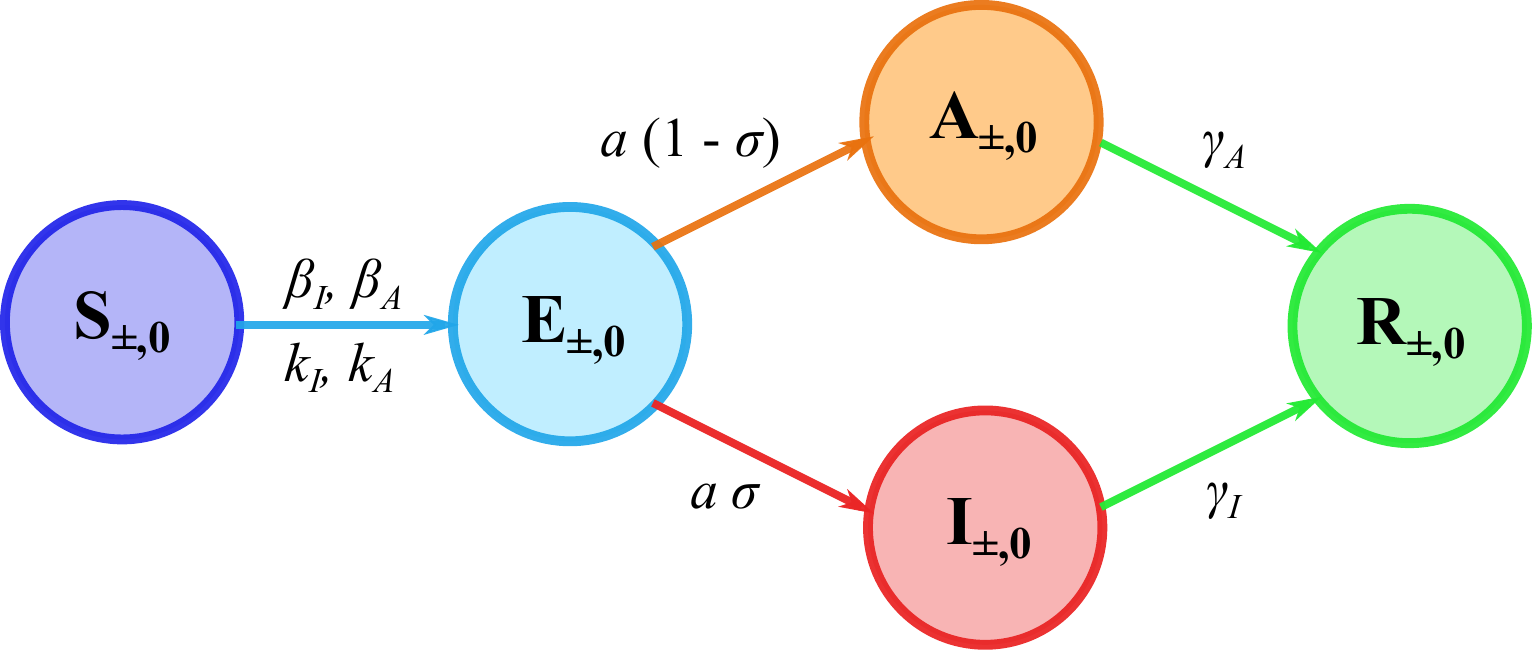}
\caption{Flow chart of the multi-population SEIAR dynamic based on five compartments: susceptible (S), exposed (E), severe symptomatic infectious (I), mildly symptomatic/asymptomatic infectious (A), and removed --healed or deceased-- population (R), each one subdivided in three classes of individuals traveling in the domain with characteristic speeds $+\lambda_i,-\lambda_i$ and $0$, with $i\in\{S,E,I,A,R\}$. The transition rates between the compartments, $\beta_j, a, \gamma_j$ are the inverse of the contact period $1/\beta_j$, the latent period $1/a$ and the infectious period $1/\gamma_j$, respectively, with distinct values for groups $j\in\{I,A\}$, except for the latent period. The rate of probability for the exposed to enter in the symptomatic and asymptomatic subgroups of the infectious population are $\sigma$ and $(1-\sigma)$, respectively. Coefficients $k_j$ affect the contact rates in response to social distancing and other control actions.}
\label{fig.SEIAR}
\end{figure}

In the above system \eqref{eq.SEIARkinetic}-\eqref{eq.SEIARkinetic_noncommuters} all the epidemic densities $S_{\pm,0}$, $E_{\pm,0}$, $I_{\pm,0}$, $A_{\pm,0}$, $R_{\pm,0}$ depend on $(x,t,\boldsymbol{z})$, where $(x,t)$ are the physical variables of space $x \in \Omega \subseteq \mathbb{R}$ and time $t>0$, while $\boldsymbol{z} = (z_1,\ldots,z_{d})^T \in \mathbb{R}^{d}$ is a random vector characterizing the possible sources of uncertainty due to independent parameters $z_1,\ldots,z_{d}$. The same applies for the contact rate function $f$, defined with respect to the infectious compartments, $I$ and $A$ respectively, as
\[
f_I(S,I)=\beta_I \frac{S I^p}{1+k_I I^p}, \qquad f_A(S,A)=\beta_A \frac{S A^p}{1+k_A A^p},\qquad p\geq 1,
\]
where $\beta_I(x,t,\boldsymbol{z})$ and $\beta_A(x,t,\boldsymbol{z})$ are the transmission rates, accounting for both number of contacts and probability of transmission, hence may vary based on the effects of government control actions, such as mandatory wearing of masks, shutdown of specific work/school activities, or full lockdowns \cite{hethcote2000,giordano2020,viguerie2020,albi2020a}; $k_I(x,t,\boldsymbol{z})$ and $k_A(x,t,\boldsymbol{z})$ act as incidence damping coefficients based on the self-protective behavior of the individual that arises from awareness of the risk associated with the epidemic \cite{bertaglia2021,franco2020}. Note that, the classic bilinear case corresponds to $p = 1$ and $k_I = k_A = 0$, even though it has been observed that an incidence rate that increases more than linearly with respect to the number of infectious can occur under certain circumstances \cite{barbera2013,capasso1978,korobeinikov2005}. Parameters $\gamma_I(x,t,\boldsymbol{z})$ and $\gamma_A(x,t,\boldsymbol{z})$ are the recovery rates of highly symptomatic infected and of asymptomatic or mildly infected (inverse of the infectious periods), respectively, while $a(x,t,\boldsymbol{z})$ represents the inverse of the latency period and $\sigma(x,t,\boldsymbol{z})$ is the probability rate of developing severe symptoms \cite{tang2020,gatto2020,buonomo2020}. The flow chart of the SEIAR model considered in this work is also illustrated in Fig. \ref{fig.SEIAR}, where transition rates between compartments are clearly displayed.

\begin{remark}~
\begin{itemize}
\item
The model allows to describe more realistically the typical dynamic of commuters, which regards only a small fraction of individuals, and to distinguish this from the epidemic process which, instead, involves the whole population, including non-commuters. The presence of a group of non-commuting population, indeed, permits to avoid that the whole population in a compartment moves indiscriminately in the full space originating an unrealistic mass migration effect.
\item 
The presence of uncertainty in the data included from the beginning in the modeling process could allow the compartmentalization of asymptomatic individuals to be eliminated by implicitly including them in the uncertainty about the number of infected individuals. This approach was proposed in \cite{albi2020a}. In our case, however, in order to highlight the link with similar models used in the literature, we chose to keep the asymptomatic compartment, which consequently will be the one affected by the highest level of uncertainty. 
\end{itemize}
\end{remark}

\subsection{Macroscopic formulation and diffusion limit}
Introducing now the macroscopic variables $S_c,E_c,I_c,A_c,R_c$ for the commuters, with
\[S_c = S_++S_-,\quad E_c = E_++E_-,\quad I_c = I_++I_-,\quad A_c = A_++A_-,\quad R_c = R_++R_-,\]
and defining the fluxes
\[
J_S = \lambda_S(S_+-S_-),\quad J_E = \lambda_E(E_+-E_-),\quad J_I = \lambda_I(I_+-I_-),\]
\[ J_A = \lambda_A(A_+-A_-),\quad J_R = \lambda_R(R_+-R_-),
\]
a hyperbolic model underlying the macroscopic formulation of the spatial propagation of an epidemic, equivalent to the mesoscopic one \cite{aylaj2020}, presented in system \eqref{eq.SEIARkinetic}, is obtained:
\begin{equation}
\begin{aligned}
\frac{\partial S_{c}}{\partial t} + \frac{\partial J_S}{\partial x} &= -f_I(S_c, I_T) -f_A(S_c, A_T) \\
\frac{\partial E_{c}}{\partial t} + \frac{\partial J_E}{\partial x} &= f_I(S_c, I_T) + f_A(S_c, A_T) -a E_{c} \\
\frac{\partial I_{c}}{\partial t} + \frac{\partial J_I}{\partial x} &= a \sigma E_{c} - \gamma_I I_{c}\\
\frac{\partial A_{c}}{\partial t} + \frac{\partial J_A}{\partial x} &= a(1- \sigma) E_{c} - \gamma_A A_{c}\\
\frac{\partial R_{c}}{\partial t} + \frac{\partial J_R}{\partial x} &= \gamma_I I_{c} + \gamma_A A_{c}\\
\frac{\partial J_S}{\partial t} + \lambda_S^2 \frac{\partial S_c}{\partial x} &= -f_I(J_S, I_T) -f_A(J_S, A_T) - \frac1{\tau_S} J_S\\
\frac{\partial J_E}{\partial t} + \lambda_E^2 \frac{\partial E_c}{\partial x} &= \frac{\lambda_E}{\lambda_S}\left(f_I(J_S, I_T) + f_A(J_S, A_T)\right) - a J_E - \frac1{\tau_E} J_E\\
\frac{\partial J_I}{\partial t} + \lambda_I^2 \frac{\partial I_c}{\partial x} &=  \frac{\lambda_I}{\lambda_E} a \sigma J_E  - \gamma_I J_I-\frac1{\tau_I} J_I\\
\frac{\partial J_A}{\partial t} + \lambda_A^2 \frac{\partial A_c}{\partial x} &=  \frac{\lambda_A}{\lambda_E} a (1- \sigma) J_E  - \gamma_A J_A-\frac1{\tau_A} J_A\\
\frac{\partial J_R}{\partial t} + \lambda_R^2 \frac{\partial R_c}{\partial x} &= \frac{\lambda_R}{\lambda_I} \gamma_I J_I + \frac{\lambda_R}{\lambda_A} \gamma_A J_A -\frac1{\tau_R} J_R
\end{aligned}
\label{eq.SEIARmacro}
\end{equation}
Note that here the above system is coupled with the equations for the non commuter population \eqref{eq.SEIARkinetic_noncommuters} through identities \eqref{eq:seiart}.

Formally, system \eqref{eq.SEIARmacro} is a 1D system of stochastic balance laws which can be rewritten in compact form as:
\begin{equation}
\begin{aligned}
	\partial_t \U_c(x,t,\boldsymbol{z}) + \partial_x \V (x,t,\boldsymbol{z}) &= \F_c(\U_T(x,t,\boldsymbol{z}),\U_c(x,t,\boldsymbol{z})) \\
	\partial_t \V (x,t,\boldsymbol{z}) + \boldsymbol{\Lambda}^2(x,t)\, \partial_x \U_c (x,t,\boldsymbol{z}) &= \G(\U_T(x,t,\boldsymbol{z}),\V(x,t,\boldsymbol{z})) + \H(\V (x,t,\boldsymbol{z})) ,
\end{aligned}
\label{systcompactform}
\end{equation}
in which
	\[\U_T =
\begin{pmatrix} 
  	S_T \\ E_T \\ I_T \\ A_T \\ R_T 
\end{pmatrix}, \quad
	\U_c =
\begin{pmatrix} 
  	S_c \\ E_c \\ I_c \\ A_c \\ R_c 
\end{pmatrix}, \quad
	\V = 
\begin{pmatrix} 
 	J_S \\ J_E \\ J_I \\ J_A \\ J_R
\end{pmatrix}, \quad
	\boldsymbol{\Lambda} = 
\begin{pmatrix} 
	\lambda_S &0 &0 &0 &0\\ 0 &\lambda_E &0 &0 &0\\ 0 &0 &\lambda_I &0 &0\\ 0 &0 &0 &\lambda_A &0\\ 0 &0 &0 &0 &\lambda_R
\end{pmatrix}, \]\\
	\[
	\F_c(\U_T,\U_c) = 
\begin{pmatrix} 
  	-f_I(S_c,I_T) -f_A(S_c,A_T) \\ f_I(S_c, I_T) +f_A(S_c,A_T) -aE_c \\ a\sigma E_c -\gamma_I I_c \\ a(1-\sigma) E_c -\gamma_A A_c \\ \gamma_I I_c + \gamma_A A_c
\end{pmatrix}, \]\\
	\[\G(\U_T,\V) = 
\begin{pmatrix}
  	-f_I(J_S, I_T) -f_A(J_S, A_T) \\ \frac{\lambda_E}{\lambda_S}\left(f_I(J_S, I_T) + f_A(J_S, A_T)\right) - a J_E\\ \frac{\lambda_I}{\lambda_E} a\sigma J_E  - \gamma_I J_I  \\ \frac{\lambda_A}{\lambda_E} a(1-\sigma) J_E  - \gamma_A J_A \\ \frac{\lambda_R}{\lambda_I} \gamma_I J_I + \frac{\lambda_R}{\lambda_A} \gamma_A J_A
\end{pmatrix},\quad
	\H(\V) = -
\begin{pmatrix}
  	{J_S}/{\tau_S} \\ {J_E}/{\tau_E}\\ {J_I}/{\tau_I} \\ {J_A}/{\tau_A} \\ {J_R}/{\tau_R}
\end{pmatrix}.\]

It is easy to verify that system \eqref{systcompactform} is symmetric hyperbolic in the sense of Friedrichs-Lax \cite{friedrichs1971}, with real finite characteristic velocities (eigenvalues) $\lambda_i$, $i\in\{S,E,I,A,R\}$,  and a complete set of linearly independent eigenvectors. All the eigenvectors are associated with genuinely non-linear fields, defining shocks and rarefactions, and the Riemann invariants of the system correspond to the kinetic transport variables
\begin{equation}
\begin{split}
S^{\pm} = \frac{1}{2} \left( S_c \pm \frac{J_S}{\lambda_S} \right) , \quad 
E^{\pm} = \frac{1}{2} &\left( E_c \pm \frac{J_E}{\lambda_E} \right) , \quad 
I^{\pm} = \frac{1}{2} \left( I_c \pm \frac{J_I}{\lambda_I} \right) , \\ 
A^{\pm} = \frac{1}{2} \left( A_c \pm \frac{J_A}{\lambda_A} \right) &,\quad 
R^{\pm} = \frac{1}{2} \left( R_c \pm \frac{J_R}{\lambda_R} \right) \, .
\end{split}
\label{eq.RI}
\end{equation}
\subsubsection{Reproduction number}
\label{sect:R0}
The standard threshold of epidemic models is the well-known basic reproduction number $R_0$, also called the basic reproduction ratio or the basic reproductive rate, which defines the expected number of secondary cases produced, in a completely susceptible population, by a typical infected individual during its entire period of infectiousness \cite{diekmann1990,hethcote2000}. For many deterministic infectious disease models, this number determines whether an infection can invade and persist in a new host population ($R_0 > 1$) or cannot ($R_0 < 1$). Its definition in the case of spatially dependent dynamics, as already noted in \cite{viguerie2020}, is not straightforward particularly when considering its spatial dependence. In the following we will consider the following definition for the average value of the reproduction number on the domain $\Omega$ for $t>0$:
\begin{equation}
\begin{aligned}
R_0 (t) &= \frac{\int_\Omega f_I(S_T,I_T) \,dx}{\int_\Omega \gamma_I(x) I_T(x,t) \,dx} \cdot \frac{\int_\Omega a(x)\sigma (x) E_T(x,t) \,dx}{\int_\Omega a(x) E_T(x,t) \,dx} \\&+ \frac{\int_\Omega f_A(S_T,A_T) \,dx}{\int_\Omega \gamma_A(x) A_T(x,t)\, dx} \cdot \frac{\int_\Omega a(x)(1-\sigma(x)) E_T(x,t) \,dx}{\int_\Omega a(x) E_T(x,t)\, dx} \,.
\end{aligned}
\label{eq.R0_2}
\end{equation}
The derivation of the above expression for $R_0(t)$, computed following the \textit{next-generation matrix} approach \cite{diekmann1990}, is presented in detail in Appendix \ref{appendix:R0}.

In the two contributions of eq. \eqref{eq.R0_2} it is possible to identify:
\begin{itemize}
\item the transmission rates for compartment $I$ and $A$: $f_I(S_T,I_T)$ and $f_A(S_T,A_T)$, respectively;
\item the mean time of staying in compartment $I$ and $A$: $\gamma_I^{-1}$ and $\gamma_A^{-1}$, respectively;
\item the fraction of individuals passing from compartment $E$ to $I$ and $A$: $a\sigma/a  = \sigma$ and  $a(1-\sigma)/a  = 1- \sigma$, respectively. 
\end{itemize}

It is worth to underline that from definition \eqref{eq.R0_2} it can be deduced that it is a combination of the growth of $E_T,I_T$ and $A_T$ that determines the persistence of the epidemic, not solely the growth of $E_T$ in time, neither the growth of the simple sum $E_T+I_T+A_T$.

\begin{remark}
If no spatial dependence is assigned to variables and parameters, hence the ODE version of system \eqref{eq.SEIARmacro} is considered, and no social distancing effects are taken into account (i.e. $k_I=k_A = 0$) the reproduction number results in accordance with \cite{tang2020,tang2020a,gatto2020}:
\begin{equation*}
R_0 (t) = \sigma \frac{\beta_I S_T}{\gamma_I} + (1-\sigma)\frac{\beta_A S_T}{\gamma_A} \, .
\end{equation*}
\end{remark}

\subsubsection{Diffusion limit}
From a formal viewpoint, it can be shown that the proposed model recovers the parabolic behavior expected from standard space-dependent epidemic models in the diffusion limit \cite{barbera2013,bertaglia2021}.
Introducing the diffusion coefficients
\[
D_i=\lambda_i^2 \tau_i, \quad i\in\{S,E,I,A,R\}
\]
that characterize the diffusive transport mechanism of $S,E,I,A,R$ respectively, and letting $\tau_i \to 0$, $i\in\{S,E,I,A,R\}$, while keeping the diffusion coefficients finite, from the last five equations of system \eqref{eq.SEIARmacro} we recover Fick’s law getting
\[
\begin{split}
J_S = - D_S\frac{\partial S_c}{\partial x}, \quad J_E = - D_E\frac{\partial E_c}{\partial x}, \quad J_I = - D_I\frac{\partial I_c}{\partial x}, \quad J_A = - D_A\frac{\partial A_c}{\partial x}, \quad J_R = - D_R\frac{\partial R_c}{\partial x}.
\end{split}
\]
Finally, inserting these results in the rest of the equations of system \eqref{eq.SEIARmacro} yields the following parabolic reaction-diffusion system for the commuters
\begin{equation}
\begin{split}
\frac{\partial S_{c}}{\partial t}  &= \frac{\partial}{\partial x}\left(D_S \frac{\partial }{\partial x}S_c \right) - f_I(S_{c}, I_T) - f_A(S_{c}, A_T)\\
\frac{\partial E_{c}}{\partial t}  &= \frac{\partial}{\partial x}\left(D_E \frac{\partial }{\partial x}E_c \right) +f_I(S_{c}, I_T) +f_A(S_{c},A_T) -aE_{c} \\
\frac{\partial I_{c}}{\partial t}  &= \frac{\partial}{\partial x}\left(D_I \frac{\partial }{\partial x}I_c \right)+ a\sigma E_{c} -\gamma_I I_{c}\\
\frac{\partial A_{c}}{\partial t}  &= \frac{\partial}{\partial x}\left(D_A \frac{\partial }{\partial x}A_c \right)+ a(1-\sigma) E_{c} -\gamma_A A_{c}\\
\frac{\partial R_{c}}{\partial t}  &= \frac{\partial}{\partial x}\left(D_R \frac{\partial }{\partial x}R_c \right)+\gamma_I I_{c} +\gamma_A A_{c}\,.
\end{split}
\label{eq.SEIARdiffusive}
\end{equation}
Therefore, the relaxation times can modify the nature of the behavior of the solution \cite{bertaglia2021,barbera2013,boscarino2017}, which can result either hyperbolic or parabolic (when considering small relaxation times and large speeds). This feature of the model makes it particularly suitable for the description of the dynamics of human populations, which are characterized by movement at different spatial scales \cite{boscheri2020}. It is therefore natural to assume $\tau_i=\tau_i(x)$, $i \in \{S,E,I,A,R\}$, since in geographic areas densely populated we can assume a diffusive dynamics while in other areas or along the main arteries of communication a hyperbolic description will be more appropriate avoiding propagation of information at infinite speed.  
\begin{remark}
In the model description adopted in this Section, the relaxation times are assumed to be space dependent but independent of the state variables. More generally, it is possible to consider the relaxation times 
\[
\tau_i = \tilde\tau_i/\kappa(i_c,i_T),\qquad i\in\{S,E,I,A,R\},
\]  
leading, as $\tilde \tau_i \to 0$ and taking $\lambda_i^2=\tilde D_i/\tilde \tau_i$, to nonlinear diffusion equations in the form \eqref{eq.SEIARdiffusive} where the diffusion coefficients $D_i=\tilde D_i \kappa(i_c,i_T)$, $i\in\{S,E,I,A,R\}$, depend on the state variables.

A classical example is represented by the choice 
\[
\kappa(i_c,i_T) = i_c^\alpha,\qquad i\in\{S,E,I,A,R\}
\]
which corresponds to a generalization of Carleman model. In the limit $\tilde\tau_i\to 0$, for $\alpha=0$ we recover again the linear diffusion model whereas assuming $\alpha \in (-1,0)$ we obtain a fast diffusion process considered by other authors in epidemiology \cite{berres2013}. 
From a mathematical viewpoint we refer to \cite{Lions1997} for rigorous results concerning these kind of diffusion limits for generalized Carlemann models. Although interesting, here we do not explore further this direction. 
\end{remark}

\subsection{Extension to network modeling}
The transport model here proposed can be extended to network approaches in the sense of those presented in \cite{bretti2006,bretti2014,piccoli2006,Piccioli2021}. In the sequel we summarize the details of the network structure we adopted to characterize arcs and nodes with different sizes.

Following \cite{bertaglia2021}, it is possible to structure a 1D network considering that the nodes of the network identify locations of interest such as municipalities, provinces or, in a wider scale, regions or nations, while the arcs, enclosing the 1D spatial dynamics, represent the paths linking each location to the others. In this way, the epidemic state of each node evolves in time influenced by the mobility of the commuting individuals, moving from the other locations included in the network, always considering a part of the population composed by non-commuting individuals which remain at the origin node.


In order to prescribe the proper coupling between nodes and arcs, ensuring the conservation of total density (population) in the network and of fluxes at the interface, it is necessary to impose appropriate transmission conditions at each arc-node interface. 

\subsubsection{Transmission conditions at nodes}
\label{section_node_conditions}
A network or a connected graph $\mathcal{G = (N,A)}$ is composed of a finite set of $N$ nodes (or vertices) $\mathcal{N}$ and a finite set of $A$ bidirectional arcs (or edges) $\mathcal{A}$, such that an arc connects a pair of nodes \cite{piccoli2006}.
Let us parametrize the $A$ arcs of the network as intervals $a_i =[0, L_i], i =1,\ldots,A$. Arcs are bidirectional, as the network is non-oriented, but an artificial orientation needs to be fixed in order to define a sign for the velocities and therefore the fluxes. For an incoming arc, $L_i$ is the abscissa of the node, whereas for an outgoing arc the same abscissa is $0$.

To define transmission conditions at a generic node $n \in \mathcal{N}$ having $a_i \in \mathcal{A}, i~=~1,\ldots,N_{a,L}$ incoming arcs and $a_j \in \mathcal{A}, j~=~1,\ldots,N_{a,0}$ outgoing arcs, we need to consider two kind of interfaces at the node: the interfaces neighboring incoming arcs ($L,i$) and the interface neighboring outgoing arcs ($0,j$). If variables are discontinuous across these interfaces, at time $t+\Delta t$, for each of them $\left(1 + N_{a,L}\right)$ new states originate at interfaces ($L,i$) and $\left(1 + N_{a,0}\right)$ new states originate at interfaces ($0,j$) \cite{piccoli2006}. To compute them, we need to solve $\left(2 + N_{a,L} + N_{a,0}\right)$ Riemann problems, using the Riemann Invariants (kinetic variables) of the system, defined in Eqs.~\eqref{eq.RI}, and the principle of conservation of fluxes at interfaces \cite{bretti2006,bretti2014}.

For each one of the compartments of commuting individuals of the SEIAR model here discussed, let us indicate, for ease of notation, with $u$ the number of individuals of the compartment, with $v$ the corresponding analytical flux, with $\lambda$ its characteristic velocity, and with $u^{\pm}$ the Riemann Invariants. Introducing transmission coefficients, $\alpha_{i,j} \in [0,1]$, which represent the probability that an individual at a generic arc-node interface decides to move across that interface, from the $j$-$th$ location to the $i$-$th$ location, transmission conditions at the interfaces with an incoming arc ($L,i$) results for the arcs side
\begin{equation}
\begin{split}
	&u^*_{L,i} = u^+_{L,i} + \sum_{k=1}^{N_{a,L}} \alpha_{i,k} u_{L,k}^+ + \alpha_{i,n} u_{n}^- \\
	&v^*_{L,i} = \lambda_i \left(u^+_{L,i} - \sum_{k=1}^{N_{a,L}} \alpha_{i,k} u_{L,k}^+ - \alpha_{i,n} u_{n}^- \right)
	\end{split}
\label{sol.junction_Li}	
\end{equation}
and for the node side (with the subscript $n$ indicating the variable --or the location, when concerning transmission coefficients-- of the node)
\begin{equation}
\begin{split}
	&u^*_{L,n} =  u_{n}^- + \sum_{k=1}^{N_{a,L}} \alpha_{n,k} u_{L,k}^+ + \alpha_{n,n} u_{n}^-  \\
	&v^*_{L,n} = - \lambda_n \left( u_{n}^- - \sum_{k=1}^{N_{a,L}} \alpha_{n,k} u_{L,k}^+ - \alpha_{n,n} u_{n}^- \right) .
\end{split}
\label{sol.junction_Ln}	
\end{equation}
On the other hand, for the transmission conditions at the interfaces with an outgoing arc ($0,j$), we have for the arcs side
\begin{equation}
\begin{split}
	&u^*_{0,j} =  u_{0,j}^-  + \sum_{k=1}^{N_{a,0}} \alpha_{j,k} u_{0,k}^- + \alpha_{j,n} u_{n}^+  \\
	&v^*_{0,j} = - \lambda_j\left( u_{0,j}^-  - \sum_{k=1}^{N_{a,0}} \alpha_{j,k} u_{0,k}^- - \alpha_{j,n} u_{n}^+\right) .
\end{split}
\label{sol.junction_0j}	
\end{equation}
and for the node side
\begin{equation}
\begin{split}
	&u^*_{0,n} =  u_{n}^+ + \sum_{k=1}^{N_{a,0}} \alpha_{n,k} u_{0,k}^- + \alpha_{n,n} u_{n}^+  \\
	&v^*_{0,n} = \lambda_n \left(u_{n}^+ - \sum_{k=1}^{N_{a,0}} \alpha_{n,k} u_{0,k}^- - \alpha_{n,n} u_{n}^+ \right)
\end{split}
\label{sol.junction_0n}	
\end{equation}
Notice that the condition differs when considering an incoming flow or an outgoing flow, due to the artificial orientation that has been set. Indeed, for each incoming arc, we need to use $u_{L,i}^+$ from the arc and $u_n^-$ from the node; while for each outgoing arc we consider $u_{0,j}^-$ from the arc and $u_n^+$ from the node \cite{bretti2014}. 

Furthermore, to guarantee the conservation of fluxes at the interface, ensuring that the global mass (population) of the system is conserved, the following must hold \cite{bretti2006,bretti2014}
\begin{equation}
v_{L,n}^*=\sum_{i=1}^{N_{a,L}} v_{L,i}^*, \qquad \qquad v_{0,n}^*=\sum_{j=1}^{N_{a,0}} v_{0,j}^* ;
\end{equation}
which is fulfilled imposing at interfaces ($L,i$)
\begin{equation}
	\lambda_i = \sum_{k=1}^{N_{a,L}} \alpha_{k,i} \lambda_k + \alpha_{n,i} \lambda_n , \qquad \qquad
	\lambda_n = \sum_{k=1}^{N_{a,L}} \alpha_{k,n} \lambda_k + \alpha_{n,n} \lambda_n ,
\label{eq.conservationFluxes_L}
\end{equation}
and at interfaces ($0,j$)
\begin{equation}
	\lambda_n = \sum_{k=1}^{N_{a,0}} \alpha_{k,n} \lambda_k + \alpha_{n,n} \lambda_n , \qquad \qquad
	\lambda_j = \sum_{k=1}^{N_{a,0}} \alpha_{k,j} \lambda_k + \alpha_{n,j} \lambda_n .
\label{eq.conservationFluxes_0}
\end{equation}

Nodes located at the inlet (outlet) end of the domain are without any incoming (outgoing) arcs. At these nodes, in order to ensure that there are no individuals entering or leaving the network (thus ensuring the preservation of the total population), we simply enforce the standard no-flux boundary condition \cite{piccoli2006}, which consists of imposing at inlet nodes
\begin{equation}
v^*_{L,n} = 0, 	\qquad \qquad u^*_{L,n} = u_n - \frac{v_{n}}{\lambda_n} ,
\label{eq.zeroflux_inlet}	
\end{equation}
and at outlet nodes
\begin{equation}
v^*_{0,n} = 0, 	\qquad \qquad	u^*_{0,n} = u_n + \frac{v_{n}}{\lambda_n} .
\label{eq.zeroflux_outlet}	
\end{equation}


\section{Numerical results}
\label{sect:numres}
The multiscale transport SEIAR model \eqref{eq.SEIARkinetic}-\eqref{eq.SEIARkinetic_noncommuters} is solved using a second-order IMEX Runge-Kutta Finite Volume Collocation method (see Appendices \ref{appendix:SCM} and \ref{appendix:IMEX} for details). In particular, we will show that the numerical scheme is capable to reach spectral accuracy in the stochastic space, if the solution is sufficiently smooth in that space, and to preserve this accuracy in the diffusive (stiff) limit (i.e. stochastic AP property) \cite{jin2015,bertaglia2021a}. 
An advantage related to the choice of the stochastic collocation method lies in its non-intrusive nature \cite{xiu2005}. This feature ensures ease of implementation, since the method requires only the evaluation of the solution of the corresponding deterministic problem, followed by a post-processing step. Thus, no major manipulation efforts of the deterministic computational code are required and the loss of important structural properties of the original problem is avoided \cite{poette2009}.


\begin{figure}[t!]
\centering
\includegraphics[width=0.5\linewidth]{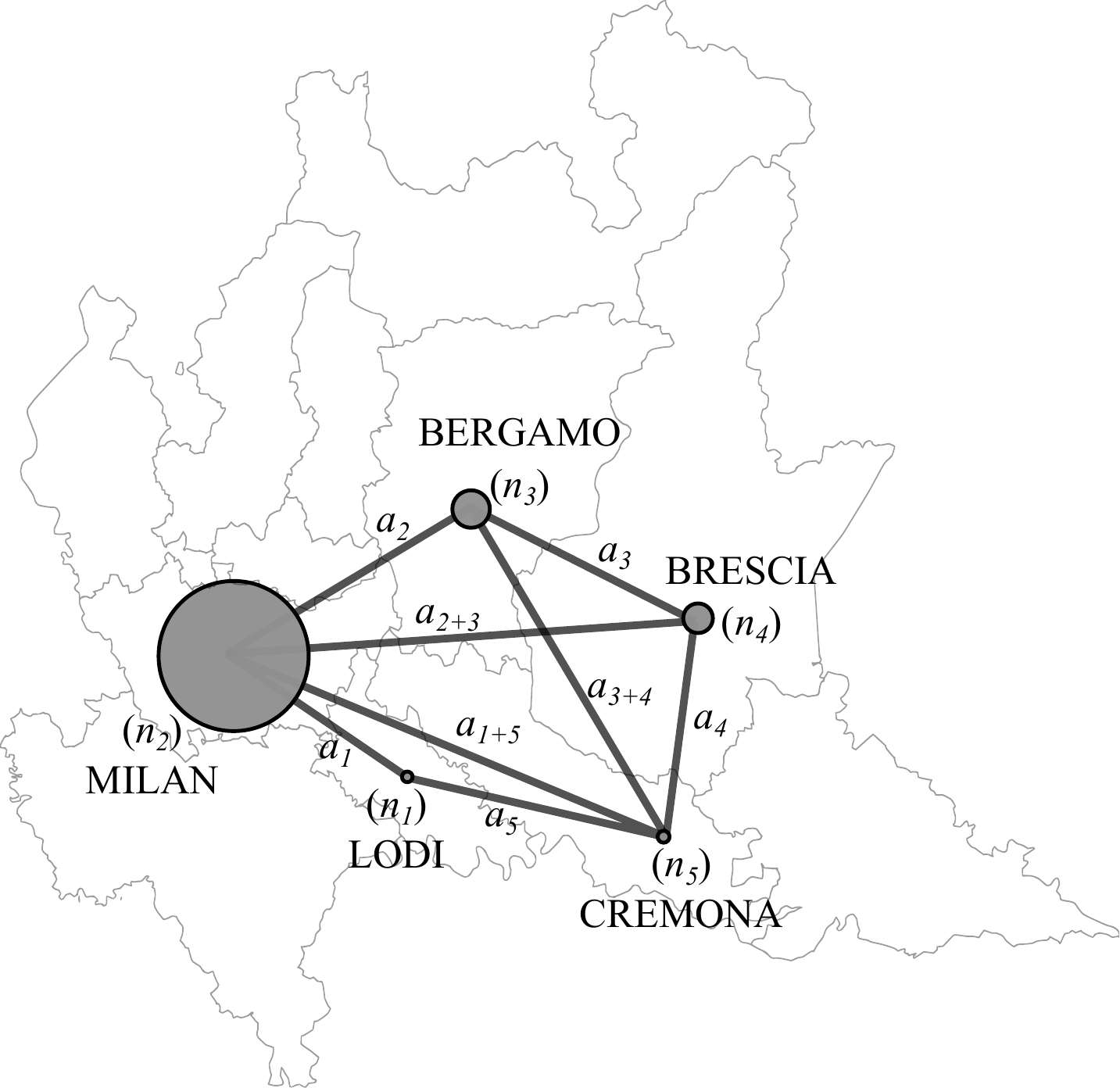}
\caption{Representation of the network of the Lombardy test case, composed of 5 nodes, corresponding to the provinces of interest (Lodi, Milan, Bergamo, Brescia and Cremona) and 5 arcs, connecting each city to the others, considering all the main paths of commuters. The dimension of the node is proportional to the dimension of the urbanized area of the province.}
\label{fig.network_lombardia}
\end{figure}
\begin{table}[b!] 
\caption{Matrix of commuters among the provinces of the Lombardy Region (Italy). Departure Provinces are listed on the first left column, while arrival Provinces are reported in the following columns (LO=Lodi, MI=Milan, BG=Bergamo, BS=Brescia, CR=Cremona). Each entry is given as number of people and percentage of individuals with respect to the total amount of commuters of the origin Province. The last column shows the amount of total commuters of the origin Province and corresponding percentage with respect to the total population of the origin Province. This matrix is extracted from the origin-destination matrix provided by the Lombardy Region for the regional fluxes of year 2020 \cite{opendataLombardia}.} \label{tab:matrix_network_lombardia} 
\centering
\vspace{0.2cm}
\begin{tabular}{ c | c c c c c || c}
From/To &LO & MI &BG &BS &CR &Total commuters\\
\hline
\multirow{2}{*}{LO} &-- &56717 &--
  &--
  &13712 &70429 \\
  &&(80.53\%) &&&(19.47\%) &(30.97\%)\\
  \hline
\multirow{2}{*}{MI} &55397 &-- &74168  &26709  &21622 &177896 \\
&(31.14\%)&&(41.69\%)&(15.01\%)&(12.15\%) &(5.45\%)\\
\hline
\multirow{2}{*}{BG} &--
 &76337  &-- &78348  &12016  &166701\\
 &&(45.79\%)&&(47.00\%)&(7.21\%)& (15.04\%)\\
 \hline
\multirow{2}{*}{BS} &--
 &26594  &70879  &-- &16967  &114440 \\
 &&(23.24\%)&(61.93\%)&&(14.83\%)&(9.12\%)\\
 \hline
\multirow{2}{*}{CR} &13264  &23142  &12025  &17681  &-- &66112 \\
&(20.06\%)&(35.00\%)&(18.19\%)&(26.75\%)&&(18.58\%)\\
\hline
\end{tabular} 
\end{table} 
\subsection{Application to the emergence of COVID-19 in Italy}
\label{sect.LombardiaTest}
To analyze the effectiveness of the proposed approach in a realistic geographical and epidemic scenario, we design a numerical setting reproducing the evolution of the first outbreak of COVID-19 in the Lombardy Region of Italy, from February 27, 2020 to March 27, 2020, with respect to uncertainties underlying the initial conditions and chosen epidemic parameters. 

\subsubsection{The Lombardy network}
A five-node network is considered, whose nodes represent the 5 main provinces interested by the epidemic outbreak in the first months of 2020: Lodi ($n_1$), Milan ($n_2$), Bergamo ($n_3$), Brescia ($n_4$) and Cremona ($n_5$). The arcs $a_{j}$ connecting each node to the others identify the main set of routes and railways viable by commuters each day. 
A schematic representation of this network is shown in Fig. \ref{fig.network_lombardia}. Routes that connect cities that are not direct neighbors and that require crossing other provinces are to be considered as the sum of the two route sections. Thus, for instance, it can be seen from Fig. \ref{fig.network_lombardia} that the Milan--Brescia connection is actually identified by the sum of the Milan--Bergamo ($a_2$) and Bergamo--Brescia ($a_3$) arcs, indeed indicated as $a_{2+3}$.
Since the space unit of measurement adopted is $1\, \mathrm{L} = 10^2\, \mathrm{km}$, the lengths of the single sections of arc result: $L_{a_1} = 0.20$, $L_{a_2} = 0.30$, $L_{a_3} = 0.35$, $L_{a_4} = L_{a_5} = 0.40\,$. These arcs are discretized with a grid size $\Delta x_a = 0.01$. The length of the spatial cell associated to each node is proportional to the dimension of the urbanized area of the corresponding province, hence: $\Delta x_{n_1} = 0.025$, $\Delta x_{n_2} = 0.420$, $\Delta x_{n_3} = 0.100$, $\Delta x_{n_4} = 0.085$, $\Delta x_{n_5} = 0.035$.

Transmission coefficients $\alpha_{i,j}$, as well as percentage of commuters belonging to each province, are imposed recurring to official national assessment of mobility flow. In particular, the matrix of commuters presented in Table \ref{tab:matrix_network_lombardia} reflects mobility data provided by the Lombardy Region for the regional fluxes of year 2020 \cite{opendataLombardia}, which is in agreement with the one derived from ISTAT data released in October, 2011 \cite{ISTATpendolarismo}, as also confirmed in \cite{gatto2020}.
Notice that connections Lodi--Bergamo and Lodi--Brescia are not taken into account (as observable also from Fig.  \ref{fig.network_lombardia}) because the amount of commuters along these routes is very low compared to the amount of individuals traveling the rest of the routes.

\begin{table}[b!] 
\caption{Lombardy test case data: total inhabitants of the province N and detected infectious individuals $i_0$ on February 27, 2020 in Lombardy  Region (Italy). The total population is given by ISTAT data of December 31, 2019 \cite{ISTATdemo}, while data of infectious correspond to those reported in the GitHub repository daily updated by the Civil Protection Department of Italy \cite{github_covid}.} \label{tab:IC_network_lombardia} 
\centering
\vspace{0.2cm}
\begin{tabular}{ c | c c }
City &Total population (N) & Infectious ($i_0$) \\
\hline
Lodi ($n_1$) &227412 &159 \\
Milan ($n_2$) &3265327 &15\\
Bergamo ($n_3$) &1108126 &72 \\
Brescia ($n_4$) &1255437 &10 \\
Cremona ($n_5$) &355908 &91 \\
\hline
\end{tabular} 
\end{table} 

The characteristic speed associated to each arc is fixed to permit a full round trip in each origin-destination section within a day. Since the time unit of measurement adopted is $1\, \mathrm{T} = 1\, \mathrm{day}$, for all the compartments we impose: $\lambda_{a_1}=0.4$, $\lambda_{a_2}=0.6$, $\lambda_{a_3}=0.7$, $\lambda_{a_4}=\lambda_{a_5}=0.8$, respectively for each arc. 
The characteristic speed of compartment $I$ is fixed to zero in all the nodes of the network, $\lambda_n = 0$, while for the rest of the compartments, namely $S,E,A,R$, $\lambda_n = 1\,$. In the arcs, the relaxation time is assigned so that the model recovers a hyperbolic regime, hence $\tau_{r,a} =10^3$. On the other hand, a parabolic setting is prescribed in the cities for commuters in order to correctly capture the diffusive behavior of the disease spread which typically occurs in highly urbanized zones, with $\tau_{r,n}=10^{-3}$. The reader is invited to refer to \cite{bertaglia2021} for further details on the sensitivity of the model to relaxation and speed parameters.

\subsubsection{Data fitting and uncertainty}
We simulate the spread of COVID-19 starting from February 27, 2020 until March 27, 2020, included.
At the beginning of the pandemic, tracking of positive individuals cannot be considered reliable. Thus, the initial amount of infected people is the first information affected by uncertainty. To this aim, we introduce a single source of uncertainty $z$ having uniform distribution, $z \sim \mathcal{U}(0, 1)$ and the initial conditions for compartment $I$, at each node, are prescribed as
\begin{equation*}
I(x,0,z) = i_0(1+\mu_1 z)\,,
\end{equation*}
with $i_0$ density of infectious people on February 27, 2020, as given by data recorded by the Civil Protection Department of Italy \cite{github_covid}, reported in Table \ref{tab:IC_network_lombardia} for each node, together with the total inhabitants of each province given by ISTAT data of 2019 \cite{ISTATdemo}. 
We chose to associate all infected persons detected to the $I$ compartment. This choice is justified by the ongoing screening policy. Indeed, during the first wave of this pandemic in Italy, tests to assess the presence of SARS-CoV-2 virus were performed almost only on patients with consistent symptoms and fever. We select $\mu_1=1$, assuming no more than half of the actual highly symptomatic infected were detected at the very beginning of the outbreak of the pandemic.

According to values used in \cite{gatto2020,buonomo2020}, we fix $\gamma_I = 1/14$, $\gamma_A = 2\gamma_I = 1/7$, $a = 1/3$, considering these clinical parameters deterministic. We also assume the probability rate of developing severe symptoms $\sigma = 1/12.5$, as in \cite{buonomo2020,kantner2020}. Since the first day of the simulation, the public was aware of the epidemic outbreak and recommendations such as washing hands often, not touching the face, avoiding handshakes and keeping distance had already been disseminated. Therefore, we initially set coefficients $k_I=k_A=30$.
The initial value of the transmission rate of asymptomatic/mildly infectious people is calibrated as the result of a least square problem, namely the L2 norm of the difference between the observed cumulative number of infected $I(t)$ and the numerical evolution of the same compartment, through a simple deterministic SEIAR ODE model set up for the whole Lombardy Region, with the result $\beta_A = 0.545$. Since after February 23, 2020, Codogno, city in the province of Lodi, is put under strict lockdown as ``red zone'' \cite{gatto2020}, a reduced contact rate is considered for the node of Lodi, with $\beta_A = 0.50$. In the above fitting, the expected initial amount of exposed and asymptomatic/mildly symptomatic individuals is also estimated imposing an initial condition of one single exposed individual, obtaining $e_0 \approx 10.23\, i_0$ and $a_0 \approx 9.15\, i_0$. In particular this permits to have a rough estimate of the presumed day of outbreak of the epidemic which, in our case, would appear to have started on January 14, 2020. 
With the above setup, we obtain an initial expected value of the basic reproduction number in the whole network (as from definition given in Section \ref{sect:R0}) $R_0~=~3.6$, which is in agreement with estimations reported in \cite{gatto2020,buonomo2020,vollmer2020}.

On the other hand, the transmission rate of compartment $I$, $\beta_I$, is considered a random parameter, given that, at each node, the initial amount of severe symptomatic subjects $I$ is affected by uncertainty, as previously discussed. Therefore we impose:
\begin{equation*}
\beta_I(0,z) = \beta_{I,0}(1+\mu_1\mu_2 z)\, .
\end{equation*}
Assuming that highly infectious subjects are mostly detected in the most optimistic scenario, being subsequently quarantined or hospitalized, the minimum value of the transmission rate of $I$ is set $\beta_{I,0} = 0.03\,\beta_A$, as in \cite{gatto2020,buonomo2020}. Furthermore, $\mu_2 = 0.06^{-1}$, indicating that the more the number of infected in $I$ increases relative to the observed value $i_0$, the more the transmission rate of this compartment tends to that of compartment $A$, proportionally to the error in $I$. 

As a consequence, also initial conditions for compartments $E$, $A$ and $S$ are stochastic, depending on the initial amount of severe infectious at each location:
\[E(x,0,z) = 10.23\,I(x,0,z) \, , \qquad A(x,0,z) = 9.15\,I(x,0,z)\, , \]
\[ S(x,0,z) = N(x) - E(x,0,z) - I(x,0,z) - A(x,0,z)\,.\]
Finally, removed individuals are considered initially null everywhere in the network, $R(x,0,z)=0.0$, with all arcs assumed empty at the beginning of the simulation. 

To model the escalation of lockdown restrictions, starting from March 9, 2020, initial day of the northern Italy lockdown, the transmission rate $\beta_A$ is reduced by 15\% (with a consequent update also of $\beta_{I,0}$ and therefore of $\beta_I$) and coefficients $k_I = k_A = 60$, due to the public being increasingly aware of the epidemic risks. Furthermore, the percentage of commuting individuals is reduced by 60\% in the whole network according to mobility data tracked through mobile phones and made available by Google \cite{aktay2020,vollmer2020}. After the additional restrictions in place as of March 22, 2020, $\beta_A$ is ulteriorly reduced by 10\% (again with a consequent rearrangement also of $\beta_{I,0}$ and $\beta_I$) and coefficients $k$ are increased to $k_I = k_A = 70$.

\begin{table}[b!]
\small
\caption{Error estimates and empirical order of accuracy of expectation $\mathbb{E}$ and variance $\mathbb{V}$ of susceptible $S$ and highly symptomatic individuals $I$ in the Lombardy network test, at node $n_1$ (Lodi). Results obtained applying the sAP-IMEX Runge-Kutta FV Collocation method. $M_p$ indicates the number of points used for the stochastic Collocation method.}
\label{tab:convergence_n1} 
\begin{tabular}{l | c c | c c | c c | c c }
\hline
\multirow{2}{*}{$M_p$}&\multicolumn{2}{c|}{$\mathbb{E}[S]$} &\multicolumn{2}{c|}{$\mathbb{V}[S]$}
&\multicolumn{2}{c|}{$\mathbb{E}[I]$} &\multicolumn{2}{c}{$\mathbb{V}[I]$}\\
\cline{2 - 9}
& $L^2$ & $\mathcal{O}\left( L^2\right)$ & $L^2$ & $\mathcal{O}\left( L^2\right)$ & $L^2$ & $\mathcal{O}\left( L^2\right)$ & $L^2$ & $\mathcal{O}\left( L^2\right)$ \\
\hline
      2 & 4.1539e-07 &         & 1.3499e-08 &          & 2.0021e-08 &         & 1.2599e-11 &         \\ 
      4 & 6.1458e-08 &  2.7568 & 1.9974e-09 &  2.7567 & 1.7945e-09 &  3.4799 & 1.1296e-12 &  3.4794 \\ 
      6 & 2.5542e-09 &  7.8443 & 8.3016e-11 &  7.8443 & 8.4784e-11 &  7.5281 & 5.3366e-14 &  7.5282 \\ 
      8 & 1.2694e-10 & 10.4343 & 4.1258e-12 & 10.4343 & 4.2423e-12 & 10.4108 & 2.6703e-15 & 10.4108 \\ 
     10 & 6.7564e-12 & 13.1451 & 2.1959e-13 & 13.1451 & 6.8746e-14 & 18.4744 & 4.3271e-17 & 18.4744 \\ 
\hline 
\end{tabular} 
\end{table} 
\begin{table}[b!]
\small
\caption{Error estimates and empirical order of accuracy of expectation $\mathbb{E}$ and variance $\mathbb{V}$ of susceptible $S$ and asymptomatic/mildly symptomatic individuals $A$ in the Lombardy network test, at node $n_3$ (Bergamo). Results obtained applying the sAP-IMEX Runge-Kutta FV Collocation method. $M_p$ indicates the number of points used for the stochastic Collocation method.}
\label{tab:convergence_n3} 
\begin{tabular}{l | c c | c c | c c | c c}
\hline
\multirow{2}{*}{$M_p$}&\multicolumn{2}{c|}{$\mathbb{E}[S]$} &\multicolumn{2}{c|}{$\mathbb{V}[S]$}
&\multicolumn{2}{c|}{$\mathbb{E}[A]$} &\multicolumn{2}{c}{$\mathbb{V}[A]$}\\
\cline{2 - 9}
& $L^2$ & $\mathcal{O}\left( L^2\right)$ & $L^2$ & $\mathcal{O}\left( L^2\right)$ & $L^2$ & $\mathcal{O}\left( L^2\right)$ & $L^2$ & $\mathcal{O}\left( L^2\right)$ \\
\hline
      2 & 4.8912e-06 &         & 1.0543e-06 &         & 2.1399e-06 &         & 4.8133e-08 &         \\ 
      4 & 3.8079e-07 &  3.6831 & 8.2083e-08 &  3.6830 & 7.5528e-08 &  4.8244 & 1.6996e-09 &  4.8237 \\ 
      6 & 1.7441e-08 &  7.6047 & 3.7596e-09 &  7.6047 & 4.2088e-09 &  7.1210 & 9.4709e-11 &  7.1211 \\ 
      8 & 9.1663e-10 & 10.2400 & 1.9759e-10 & 10.2400 & 2.4765e-10 &  9.8474 & 5.5728e-12 &  9.8474 \\ 
     10 & 4.3175e-11 & 13.6927 & 9.3069e-12 & 13.6927 & 1.4533e-11 & 12.7075 & 3.2703e-13 & 12.7075 \\ 
\hline 
\end{tabular} 
\end{table} 
\begin{figure}[t!]
\centering
\begin{subfigure}{0.49\textwidth}
\includegraphics[width=1\linewidth]{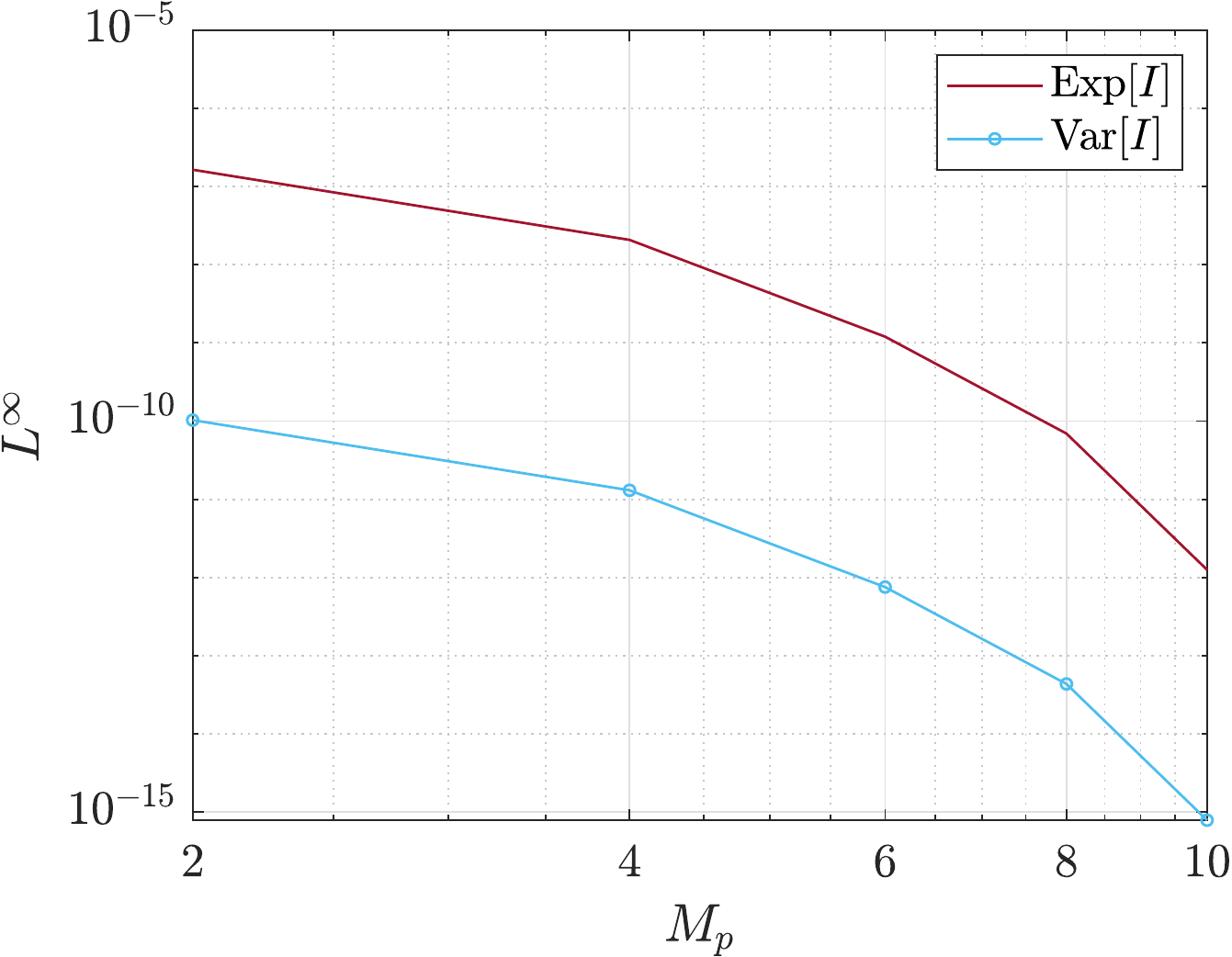}
\caption{Lodi ($n_1$)}
\label{fig.convergence_n1}
\end{subfigure}
\begin{subfigure}{0.49\textwidth}
\includegraphics[width=1\linewidth]{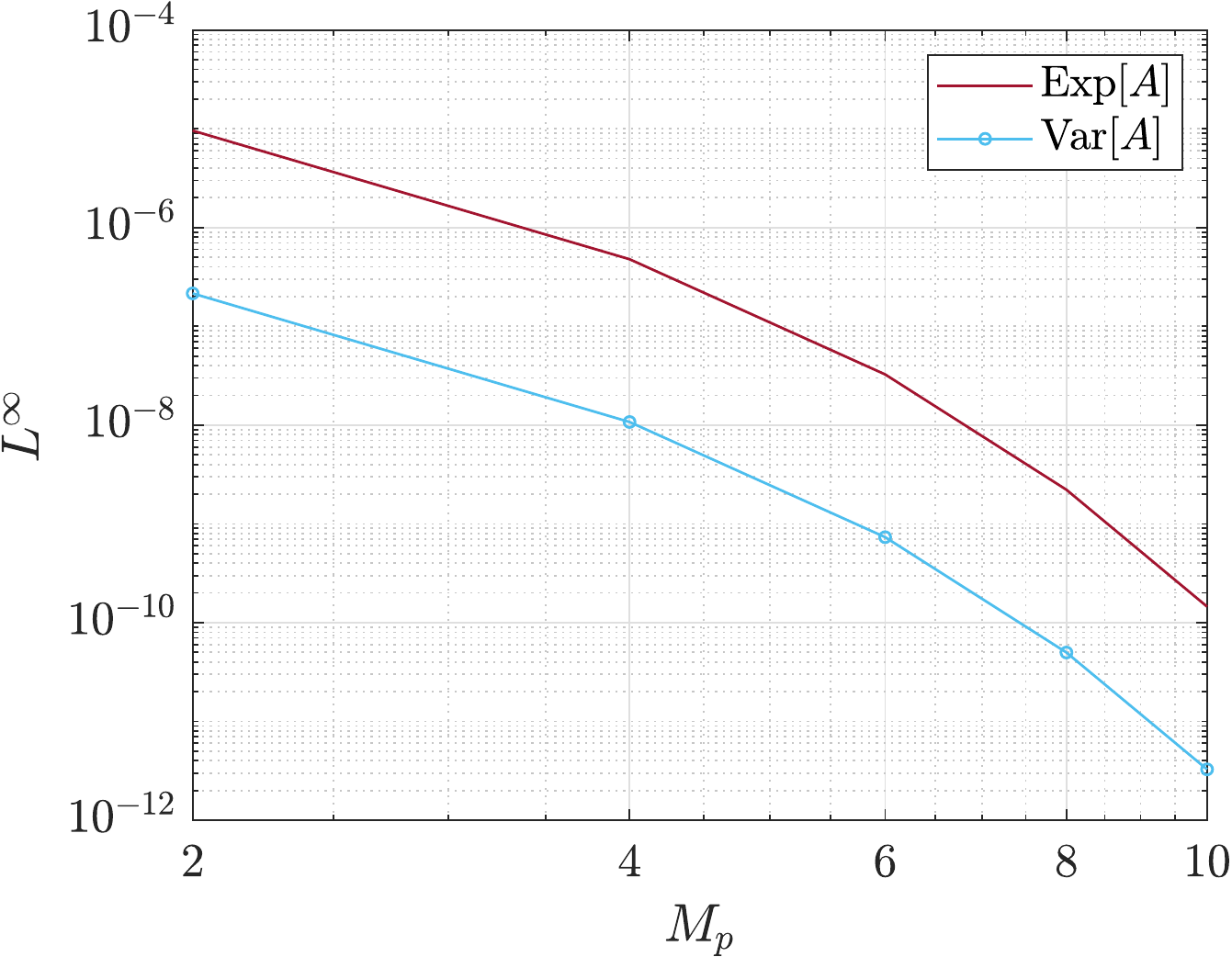}
\caption{Bergamo ($n_3$)}
\label{fig.convergence_n3}
\end{subfigure}
\caption{Exponential decay of $L^{\infty}$ norm of highly infectious $I$ at node $n_1$ (left) and of asymptomatic/weakly symptomatic individuals at node $n_3$ (right) in the Lombardy network test, as the number of collocation points $M_p$ increases.}
\label{fig.convergence}
\end{figure}

\subsubsection{Convergence analysis}
As a first numerical test we analyze the numerical convergence of the stochastic Collocation method, discussed in details in Appendix \ref{appendix:SCM}. We consider the above setting, but with a larger level of uncertainty, with $\mu_1 = 10$, to emphasize the convergence rates. The numerical results evaluated with an increased number of collocation points $M_p$ are compared to a reference solution obtained using $M_p = 50$, in terms of expectation and variance of the state variables. The expected exponential convergence is shown in Tables \ref{tab:convergence_n1}--\ref{tab:convergence_n3} for chosen state variables, respectively for nodes $n_1$ and $n_3$ (taken as representative nodes), where $L^2$ error norms and the related order of accuracy are presented. The result is highlighted by Fig. \ref{fig.convergence}, in which the spectral decay of the $L^{\infty}$ norm is observed in terms of both expected value and variance as the number of collocation points increases.
\begin{figure}[p!]
\centering
\begin{subfigure}{0.37\textwidth}
\includegraphics[width=1\linewidth]{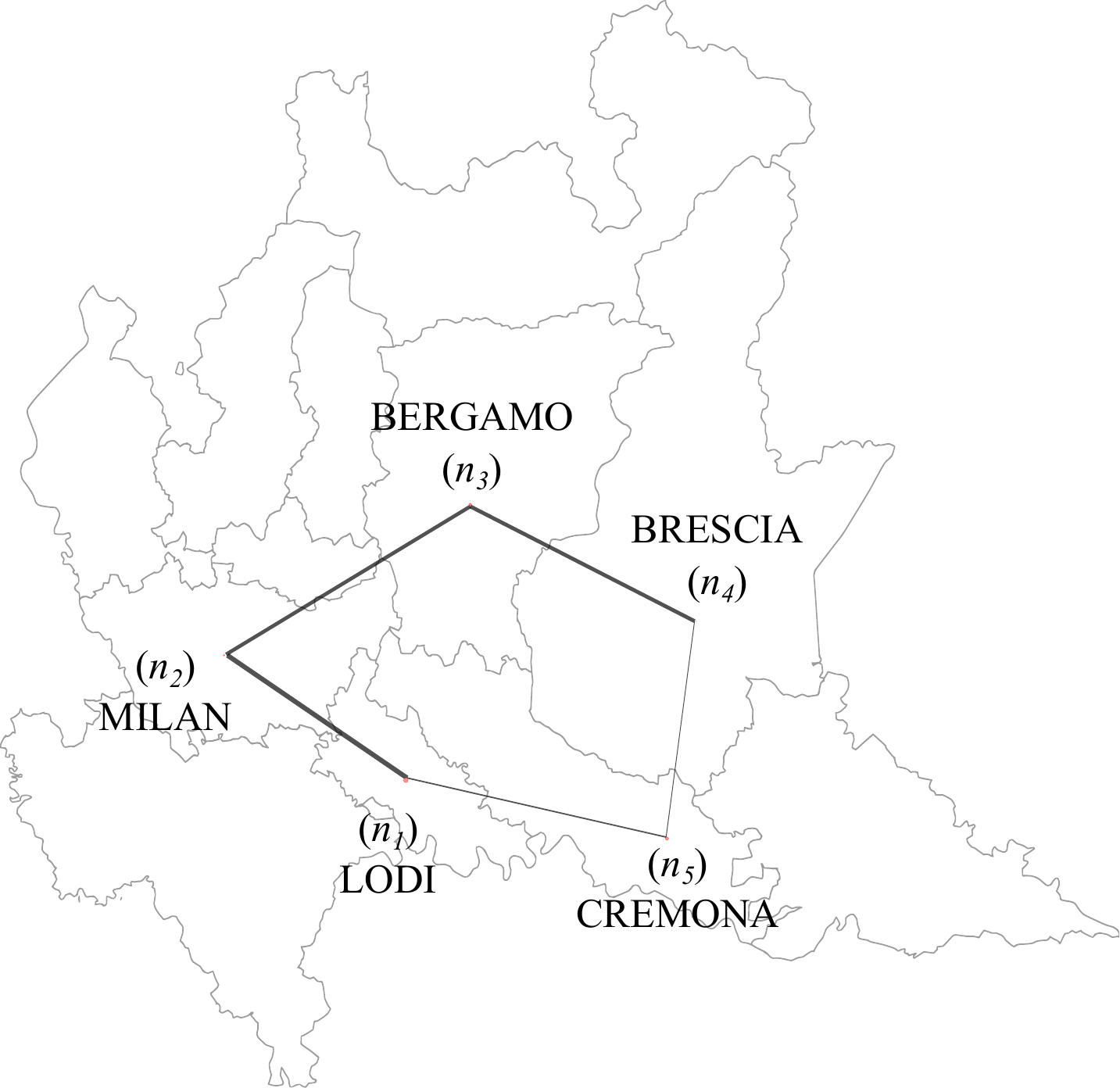}
\caption{March 2, 2020}
\label{fig.lombardia_results_t5}
\end{subfigure}
\hspace{1cm}
\begin{subfigure}{0.37\textwidth}
\includegraphics[width=1\linewidth]{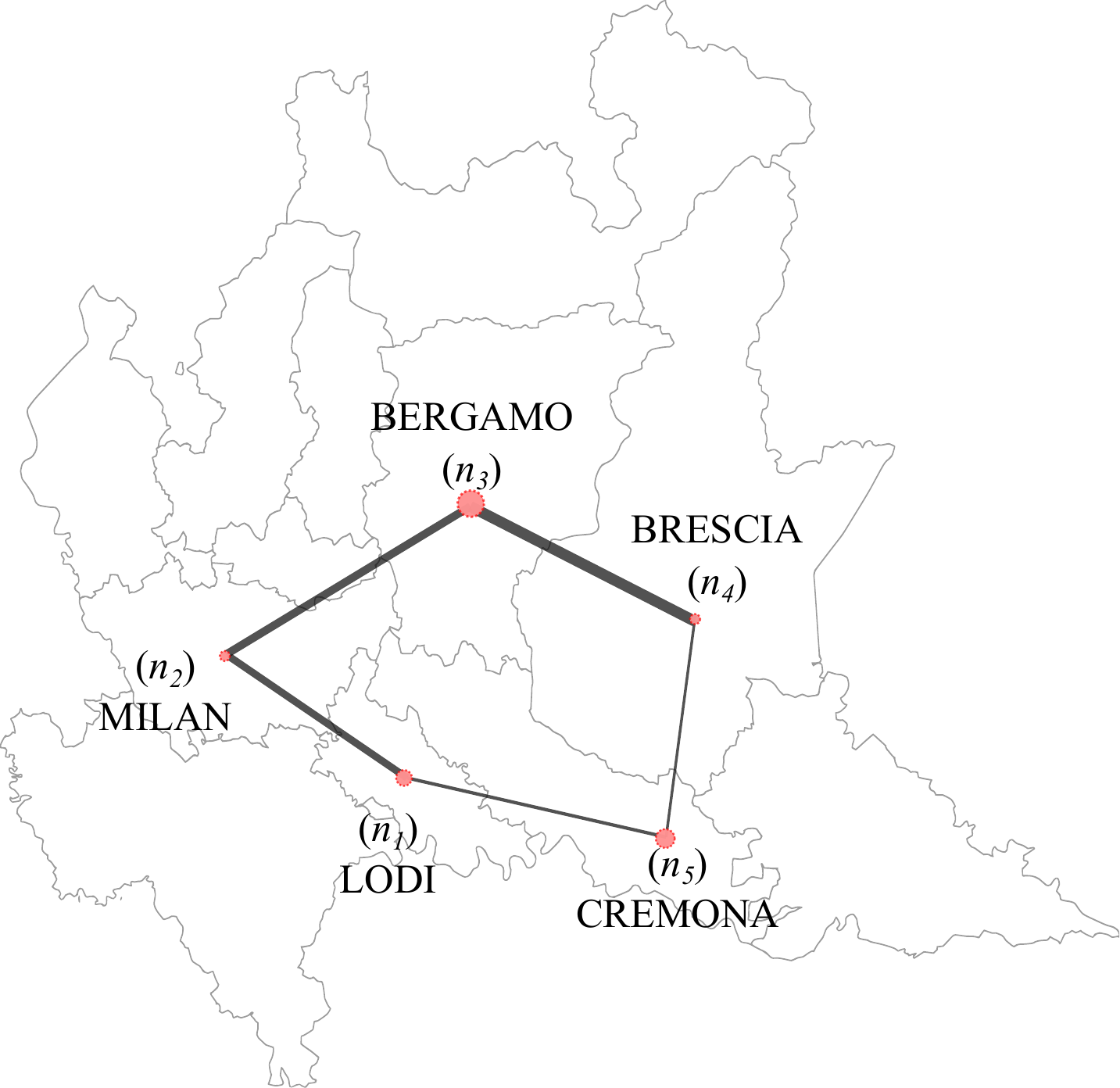}
\caption{March 7, 2020}
\label{fig.lombardia_results_t10}
\end{subfigure}
\begin{subfigure}{0.37\textwidth}
\includegraphics[width=1\linewidth]{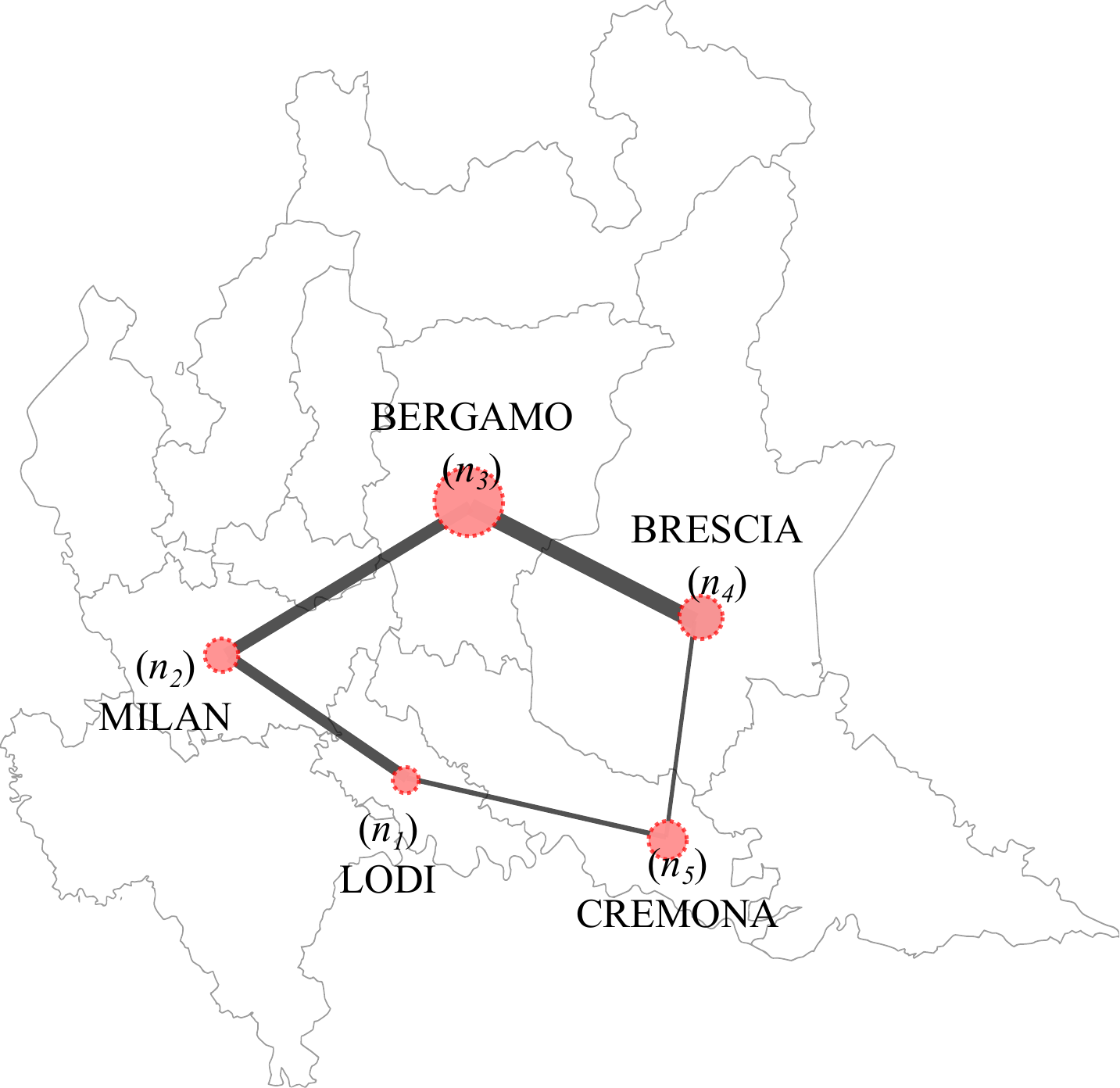}
\caption{March 12, 2020}
\label{fig.lombardia_results_t15}
\end{subfigure}
\hspace{1cm}
\begin{subfigure}{0.37\textwidth}
\includegraphics[width=1\linewidth]{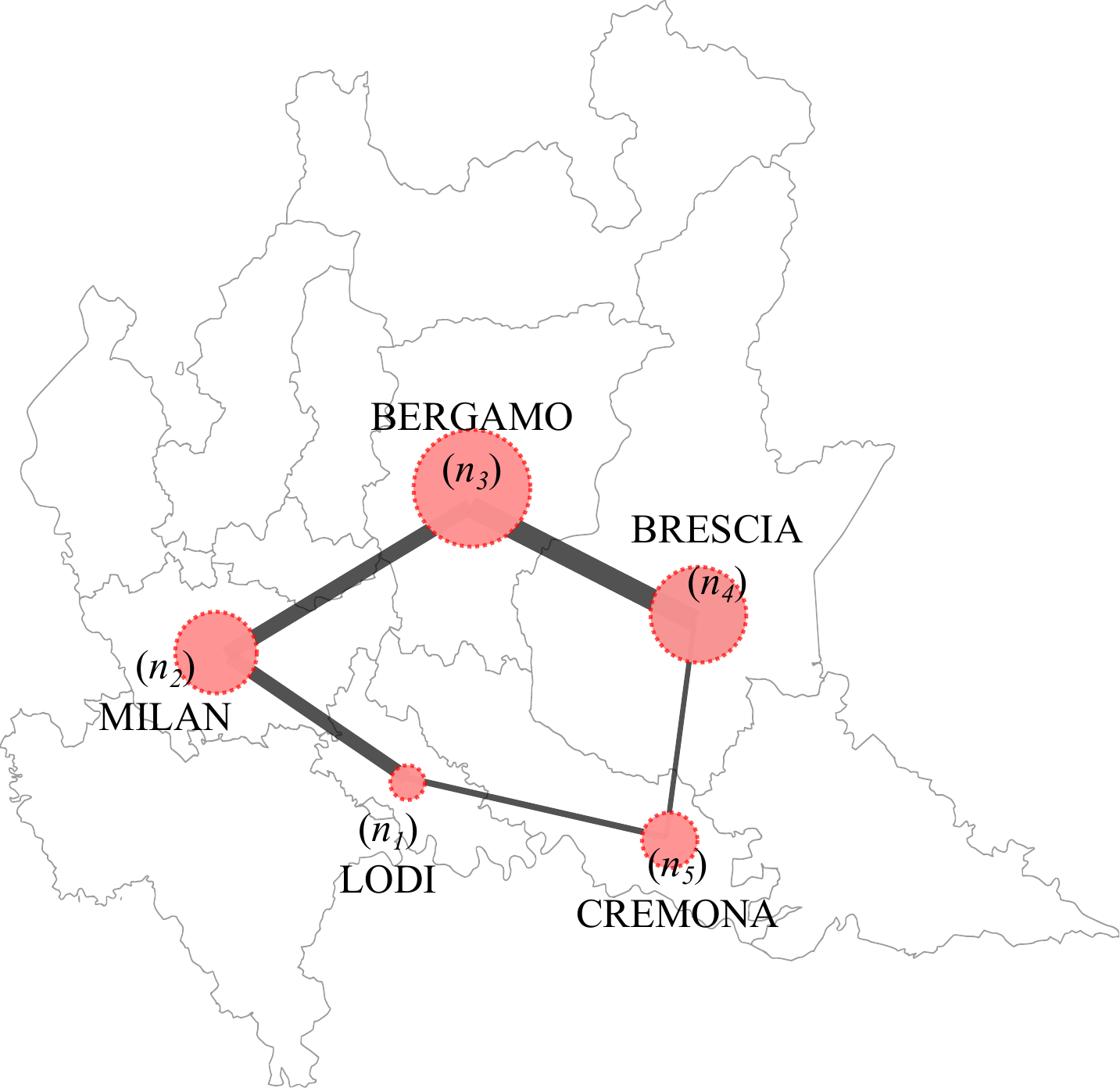}
\caption{March 17, 2020}
\label{fig.lombardia_results_t20}
\end{subfigure}
\begin{subfigure}{0.37\textwidth}
\includegraphics[width=1\linewidth]{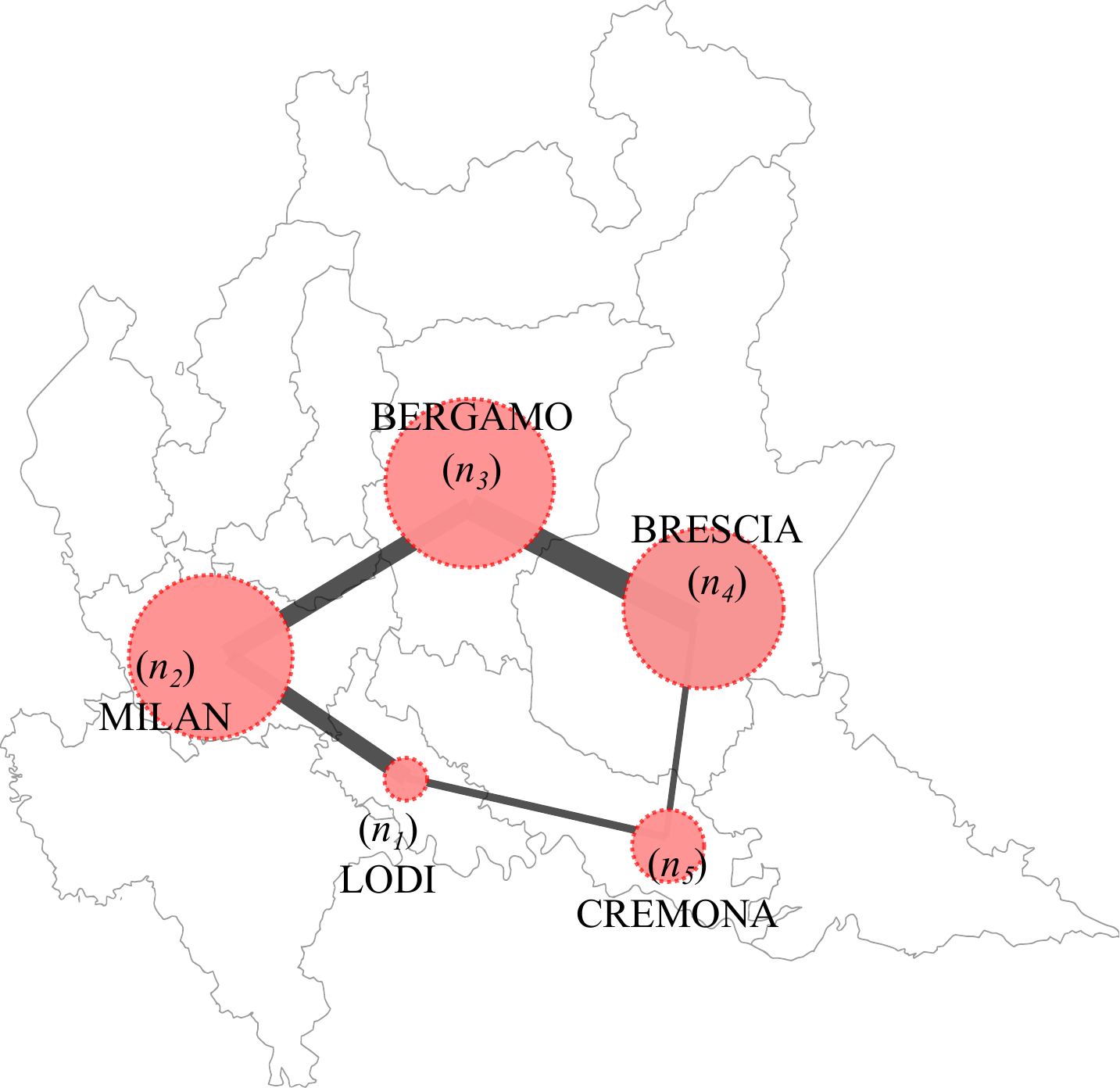}
\caption{March 22, 2020}
\label{fig.lombardia_results_t25}
\end{subfigure}
\hspace{1cm}
\begin{subfigure}{0.37\textwidth}
\includegraphics[width=1\linewidth]{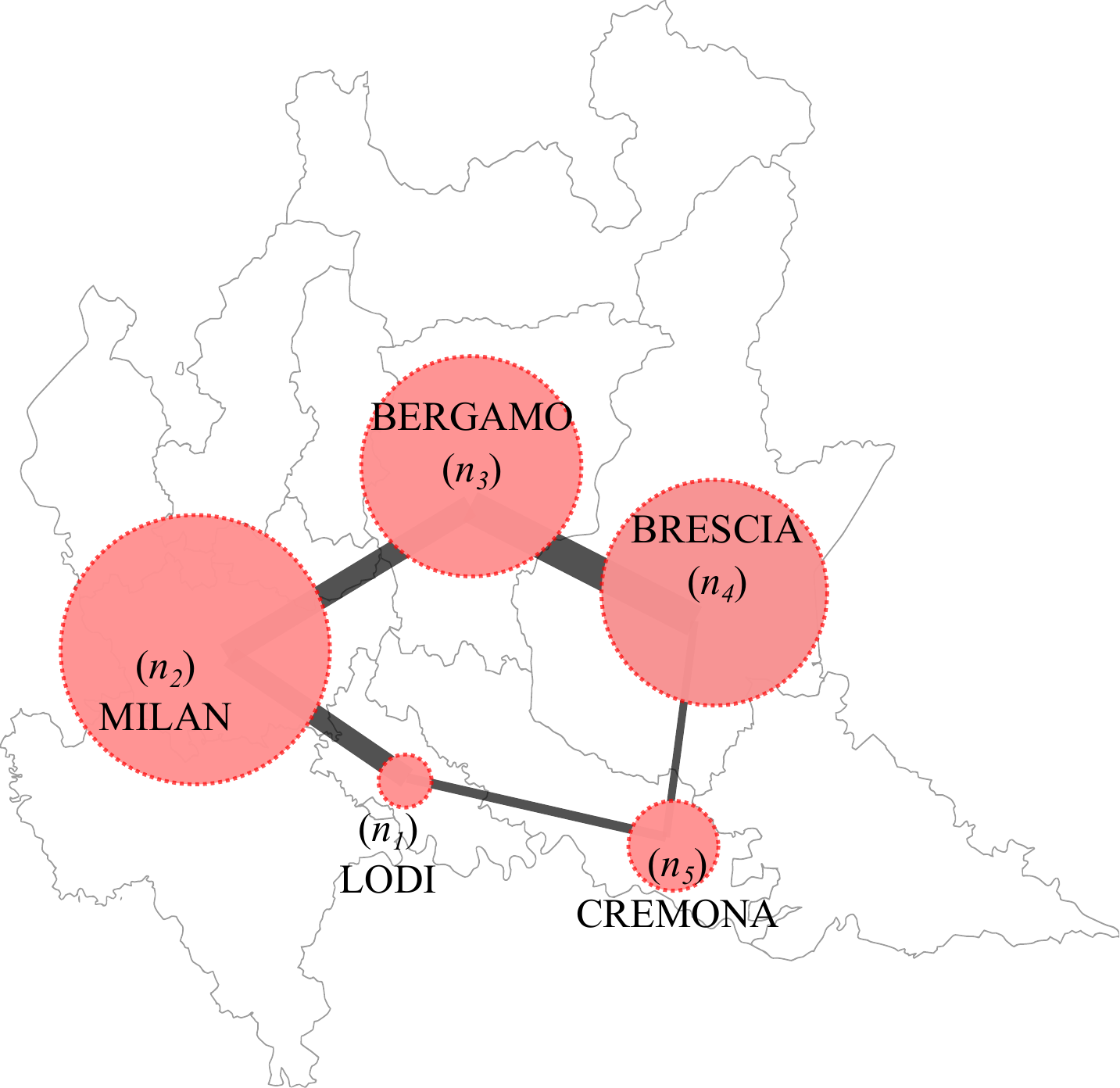}
\caption{March 27, 2020}
\label{fig.lombardia_results_t30}
\end{subfigure}
\caption{Expected temporal and spatial trend of the spread of COVID-19 in Lombardy (Italy) during the first wave of the virus. The radius of the nodes in the network and the width of the arcs is proportional to the expected cumulative amount of total infectious people in the location, including asymptomatic and mildly symptomatic individuals ($I+A+R$).}
\label{fig.lombardia_results}
\end{figure}

\subsection{Simulation of different scenarios of the spread of COVID-19}
In the following, we consider several scenarios regarding the spread of the COVID-19 epidemic in the Lombardy region based on the data and parameters determined in the previous section. In particular, we will consider the baseline scenario, corresponding to the actual spread of the epidemic observed from the data, and two hypothetical scenarios corresponding to the absence of long-range mobility and the absence of restrictions.

\subsubsection{Baseline scenario}
\begin{figure}[p!]
\centering
\begin{subfigure}{0.45\textwidth}
\includegraphics[width=1\linewidth]{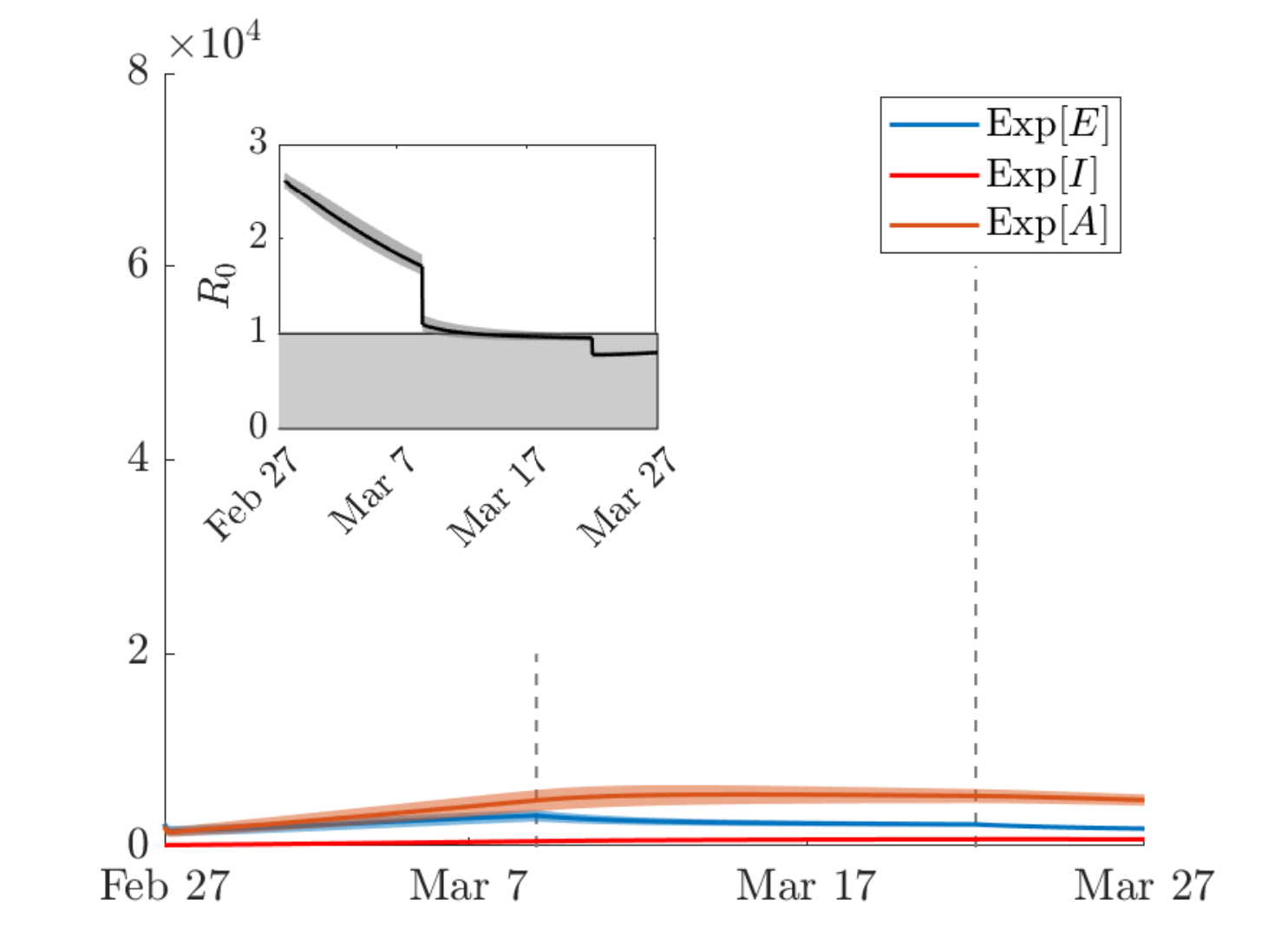}
\caption{Lodi ($n_1$)}
\label{fig.node1}
\end{subfigure}
\hspace{0.5 cm}
\begin{subfigure}{0.45\textwidth}
\includegraphics[width=1\linewidth]{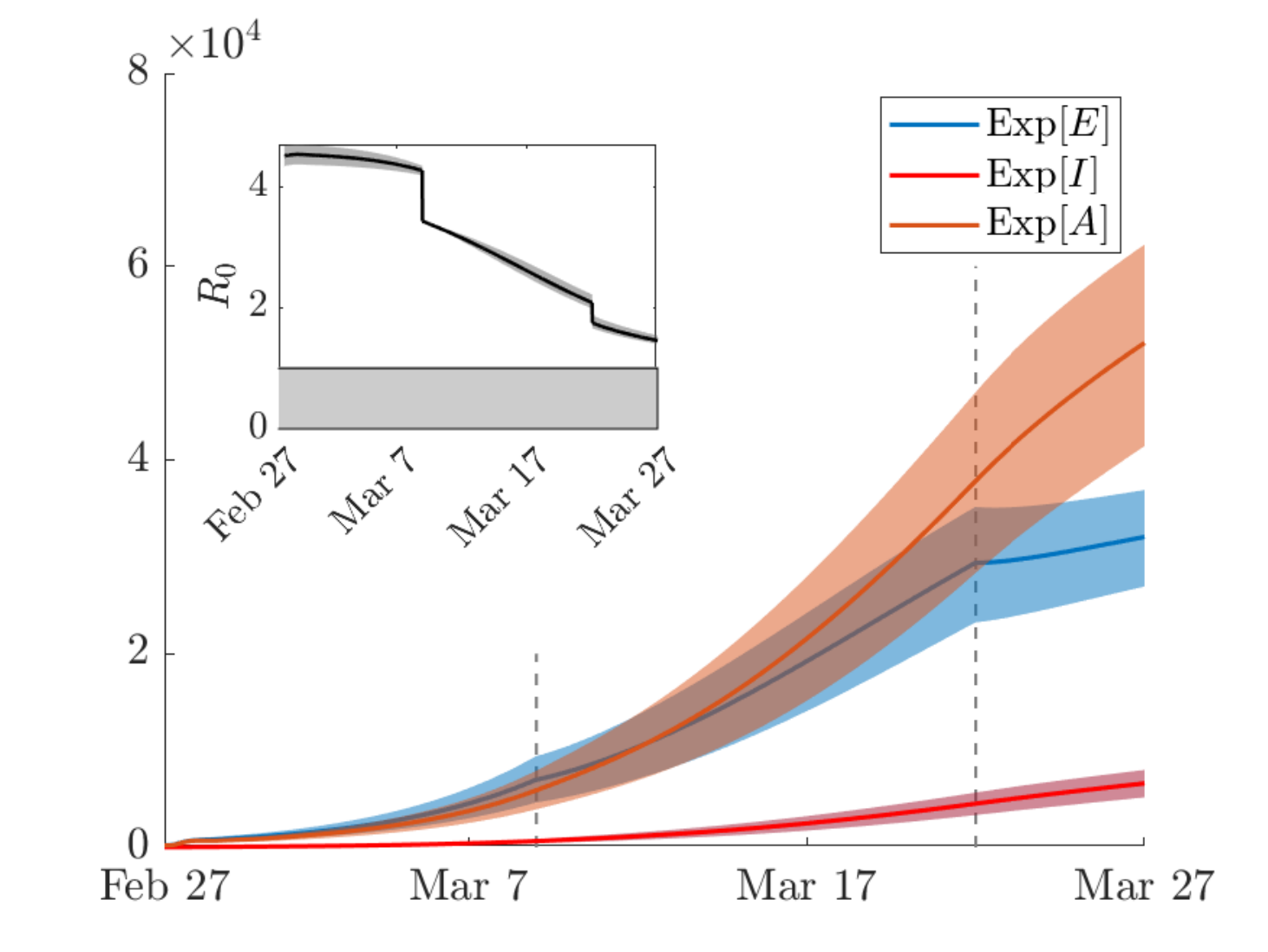}
\caption{Milan ($n_2$)}
\label{fig.node2}
\end{subfigure}
\begin{subfigure}{0.45\textwidth}
\includegraphics[width=1\linewidth]{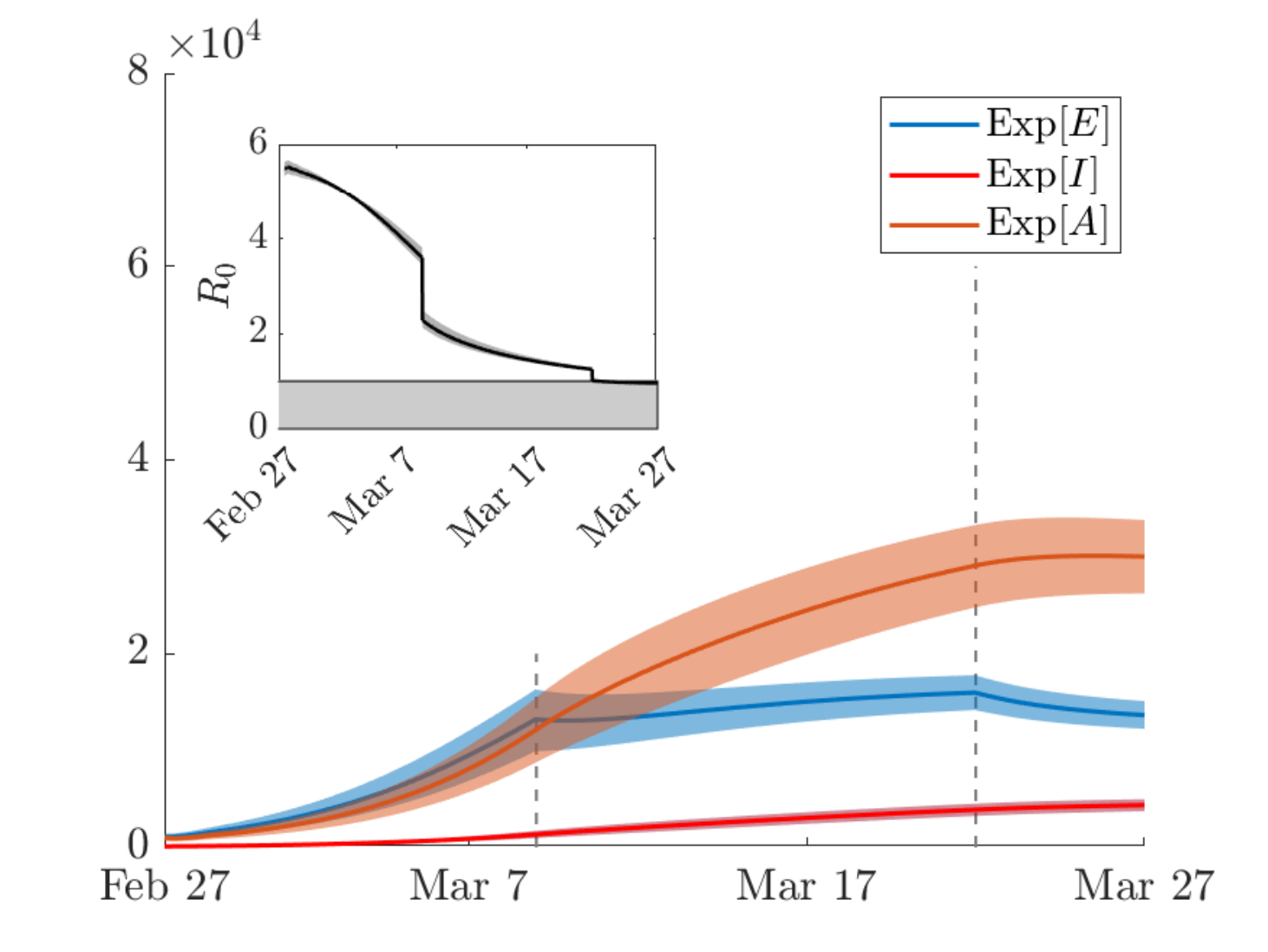}
\caption{Bergamo ($n_3$)}
\label{fig.node3}
\end{subfigure}
\hspace{0.5 cm}
\begin{subfigure}{0.45\textwidth}
\includegraphics[width=1\linewidth]{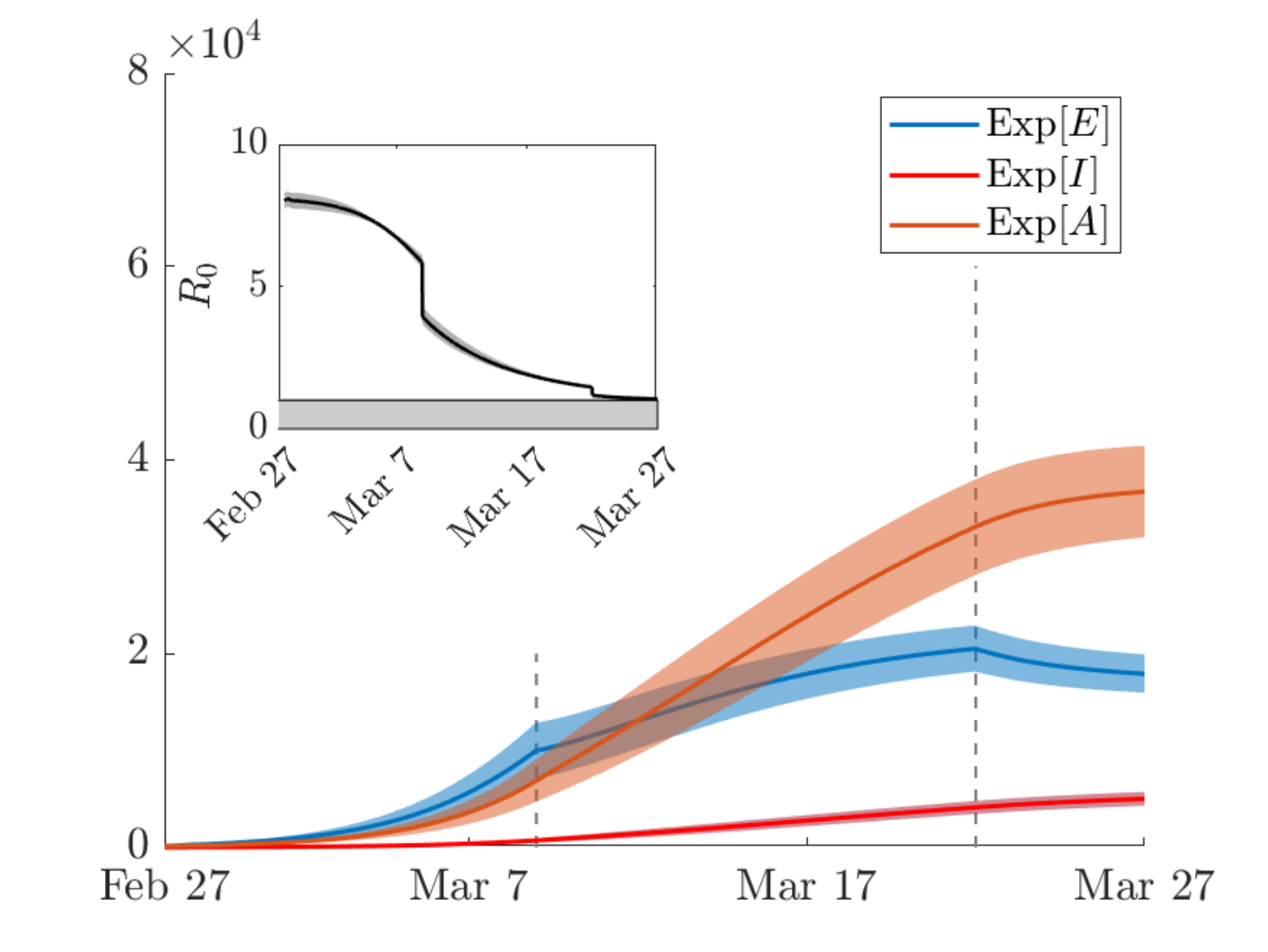}
\caption{Brescia ($n_4$)}
\label{fig.node4}
\end{subfigure}
\begin{subfigure}{0.45\textwidth}
\includegraphics[width=1\linewidth]{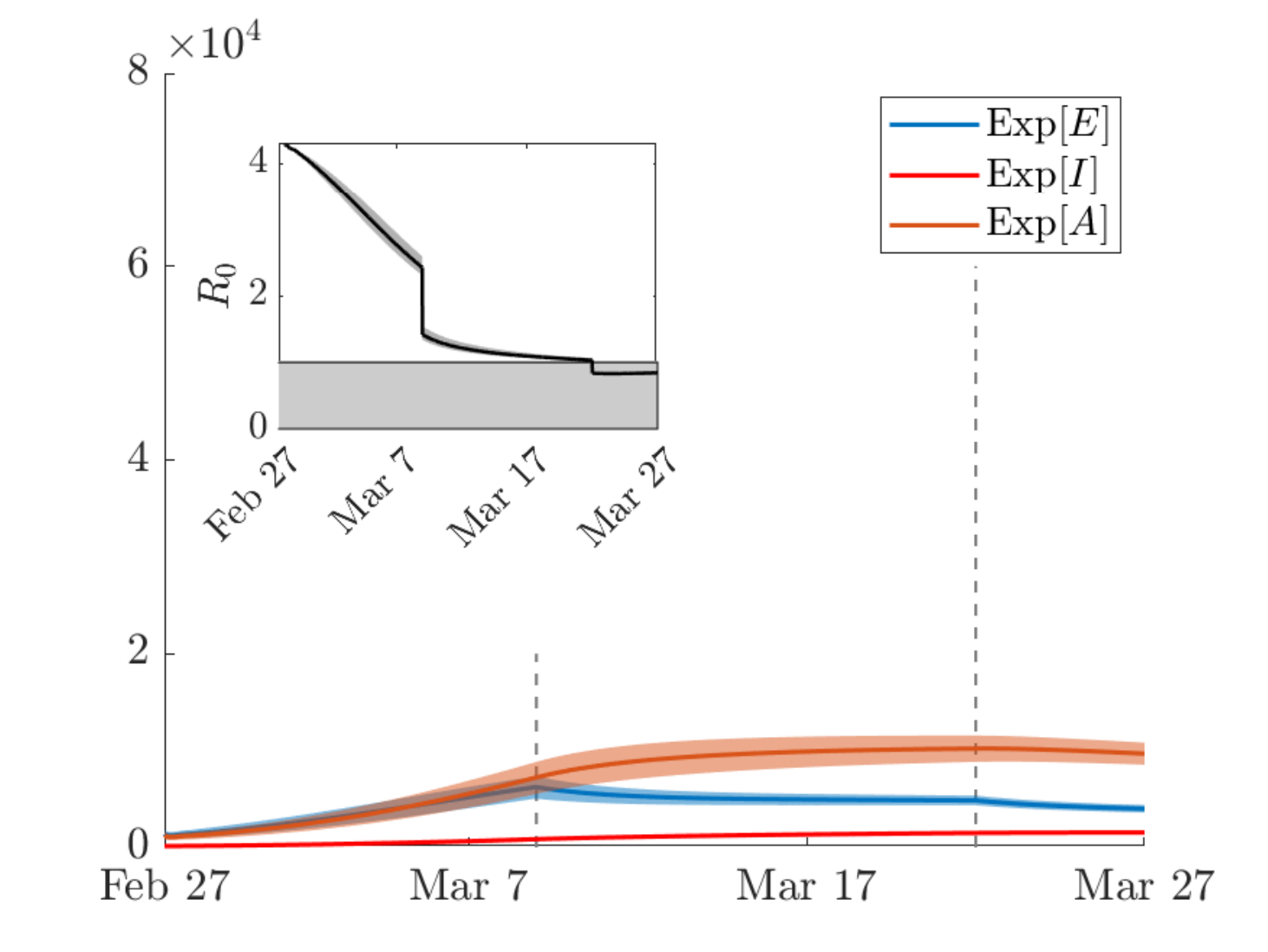}
\caption{Cremona ($n_5$)}
\label{fig.node5}
\end{subfigure}
\hspace{0.5 cm}
\begin{subfigure}{0.45\textwidth}
\includegraphics[width=1\linewidth]{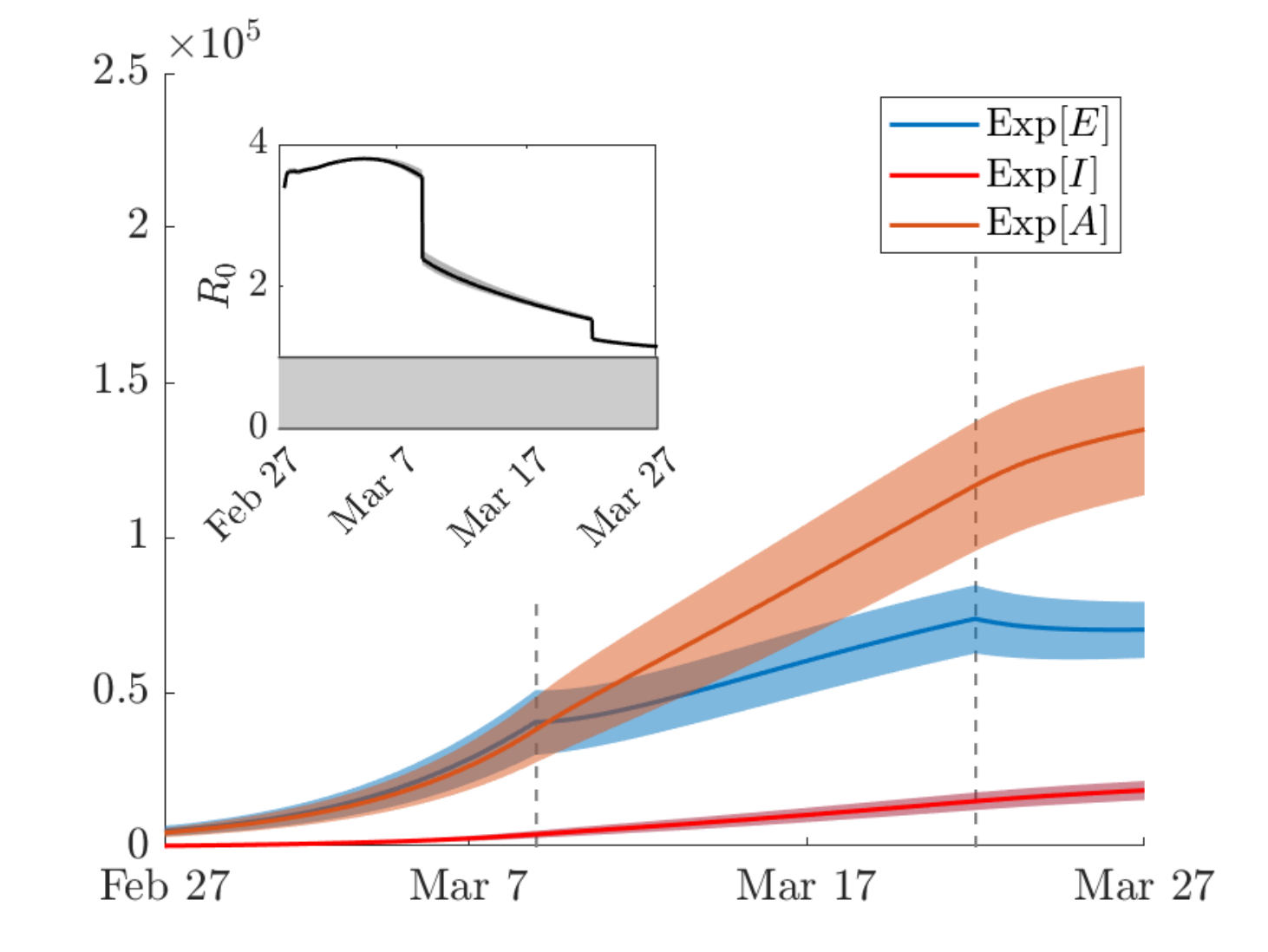}
\caption{Lombardy (total network)}
\label{fig.tot_network}
\end{subfigure}
\caption{Numerical results, with 95\% confidence intervals, of the simulation of the first outbreak of COVID-19 in Lombardy, Italy. Expected evolution in time of compartments $E$, $A$, $I$, together with the basic reproduction number $R_0$. Vertical dashed lines identify the onset of governmental lockdown restrictions.}
\label{fig.NetworkLombardia}
\end{figure}

\begin{figure}[p!]
\centering
\begin{subfigure}{0.45\textwidth}
\includegraphics[width=1\linewidth]{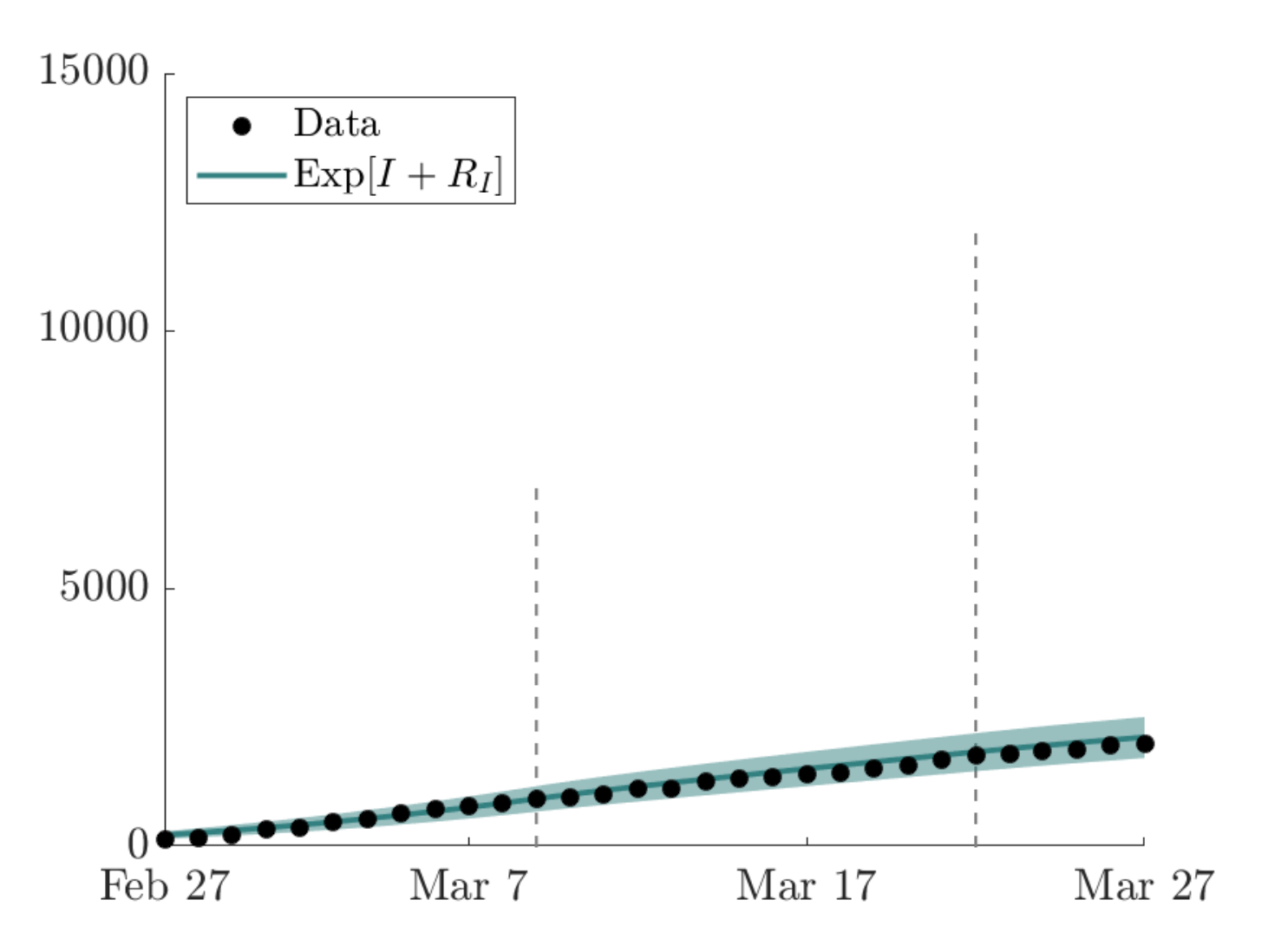}
\caption{Lodi ($n_1$)}
\label{fig.node1_data}
\end{subfigure}
\hspace{0.5 cm}
\begin{subfigure}{0.45\textwidth}
\includegraphics[width=1\linewidth]{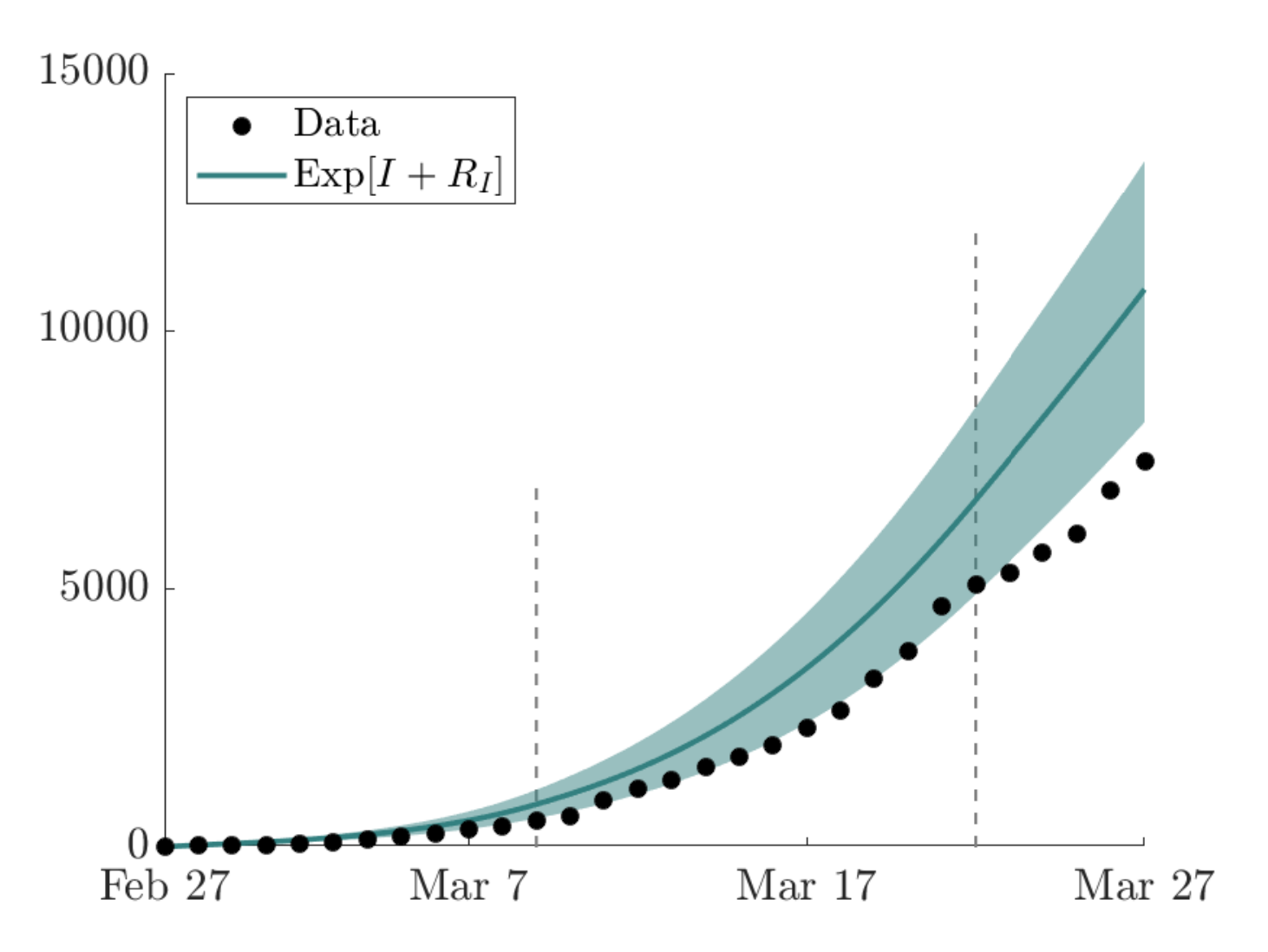}
\caption{Milan ($n_2$)}
\label{fig.node2_data}
\end{subfigure}
\begin{subfigure}{0.45\textwidth}
\includegraphics[width=1\linewidth]{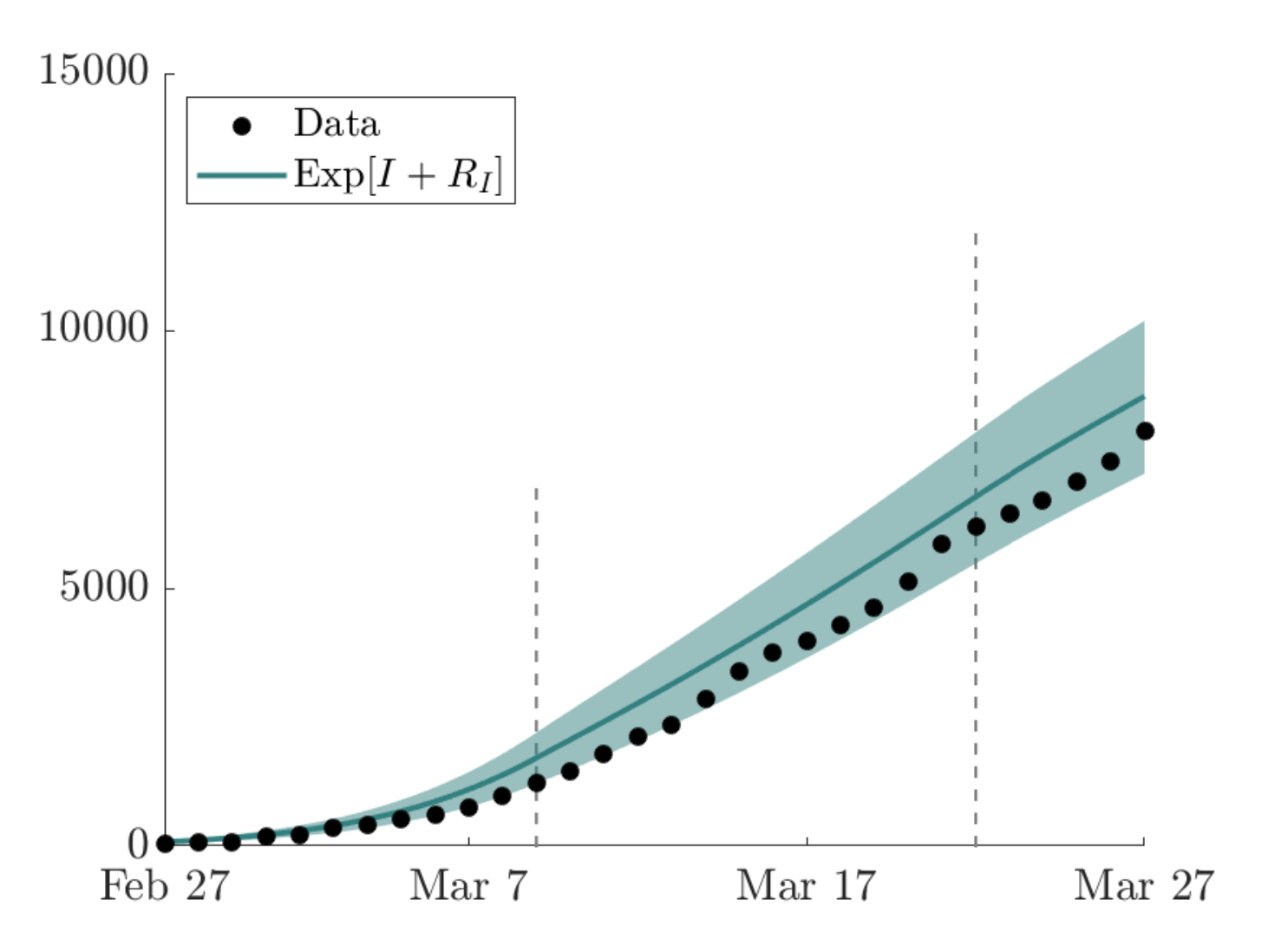}
\caption{Bergamo ($n_3$)}
\label{fig.node3_data}
\end{subfigure}
\hspace{0.5 cm}
\begin{subfigure}{0.45\textwidth}
\includegraphics[width=1\linewidth]{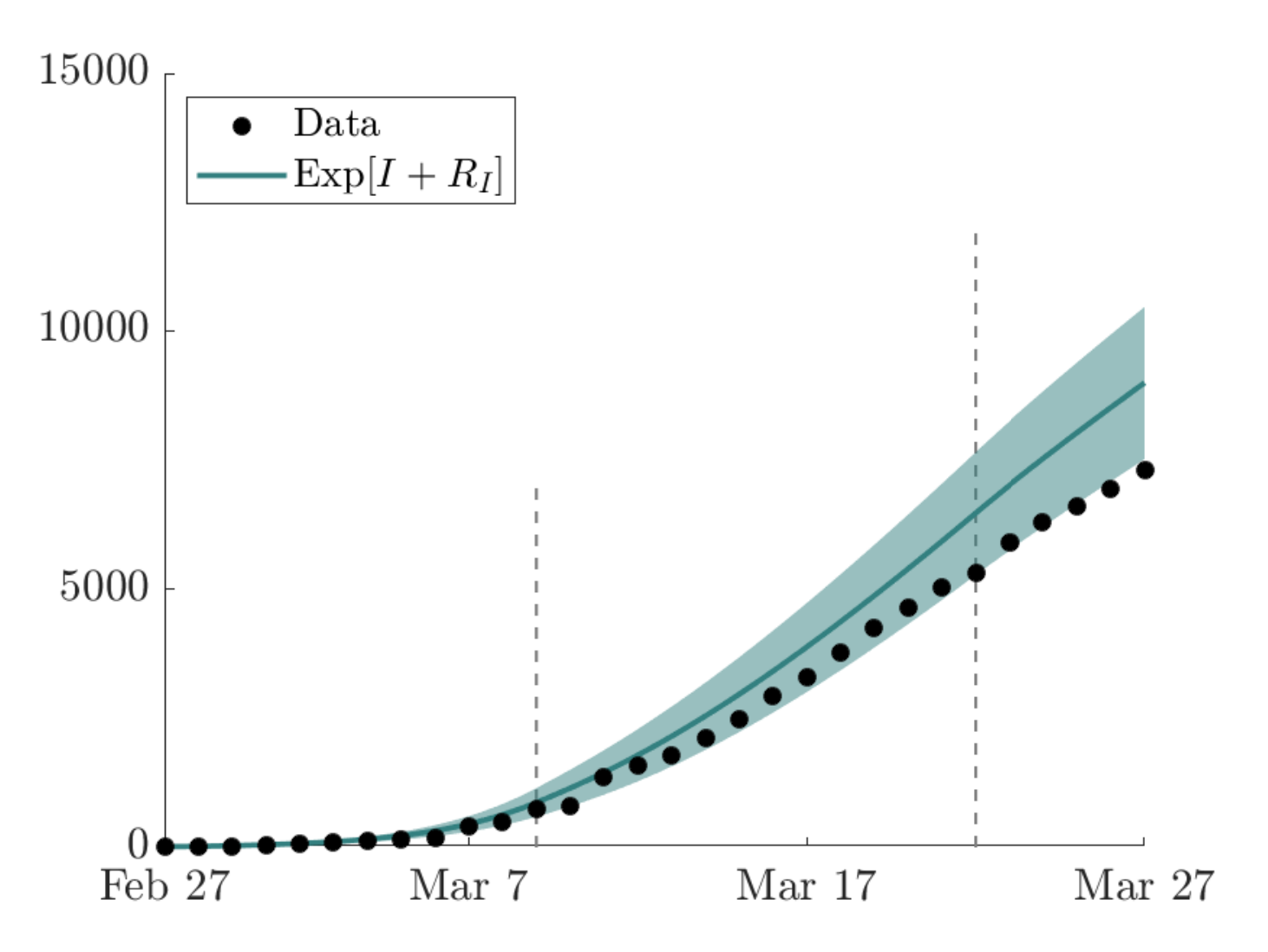}
\caption{Brescia ($n_4$)}
\label{fig.node4_data}
\end{subfigure}
\begin{subfigure}{0.45\textwidth}
\includegraphics[width=1\linewidth]{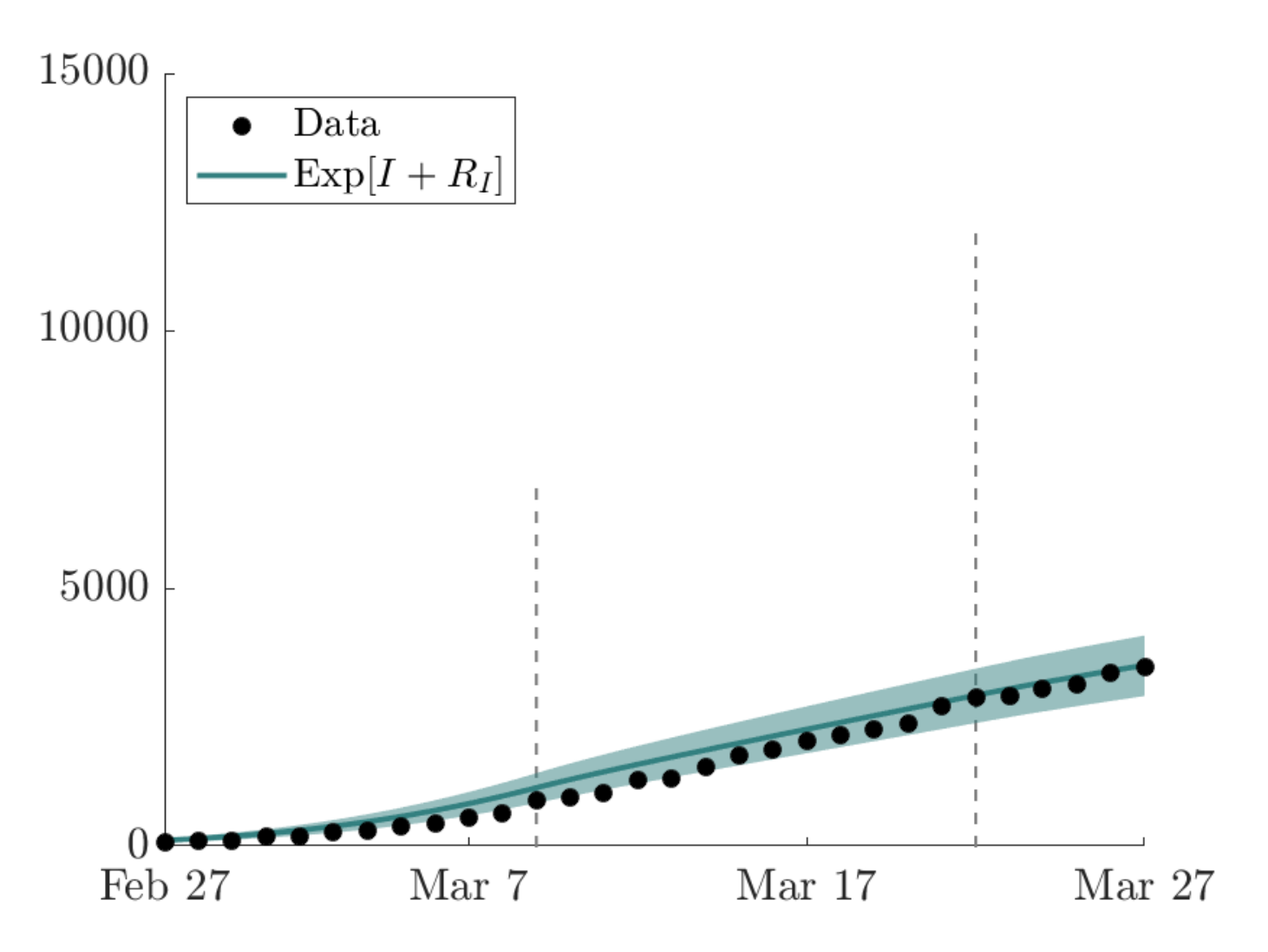}
\caption{Cremona ($n_5$)}
\label{fig.node5_data}
\end{subfigure}
\hspace{0.5 cm}
\begin{subfigure}{0.45\textwidth}
\includegraphics[width=1\linewidth]{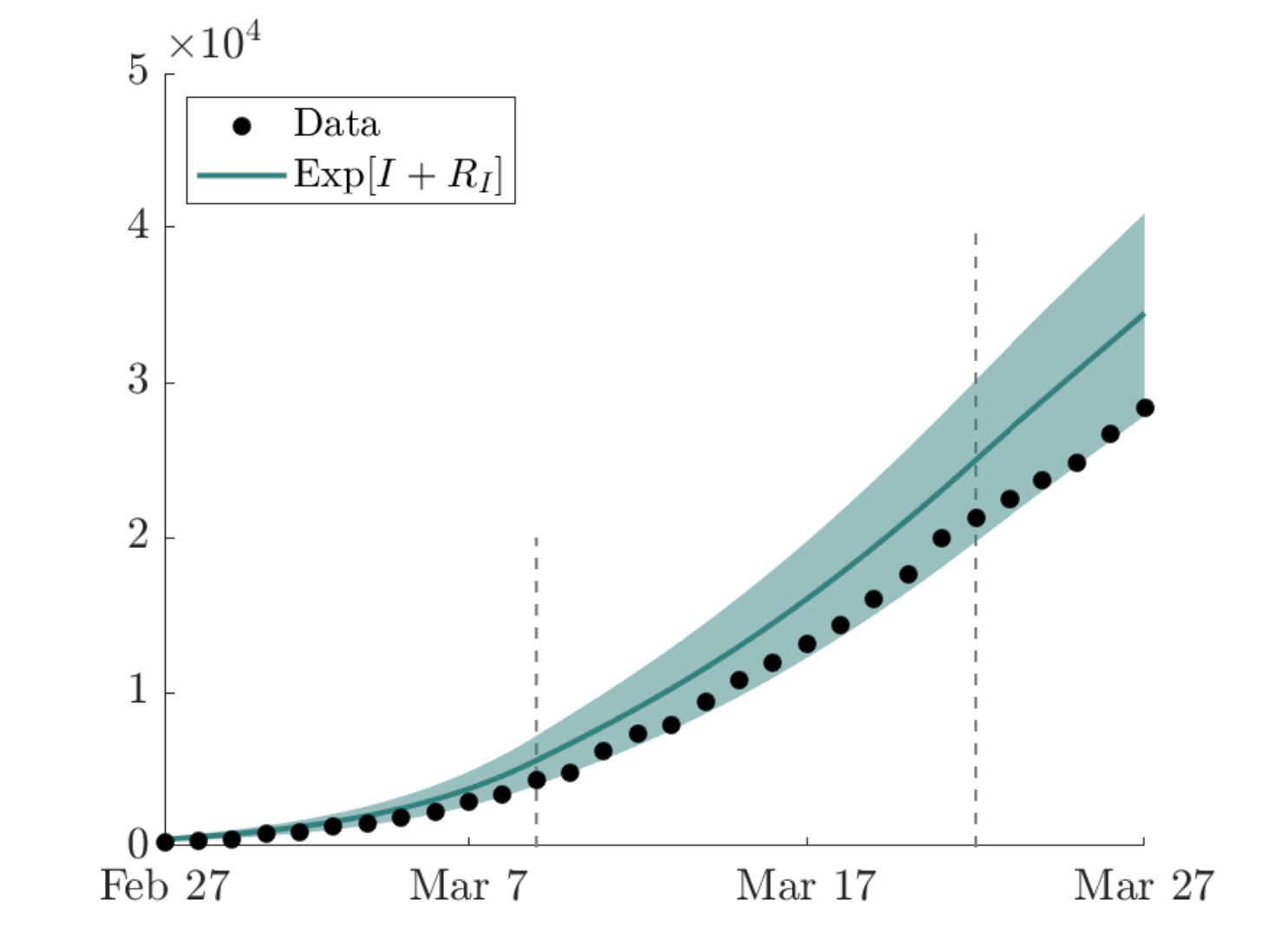}
\caption{Lombardy (total network)}
\label{fig.tot_network_data}
\end{subfigure}
\caption{Numerical results, with 95\% confidence intervals, of the simulation of the first outbreak of COVID-19 in Lombardy, Italy. Expected evolution in time of the cumulative amount of severe infectious ($I+R_I$) compared with data of cumulative infectious taken from the COVID-19 repository of the Civil Protection Department of Italy \cite{github_covid}. Vertical dashed lines identify the onset of governmental lockdown restrictions.}
\label{fig.NetworkLombardia_data}
\end{figure}

\begin{figure}[p!]
\centering
\begin{subfigure}{0.45\textwidth}
\includegraphics[width=1\linewidth]{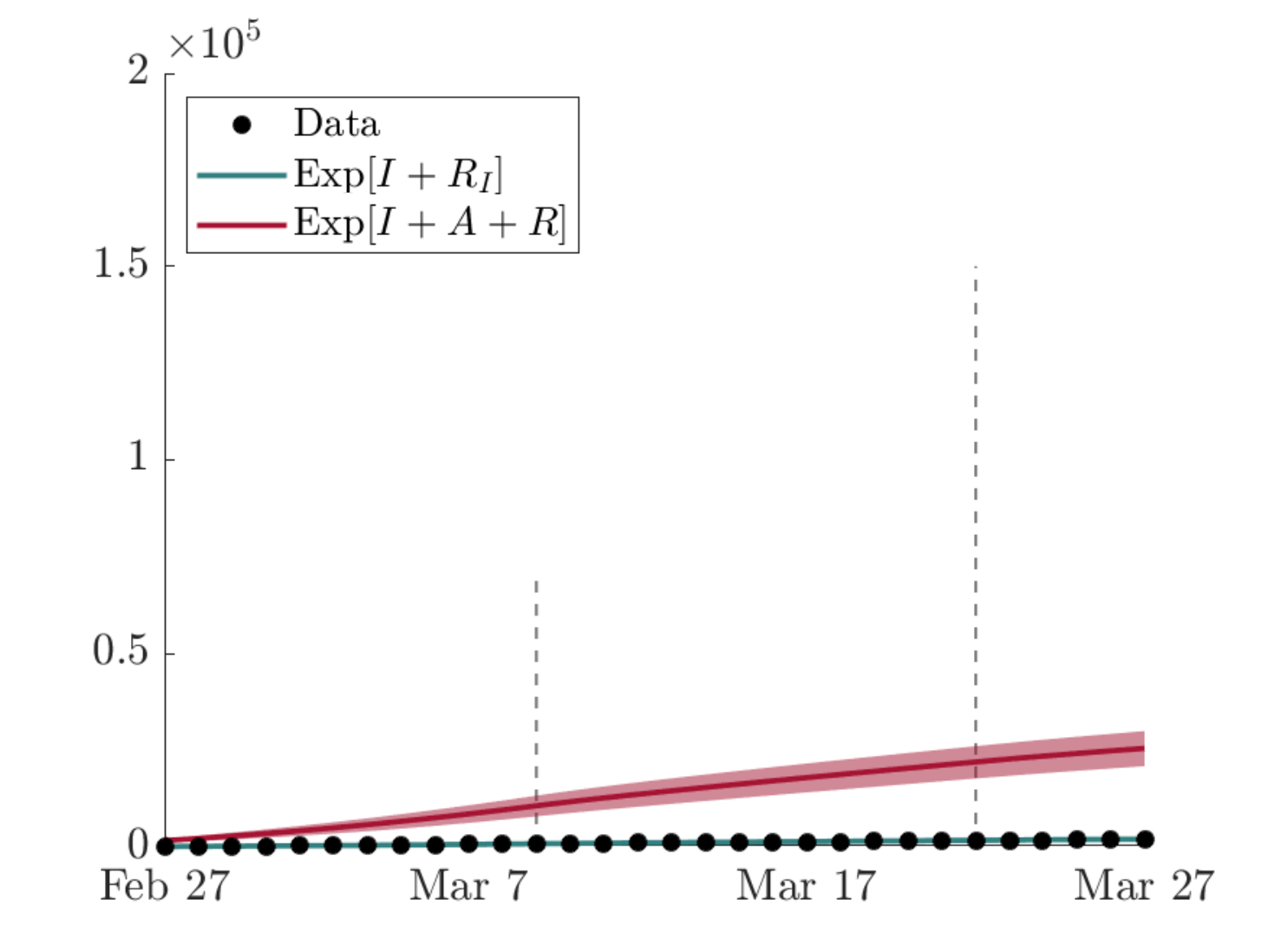}
\caption{Lodi ($n_1$)}
\label{fig.node1_data1}
\end{subfigure}
\hspace{0.5 cm}
\begin{subfigure}{0.45\textwidth}
\includegraphics[width=1\linewidth]{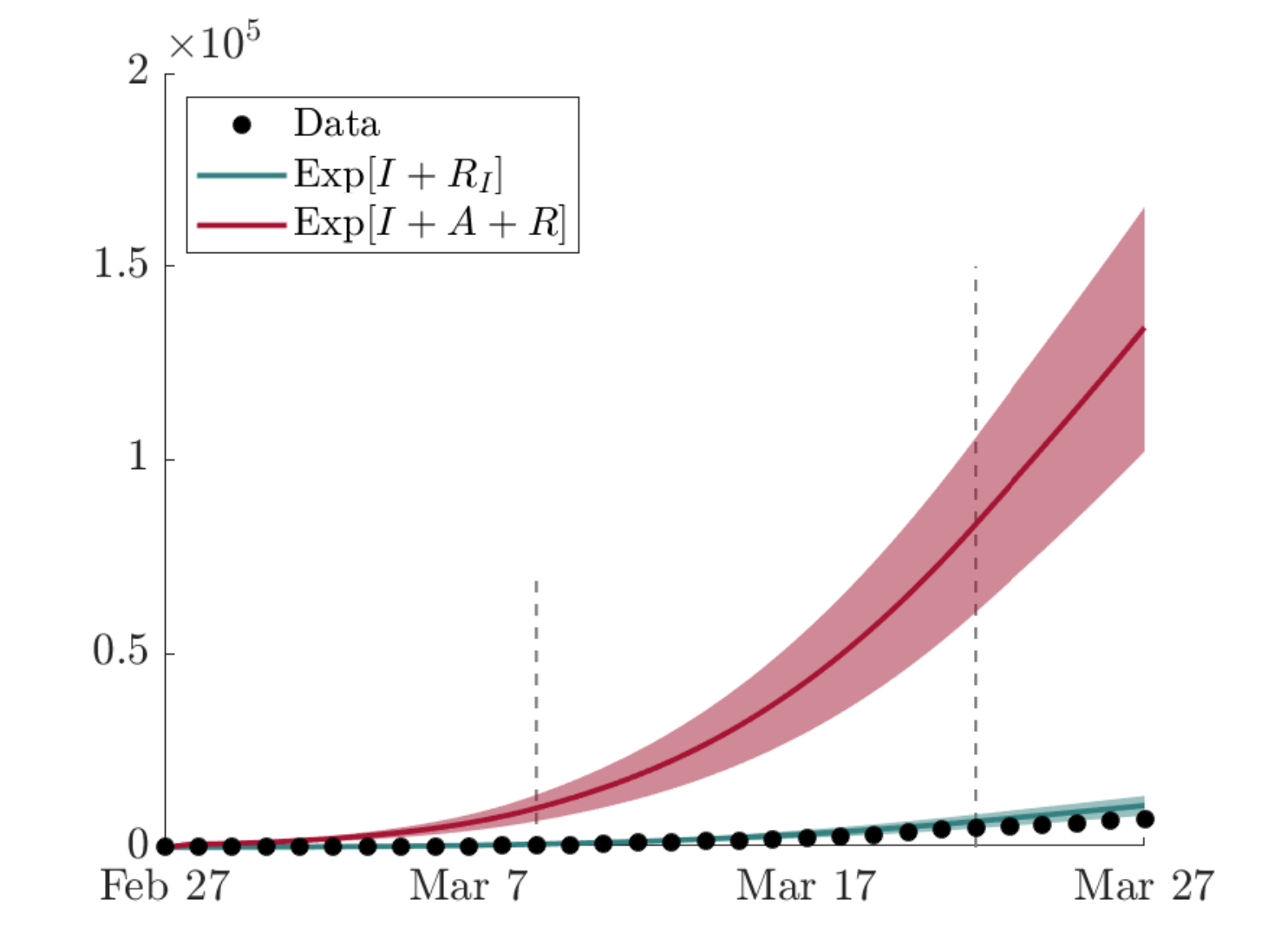}
\caption{Milan ($n_2$)}
\label{fig.node2_data1}
\end{subfigure}
\begin{subfigure}{0.45\textwidth}
\includegraphics[width=1\linewidth]{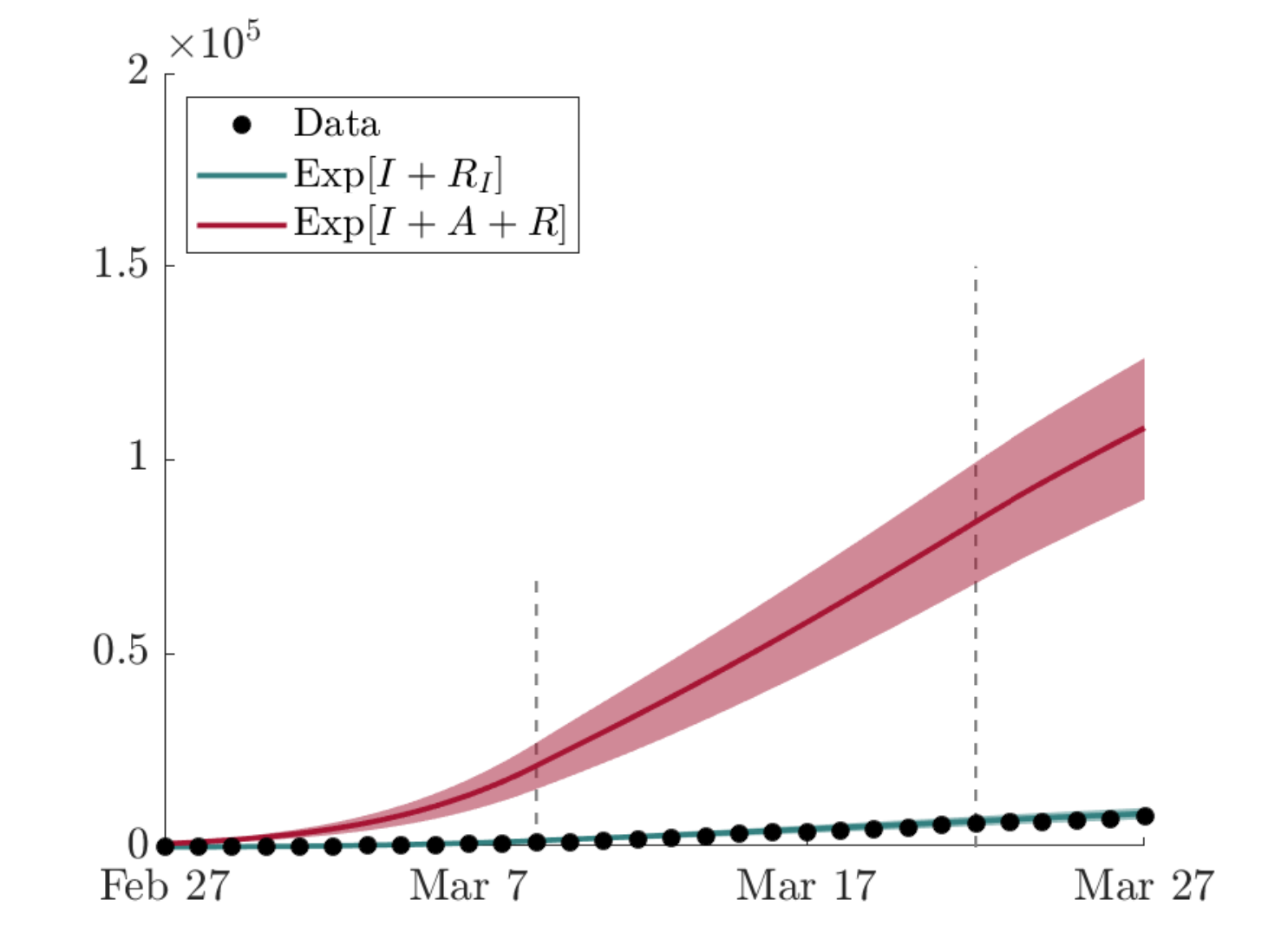}
\caption{Bergamo ($n_3$)}
\label{fig.node3_data1}
\end{subfigure}
\hspace{0.5 cm}
\begin{subfigure}{0.45\textwidth}
\includegraphics[width=1\linewidth]{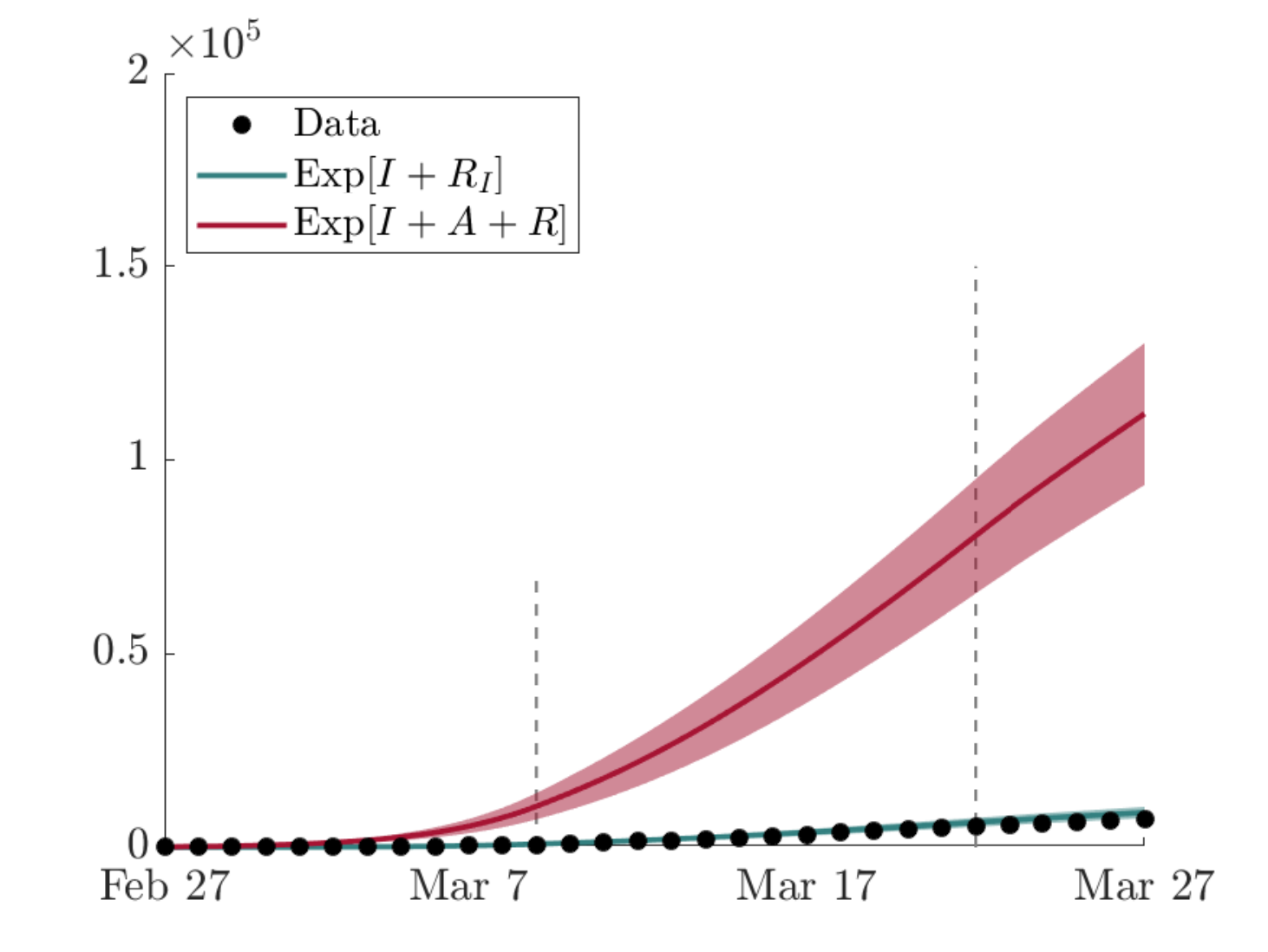}
\caption{Brescia ($n_4$)}
\label{fig.node4_data1}
\end{subfigure}
\begin{subfigure}{0.45\textwidth}
\includegraphics[width=1\linewidth]{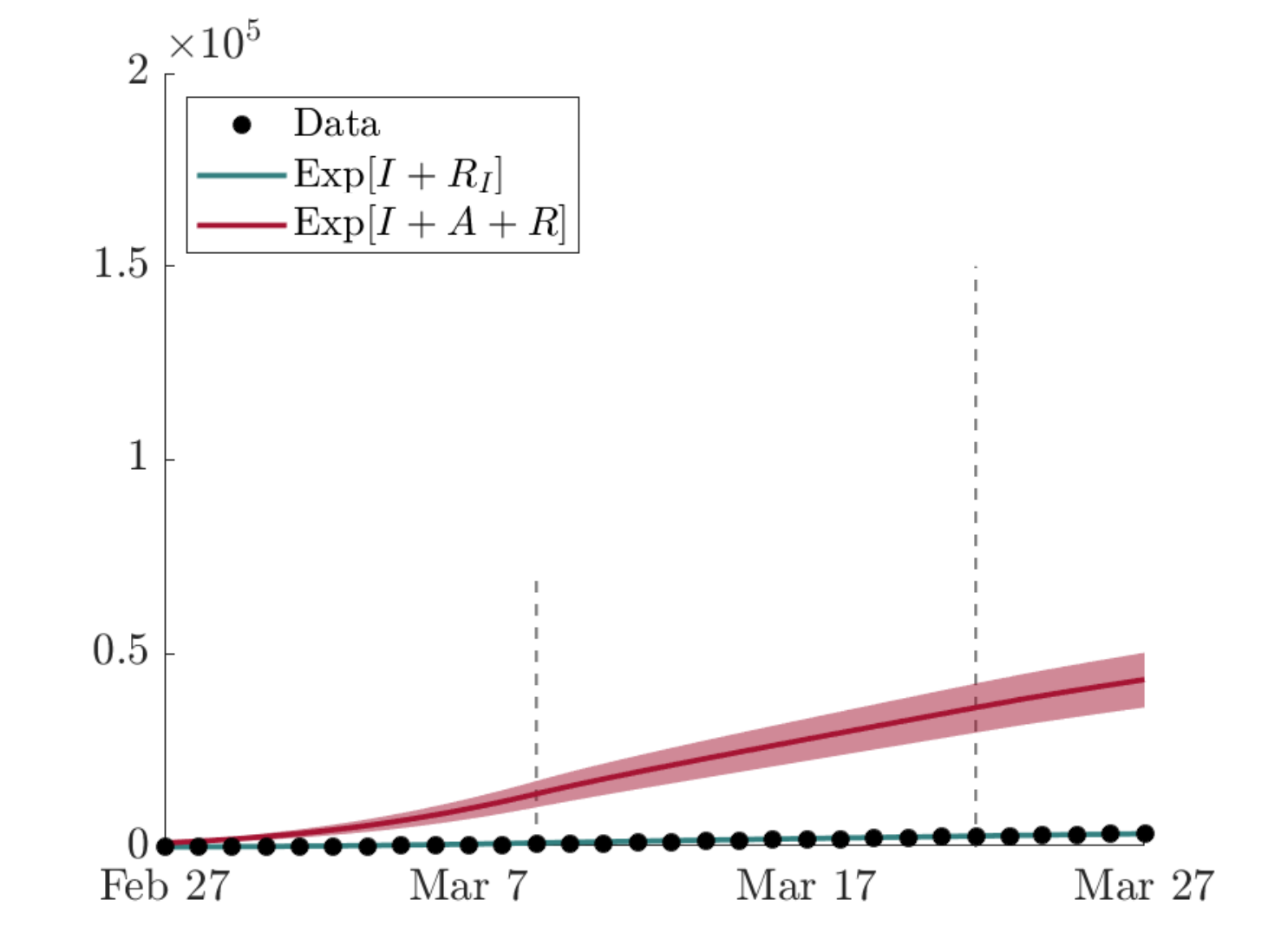}
\caption{Cremona ($n_5$)}
\label{fig.node5_data1}
\end{subfigure}
\hspace{0.5 cm}
\begin{subfigure}{0.45\textwidth}
\includegraphics[width=1\linewidth]{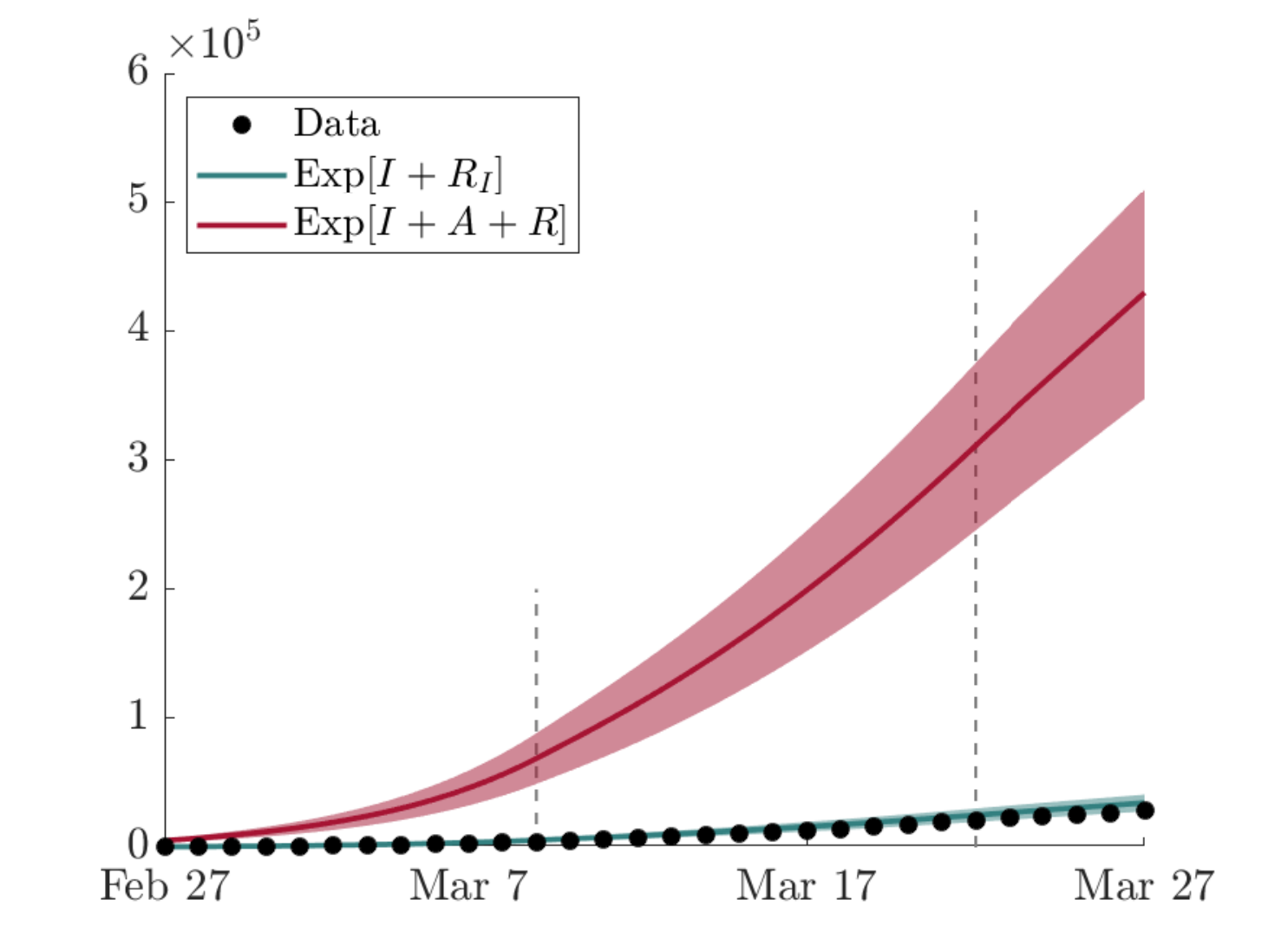}
\caption{Lombardy (total network)}
\label{fig.tot_network_data1}
\end{subfigure}
\caption{Numerical results, with 95\% confidence intervals, of the simulation of the first outbreak of COVID-19 in Lombardy, Italy. Expected evolution in time of the cumulative amount of severe infectious ($I+R_I$) with respect to the effective cumulative amount of total infectious people, including asymptomatic and mildly symptomatic individuals ($I+A+R$). Data of cumulative infectious is taken from the COVID-19 repository of the Civil Protection Department of Italy \cite{github_covid}. Vertical dashed lines identify the onset of governmental lockdown restrictions.}
\label{fig.NetworkLombardia_data1}
\end{figure}

Numerical results of the Lombardy network test, in its baseline configuration (presented in details at the beginning of Section \ref{sect.LombardiaTest}), obtained using the sAP IMEX Runge-Kutta FV Collocation method with $M_p = 6$ points, are reported in Figs.~\ref{fig.lombardia_results}--\ref{fig.NetworkLombardia_data1}.
In Fig.~\ref{fig.lombardia_results} a first qualitative trend of the spatial and temporal spread of the epidemic is presented. Here it can be observed the heterogeneity of the diffusion of the virus, which has firstly mostly affected the city of Bergamo and only in a second moment prevailing in Milan. 
In Fig.~\ref{fig.NetworkLombardia} the expected evolution in time of the infected individuals, together with 95\% confidence intervals, is shown for each node and for the whole Lombardy network, including exposed $E$, highly symptomatic subjects $I$ and asymptomatic or weakly symptomatic people $A$. Each plot is also associated with the expected temporal evolution of the reproduction number $R_0(t)$, computed as described in Section \ref{sect:R0}. 

One can see the capacity of the model to reproduce a very heterogeneous epidemic trend in the network analyzed, which is also reflected in the different ranges and patterns shown for the $R_0$ of each province. It can also be observed the agreement between the evolution of the reproduction number and the epidemic spread. In particular, it is confirmed the decline of the daily number of infected as $R_0$ reaches values below 1, as shown in the plots for Lodi, Bergamo and Cremona. On the other hand, the persistence of the virus in the complete network, and especially in Milan, is noticed until March 27, 2020 (last day of the simulation), where the reproduction number remains $R_0 > 1$, confirming the consistency of the definition proposed for $R_0(t)$. 


As visible from Fig. \ref{fig.NetworkLombardia_data}, the lower bound of the confidence band of the cumulative amount in time of highly symptomatic individuals is comparable with the observed data of the Civil Protection Department of Italy \cite{github_covid}. As expected, the mean value of the numerical result reports an higher amount of $I$, especially in Milan, the province most affected by the virus, due to the uncertainty of available data, which surely underestimate the real amount of infectious people, as discussed in Section \ref{sect.LombardiaTest}.

The comparison between the expected evolution in time of the cumulative amount of severe infectious with respect to the effective cumulative amount of total infectious people, including asymptomatic and mildly symptomatic individuals, is shown in Fig. \ref{fig.NetworkLombardia_data1}. Here, it can be noticed how much of the spread of COVID-19 has actually been lost from the data of the first outbreak in Lombardy and the impact that the presence of asymptomatic or undetected subjects has had on the epidemic evolution. By the end of the simulation, indeed, $\mathbb{E}[A+R_A]/\mathbb{E}[I+A+R] = 0.92$ in the network, indicating with $R_A$ the amount of recovered coming from the compartment $A$. This result is in line with WHO guidelines given during the first wave of COVID-19, stating that approximately 80\% of the infected population was asymptomatic \cite{WHO}. It is here remarked, indeed, that the compartment $A$ in the proposed model includes not only the asymptomatic people but also the mildly symptomatic, which would approximately be 12\% of the total cases.

\begin{remark}
In the results here presented, transmission coefficients $\alpha_{i,j}(x)$, which define the behavior of commuters in the network, are imposed as deterministic and constant in time, meaning that the amount of individuals exiting from (and entering in) each city is the same in each instant of the simulation. Clearly, this assumption leads to a simplification of the description of the phenomenon of commuting, generally characterized by a peak in the early morning hours and a drastic drop in the night hours. What we represent is indeed the mean commuting trend during the day, which tends to assume a stationary solution in time. However, a more realistic behavior, characterized for example by periodic sinusoidal functions, leads to a slightly oscillatory trend in the reported curves of each compartment and also in the reproduction number $R_0$. By selecting $\alpha_{i,j}(x)$ values that define an average trend over the day, allows us to avoid this misleading representation, while maintaining the consistency of the results.
\end{remark}

\subsubsection{No-mobility scenario}
\begin{figure}[t!]
\centering
\begin{subfigure}{0.45\textwidth}
\includegraphics[width=1\linewidth]{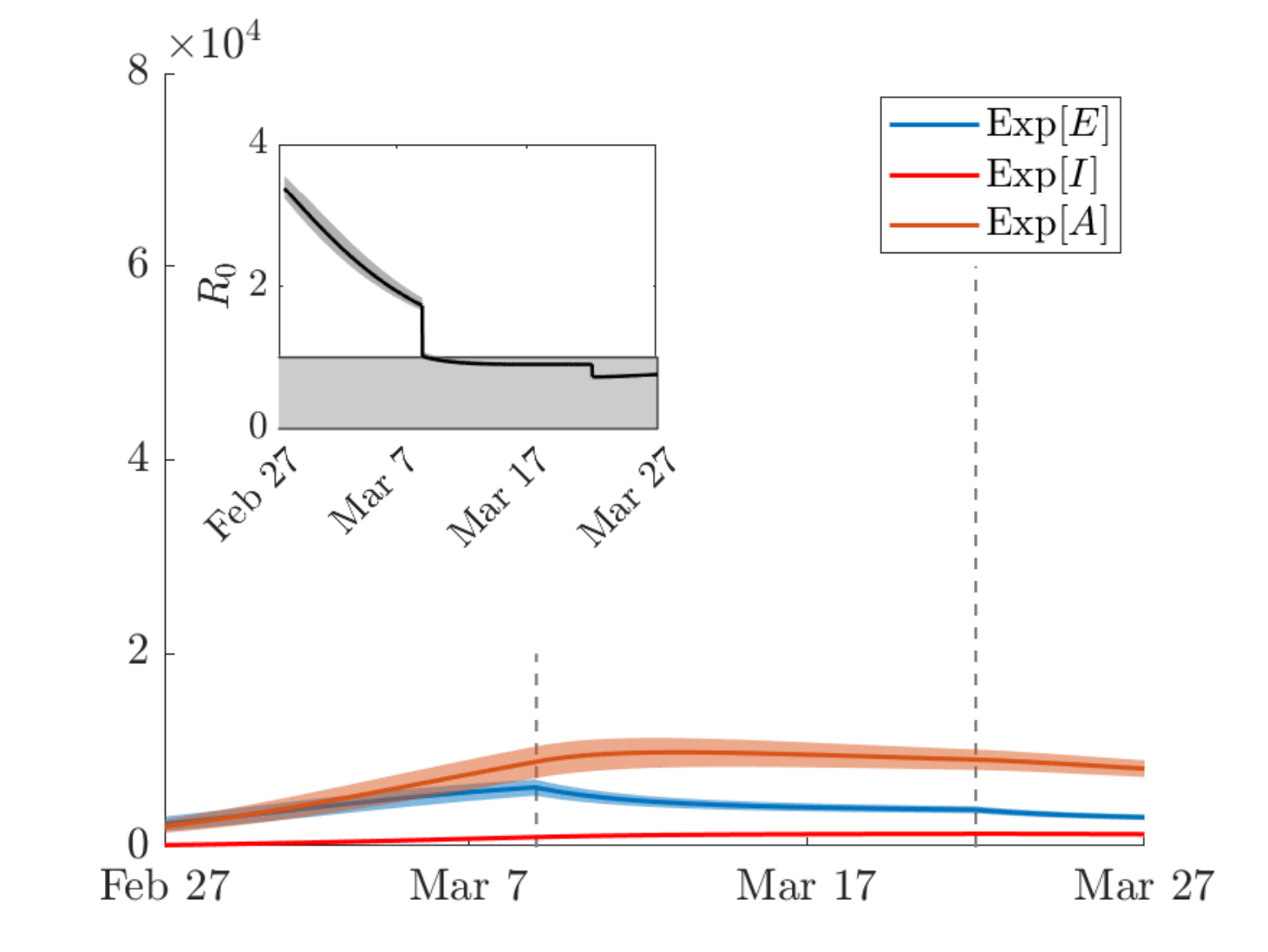}
\caption{Lodi ($n_1$)}
\label{fig.node1_nomobility}
\end{subfigure}
\hspace{0.5 cm}
\begin{subfigure}{0.45\textwidth}
\includegraphics[width=1\linewidth]{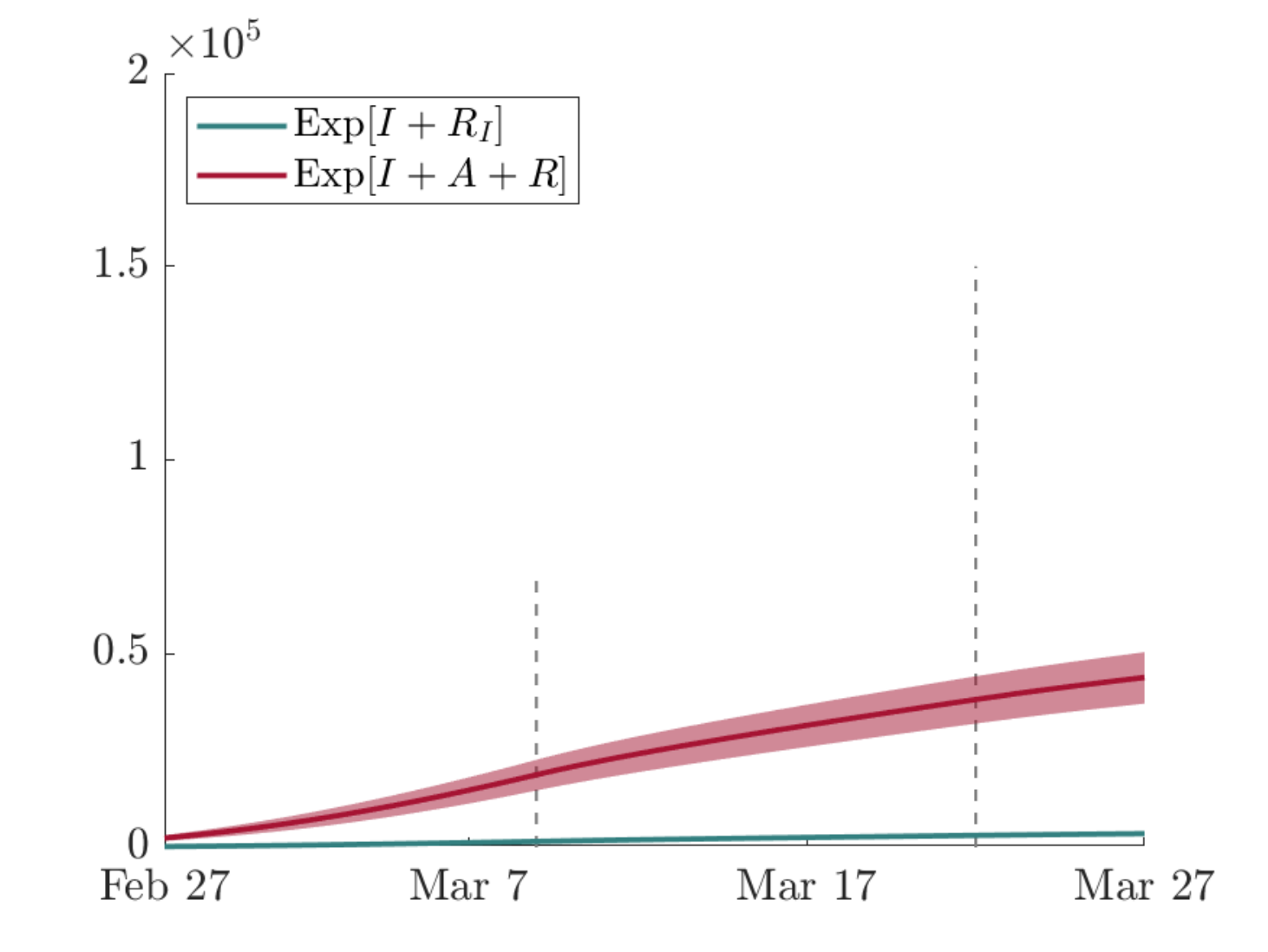}
\caption{Lodi ($n_1$)}
\label{fig.node1_data1_nomobility}
\end{subfigure}
\begin{subfigure}{0.45\textwidth}
\includegraphics[width=1\linewidth]{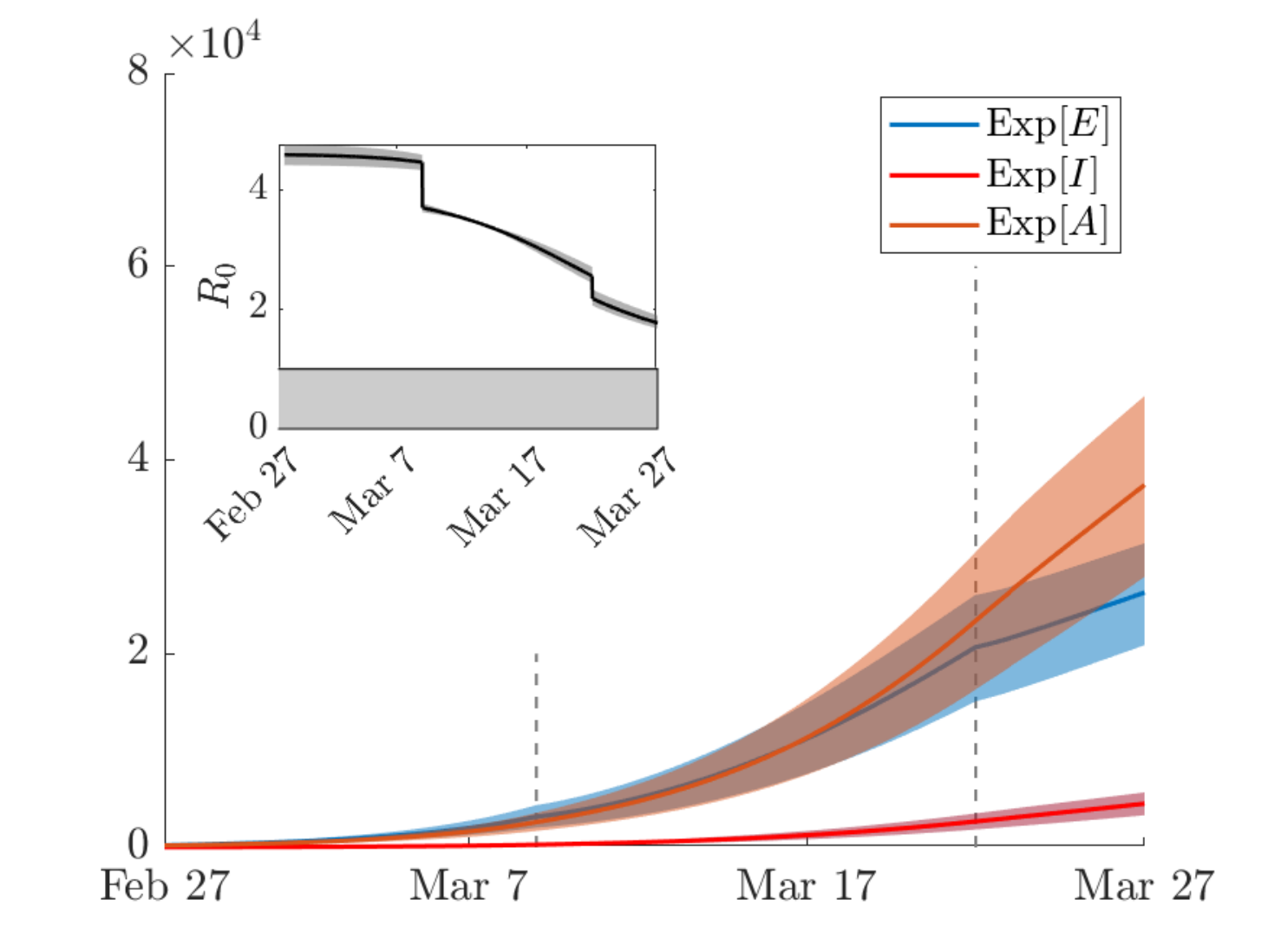}
\caption{Milan ($n_2$)}
\label{fig.node2_nomobility}
\end{subfigure}
\hspace{0.5 cm}
\begin{subfigure}{0.45\textwidth}
\includegraphics[width=1\linewidth]{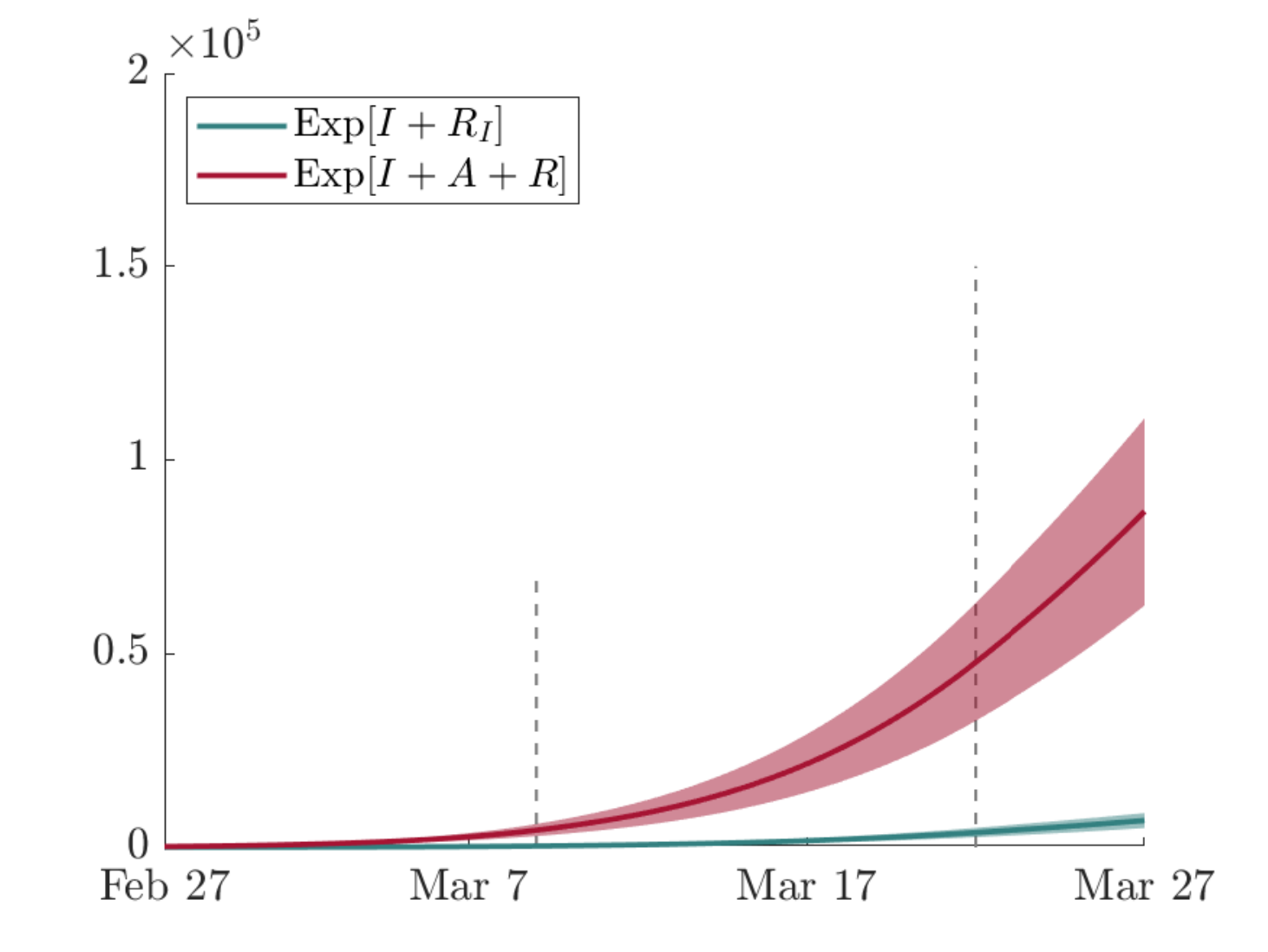}
\caption{Milan ($n_2$)}
\label{fig.node2_data1_nomobility}
\end{subfigure}
\caption{Numerical results, with 95\% confidence intervals, of the simulation of the first outbreak of COVID-19 in Lombardy, Italy, if the entire population was non-commuting, thus removing network mobility. Results are presented for cities Lodi and Milan. Expected evolution in time of compartments $E$, $A$, $I$, together with the basic reproduction number $R_0$ (first column); expected time evolution of the cumulative amount of severe infectious ($I+R_I$) with respect to the effective cumulative amount of total infectious people ($I+A+R$), including asymptomatic and mildly symptomatic individuals (second column). Vertical dashed lines identify the onset of governmental lockdown restrictions.}
\label{fig.NetworkLombardia_nomobility}
\end{figure}

To assess the effects of the mobility of commuters, a second scenario is investigated in which the whole population is not allowed to move out of the residence city. Numerical results of this test are reported in Fig. \ref{fig.NetworkLombardia_nomobility} for two representative provinces of the network: Lodi and Milan. It can be observed the different evolution of the spread of the epidemic comparing trends presented in Fig. \ref{fig.NetworkLombardia_nomobility} with the corresponding ones in Fig. \ref{fig.NetworkLombardia} (for the first column of plots) and Fig. \ref{fig.NetworkLombardia_data1} (for the second column of plots). 

As a consequence of the absence of commuting people in the network, the spread of COVID-19 in Milan results consistently damped. The province of Milan, in fact, is the one with the highest number of daily incomes of commuting workers and students, followed by Bergamo and Brescia, as shown in Table \ref{tab:matrix_network_lombardia}. On the other hand, Lodi is the city with the highest amount of daily  outcomes of commuters (30\% of its total population). Thus, preventing the population from leaving the province shows a slight increase in local infections with respect to the baseline scenario. This result is not intended to suggest that a constraint on the mobility of people would be disadvantageous in fighting the spread of the epidemic. In fact, one must consider that the contagions shown in these Figures are due to a still present interaction of people at the provincial level, which has not been reduced in any way with respect to the previous scenario. Similar conclusions are drawn also in \cite{espinoza2020}. These results are primarily intended to demonstrate how much the evolution of a pandemic can vary in response to changes in people's mobility.

\begin{figure}[p!]
\centering
\begin{subfigure}{0.45\textwidth}
\includegraphics[width=1\linewidth]{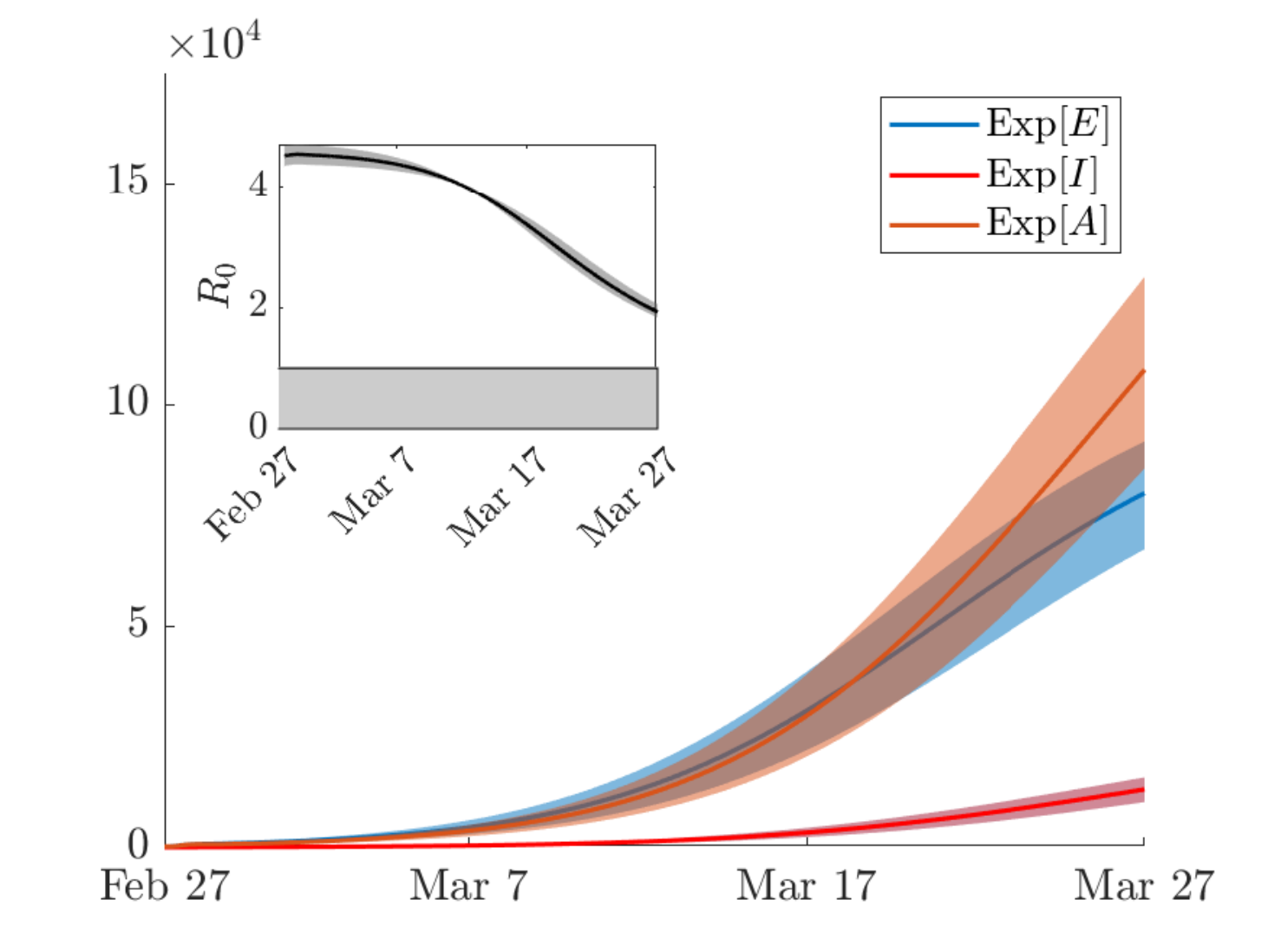}
\caption{Milan ($n_2$)}
\label{fig.node2_norestrictions}
\end{subfigure}
\hspace{0.5 cm}
\begin{subfigure}{0.45\textwidth}
\includegraphics[width=1\linewidth]{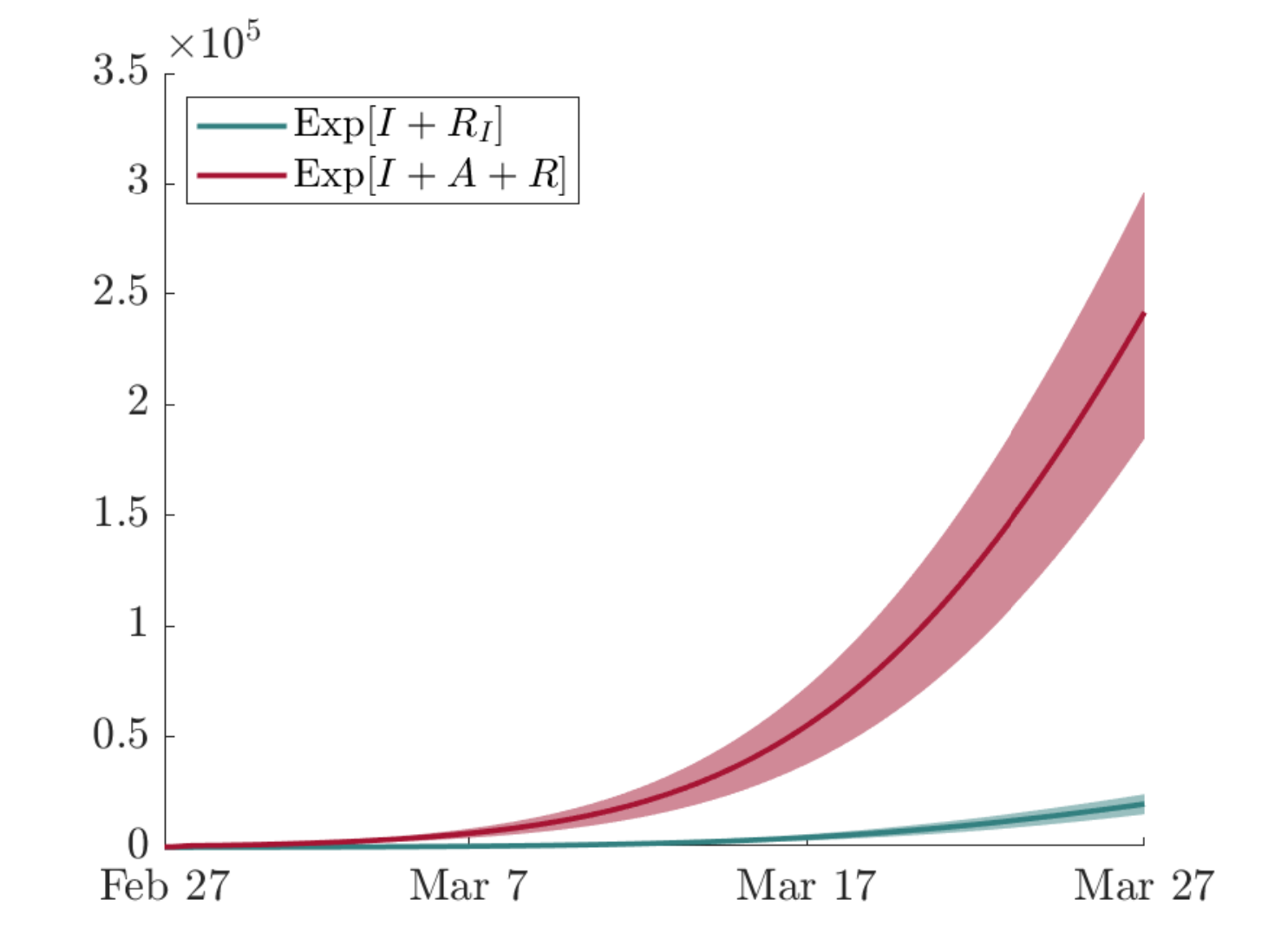}
\caption{Milan ($n_2$)}
\label{fig.node2_data1_norestrictions}
\end{subfigure}
\begin{subfigure}{0.45\textwidth}
\includegraphics[width=1\linewidth]{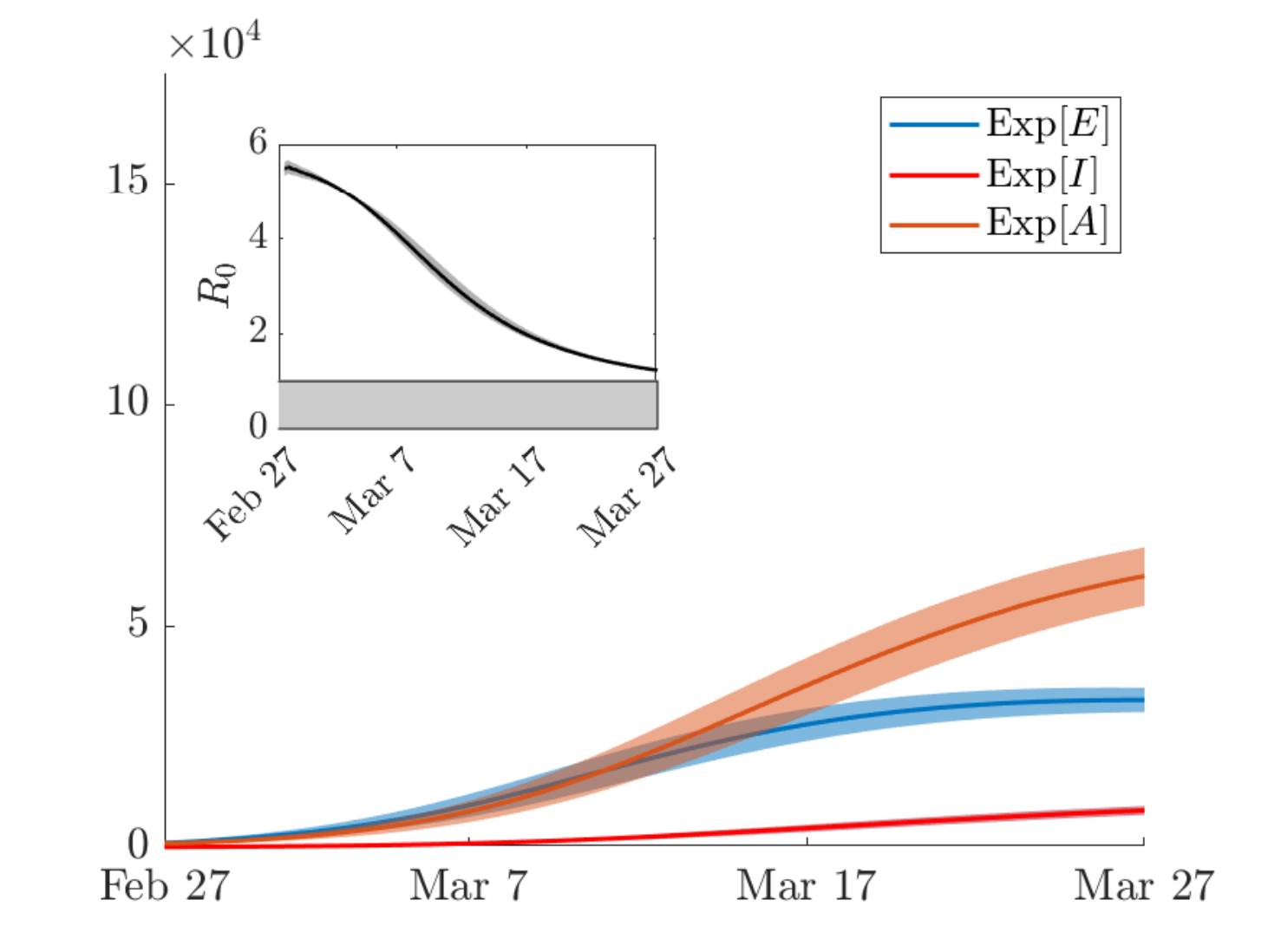}
\caption{Bergamo ($n_3$)}
\label{fig.node3_norestrictions}
\end{subfigure}
\hspace{0.5 cm}
\begin{subfigure}{0.45\textwidth}
\includegraphics[width=1\linewidth]{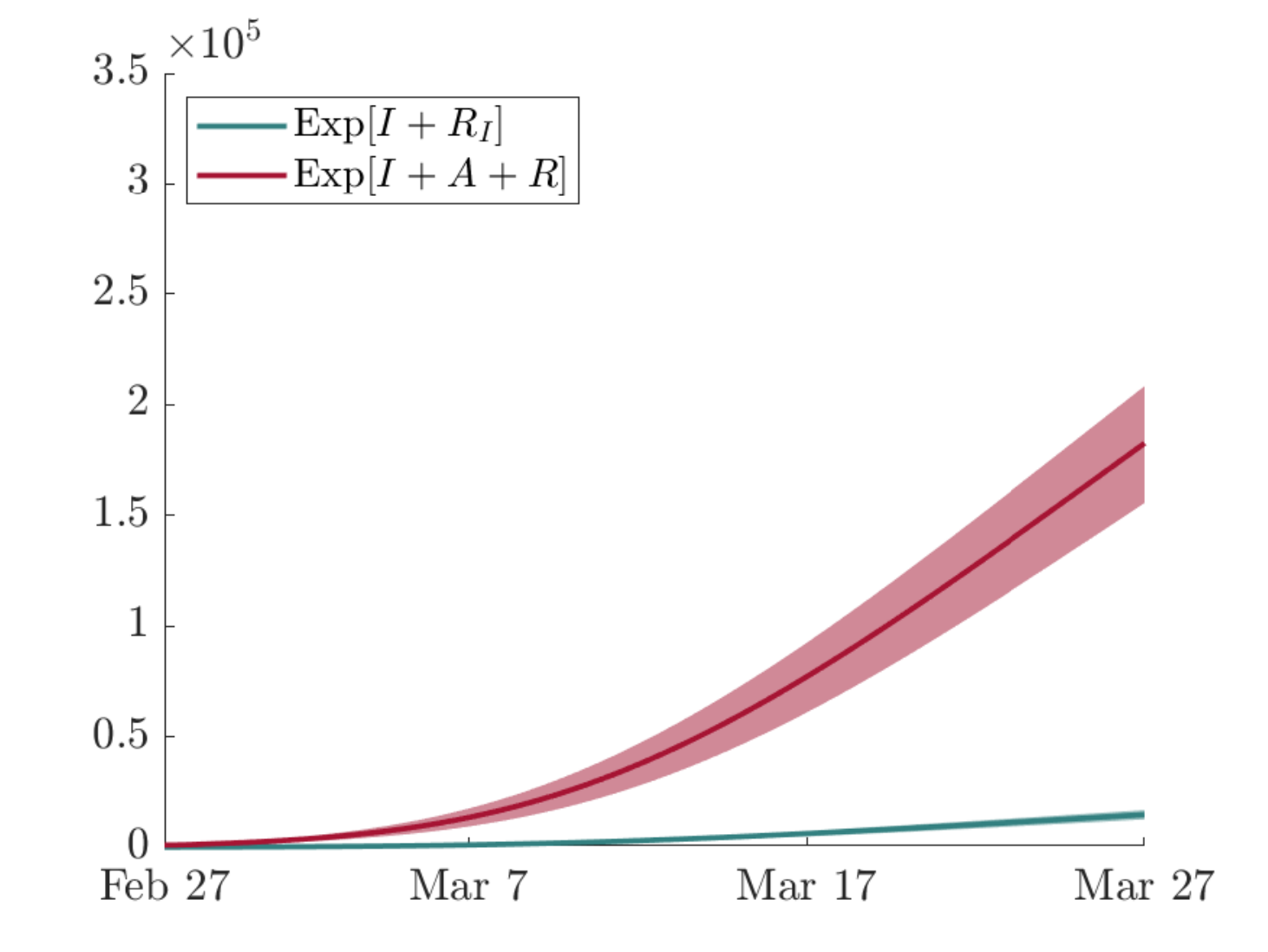}
\caption{Bergamo ($n_3$)}
\label{fig.node3_data1_norestrictions}
\end{subfigure}
\begin{subfigure}{0.45\textwidth}
\includegraphics[width=1\linewidth]{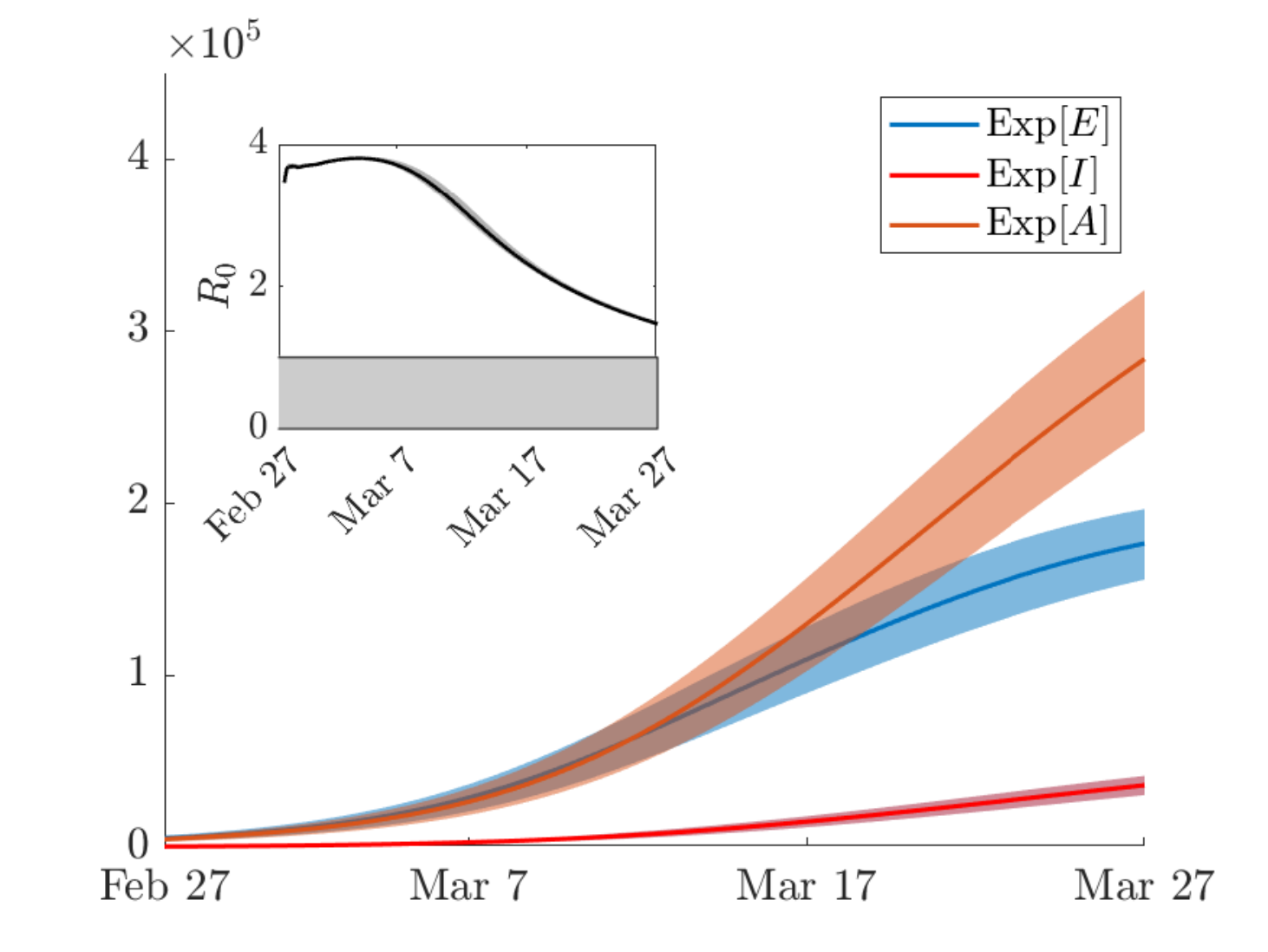}
\caption{Lombardy (total network)}
\label{fig.tot_network_norestrictions}
\end{subfigure}
\hspace{0.5 cm}
\begin{subfigure}{0.45\textwidth}
\includegraphics[width=1\linewidth]{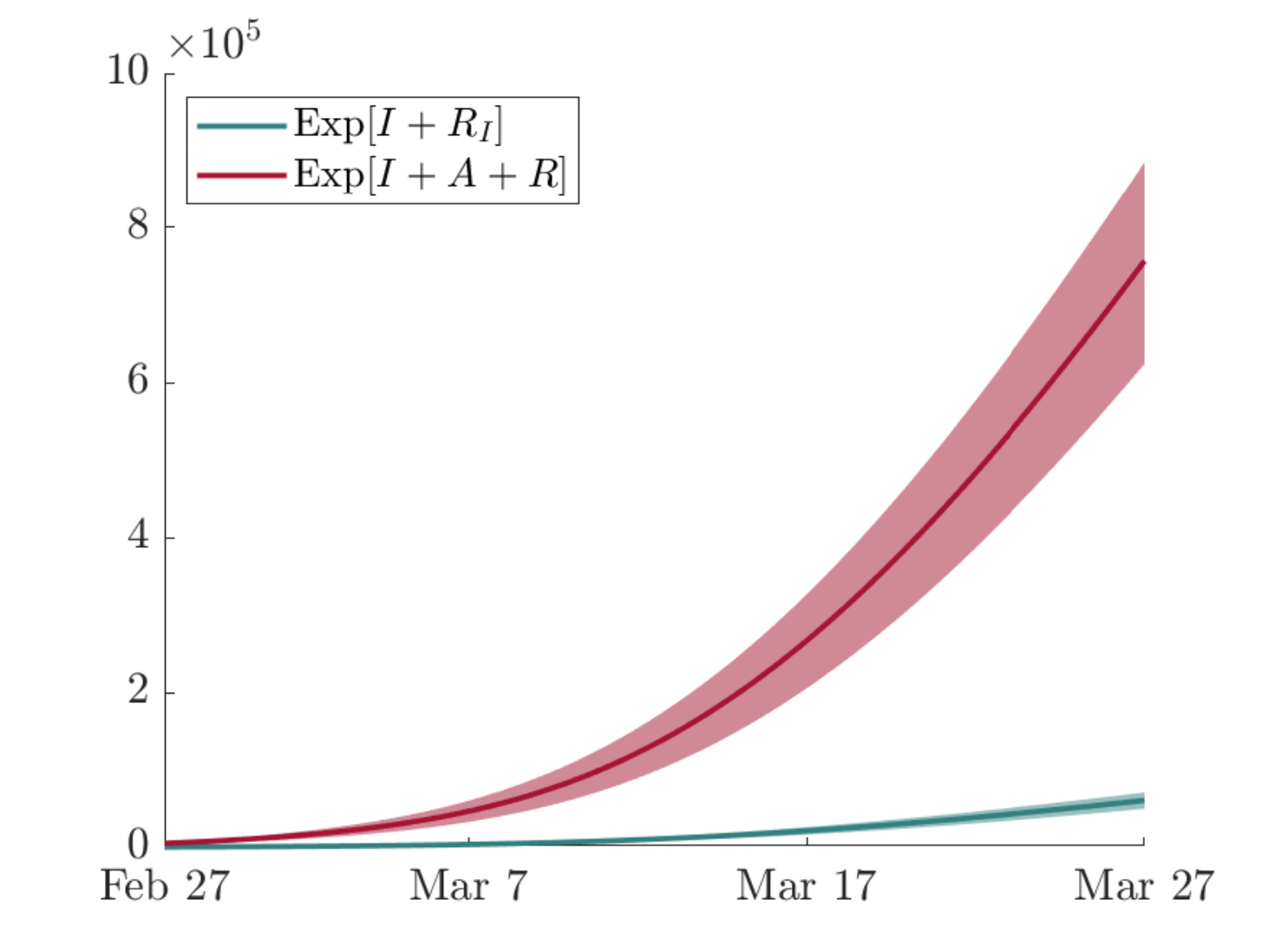}
\caption{Lombardy (total network)}
\label{fig.tot_network_data1_norestrictions}
\end{subfigure}
\caption{Numerical results, with 95\% confidence intervals, of the simulation of the first outbreak of COVID-19 in Lombardy, Italy, in the event that no restrictions of any kind were applied by the government. Results are presented for Milan, Bergamo and the whole network. Expected evolution in time of compartments $E$, $A$, $I$, together with the basic reproduction number $R_0$ (first column); expected time evolution of the cumulative amount of severe infectious ($I+R_I$) with respect to the effective cumulative amount of total infectious people, including asymptomatic and mildly symptomatic individuals ($I+A+R$), compared with data of cumulative infectious taken from the COVID-19 repository of the Civil Protection Department of Italy \cite{github_covid} (second column).}
\label{fig.NetworkLombardia_norestrictions}
\end{figure}

\subsubsection{No-restrictions scenario}
Finally, results in the event that no restrictions of any kind were applied by the government are presented in Fig. \ref{fig.NetworkLombardia_norestrictions} for the province of Milan, Bergamo and the complete Lombardy network. Comparing these results with the corresponding in Fig. \ref{fig.NetworkLombardia} (for the first column of plots) and Fig. \ref{fig.NetworkLombardia_data1} (for the second column of plots), it can be seen immediately the fundamental role that government restrictions have played in containing the spread of the disease (notice the different y-axis scale in the Figures). In fact, the cumulative number of infected people across the region in this scenario is almost 1.8 times higher than in the baseline scenario at the end of the simulation. This value indicates that the restrictions imposed during the first wave of COVID-19 in Italy, and the consequent increasing risk awareness in the population, have contributed to attenuate the spread of the virus by 43\%, which is in agreement with the result presented in \cite{gatto2020}. Also the basic reproduction number $R_0$ shows a totally different evolution in time, being far from reaching the value 1 at the end of the simulation. Indeed, on March 27, 2020 the virus still persists in its propagation, with mean value $R_0 = 1.50$ when considering the whole network and, in particular, $R_0 = 1.90$ in the province of Milan.

\section{Conclusions}
\label{sect:conclusion}
In this paper, a stochastic transport model for the spread of an epidemic phenomenon described by a multi-population SEIAR compartmental dynamics on networks is presented. The starting point for the description of spatial motion and the interactions of the individuals has its roots in the kinetic theory of discrete velocity models \cite{BellGat03,Lions1997,aylaj2020}. The model is structured on spatial networks, where nodes identify the locations of interest (cities in this case) and arcs represent the set of major mobility paths (roads and railways). Individuals are divided into commuters, who move on a suburban scale, and non-commuters, who act on an urban scale. In this way, we avoid unrealistic mass migration effects in which the entire population moves indiscriminately through the network. 

Thanks to the hyperbolicity of the resulting system, the unphysical feature of instantaneous diffusive effects, which is typical of parabolic reaction-diffusion models proposed in literature, is removed. Nevertheless, we show that for small relaxation times and large characteristic speeds, in the diffusive regime, the proposed model recovers its parabolic nature.

Since the derivation and the consequent definition of the basic reproduction number is well-established in the literature for ODE epidemic models, we resort on these results to introduce a derivation of $R_0$ for spatial epidemic models in the case of no-inflow/outflow boundary conditions, validating the effectiveness of the resulting definition as an indicator of the viral growth rate.

The model has been tested for the analysis of the emergence of COVID-19 in Italy, by simulating the propagation and evolution of the virus in the months of February and March 2020 in Lombardy. Indeed, it is in these early stages of the epidemic that uncertainty in data and transport dynamics played a key role. In order to study the effects of the uncertainties of the initial conditions and of the parameters involved in the model on the solution of the problem, a second-order stochastic asymptotic-preservative IMEX Runge-Kutta Finite Volume Collocation method has been used. It is shown that the proposed numerical scheme achieves spectral accuracy in the stochastic space by achieving an exponential convergence rate even in the case where uncertainty is present not only in the initial data but also in the nonlinear interaction terms.

Numerical results of the baseline scenario were compared with observed data made available by the Italian Civil Protection Department, demonstrating that the proposed model is suitable to adapt to real settings and applications and showing its ability on capturing the heterogeneity underlying epidemics. Also alternative scenarios have been evaluated, considering both a total lockdown of extra-urban mobility and a total absence of restrictions at governmental level.

The proposed methodology has the potential to be applied to epidemiological models structured in more  compartments, such as those proposed in \cite{gatto2020,giordano2020,buonomo2020}. In this context, it was decided to keep the compartmentalization as simple as possible given the lack of more specific observed data at the provincial level. As a matter of fact, for the single provinces during the first epidemic wave, the Italian Civil Protection Department has made available only the cumulative trend of the subjects tested positive to COVID-19 \cite{github_covid}. There is no distinction between individuals simply quarantined, hospitalized or even in intensive care. Similarly, no structured data are publicly available at the level of provinces for individuals who have died. 

Future developments foresee an extension of the model to include the age structure of the population, as this is an essential feature to correctly describe mobility flows (which mainly affect the younger part of the population) and consequently a more correct spread of viruses such as COVID-19 \cite{albi2020,albi2020a}. Another interesting aspect, especially for analysis related to recent developments of the COVID-19 pandemic, is certainly the extension of the model to take into account the effects of the viral load of individuals, by analyzing the effects of the so called super-spreaders \cite{NATSS}, and the immunization of susceptible individuals through vaccination campaigns \cite{giordano2021}.
\section*{Acknowledgements}
This work was partially supported by MIUR (Ministero dell’Istruzione, dell’Università e della
Ricerca) PRIN 2017, project “\textit{Innovative numerical methods for evolutionary partial differential equations and applications}”, code 2017KKJP4X.

\appendix
\section{Derivation of the reproduction number}
\label{appendix:R0}
Recurring to the \textit{next-generation matrix} (NGM) procedure, it is possible to compute the dominant eigenvalue of a positive linear operator, called \textit{next-generation operator}, which permits to define the reproduction number when there are several compartments contributing to the spread of the infection \cite{diekmann1990,gatto2020,viguerie2020,tang2020,giordano2020,buonomo2020}. The generalization to space dependent models is non straightforward due to the nonlinear nature of the interactions. In the following we consider the situation where no-inflow/outflow boundary conditions are assumed. 

Given the spatial model defined in \eqref{eq.SEIARmacro}, let us consider the following subsystem of balance equations of the infectious compartments including both commuters and non-commuters
\begin{equation*}
\begin{aligned}
\frac{\partial E_T}{\partial t} + \frac{\partial J_E}{\partial x} &= f_I(S_T, I_T) +f_A(S_T,A_T) -aE_T \\
\frac{\partial I_T}{\partial t} + \frac{\partial J_I}{\partial x} &= a \sigma E_T -\gamma_I I_T\\
\frac{\partial A_T}{\partial t} + \frac{\partial J_A}{\partial x} &= a(1- \sigma) E_T -\gamma_A A_T\,.
\end{aligned}
\label{eq.EIA}
\end{equation*}
Considering, a deterministic framework, in which the epidemic parameters only depend on the variable $x$, thanks to our assumptions, integration over $\Omega$ leads to
\begin{equation}
\begin{aligned}
\frac{\partial}{\partial t}\int_\Omega E_T(x,t) dx &= \int_\Omega f_I(S_T, I_T) dx  + \int_\Omega f_A(S_T,A_T) dx - \int_\Omega a(x) E_T(x,t) dx \\
\frac{\partial}{\partial t} \int_\Omega I_T(x,t) dx &= \int_\Omega a(x) \sigma (x) E_T(x,t) dx - \int_\Omega \gamma_I(x) I_T(x,t) dx\\
\frac{\partial}{\partial t} \int_\Omega A_T(x,t) dx &= \int_\Omega a(x) (1- \sigma (x)) E_T(x,t) dx -\int_\Omega \gamma_A(x) A_T(x,t) dx\,.
\end{aligned}
\label{eq.EIA_int}
\end{equation}
If we define the total densities
\begin{equation*}
\bar E = \int_\Omega E_T(x,t) dx \, , \qquad \bar I = \int_\Omega I_T(x,t) dx \, , \qquad \bar A = \int_\Omega A_T(x,t) dx 
\end{equation*}
and the averaged epidemic operators,
\begin{equation*}
\bar f_I(S_T,I_T) = \frac{\int_\Omega f_I(S_T, I_T) dx}{\int_\Omega I_T(x,t) dx} \, , \qquad
\bar f_A(S_T,A_T) = \frac{\int_\Omega f_A(S_T, A_T) dx}{\int_\Omega A_T(x,t) dx} \, ,
\end{equation*}
\begin{equation*}
\bar \gamma_I = \frac{\int_\Omega \gamma_I(x) I_T(x,t) dx}{\int_\Omega I_T(x,t) dx}\, , \qquad
\bar \gamma_A = \frac{\int_\Omega \gamma_A(x) A_T(x,t) dx}{\int_\Omega A_T(x,t) dx}\, ,
\end{equation*}
\begin{equation*}
\bar a = \frac{\int_\Omega a(x) E_T(x,t) dx}{\int_\Omega E_T(x,t) dx}\, , \quad
\hat a = \frac{\int_\Omega a(x) \sigma(x) E_T(x,t) dx}{\int_\Omega E_T(x,t) dx}\, , \quad
\tilde a = \frac{\int_\Omega a(x)(1- \sigma(x)) E_T(x,t) dx}{\int_\Omega E_T(x,t) dx}\, , 
\end{equation*}
system \eqref{eq.EIA_int} can be rewritten in the following ODE form
\begin{equation}
\begin{aligned}
\dot{\bar E} &= \bar f_I(S_T,I_T) \bar I + \bar f_A(S_T,A_T) \bar A - \bar a \bar E\\
\dot{\bar I} &= \hat a \bar E - \bar \gamma_I \bar I\\
\dot{\bar A} &= \tilde a \bar E - \bar \gamma_A \bar A \,.
\end{aligned}
\label{eq.EIA_ode}
\end{equation}
Following the NGM approach \cite{diekmann1990,gatto2020}, the Jacobian matrix of the above obtained ODE system results
\[\bf{J_0} = 
\begin{pmatrix} 
	\bar a &\bar f_I(S_T,I_T) &\bar f_A(S_T,A_T) \\ \hat a &-\bar \gamma_I &0 \\ \tilde a &0 & -\bar \gamma_A 
\end{pmatrix}, \]
which can be decomposed into the following transmission matrix $\bf{T}$ and transition matrix $\bf{\Sigma}$, so that $\bf{J_0} = \bf{T} + \bf{\Sigma}$
\[\bf{T} = 
\begin{pmatrix} 
	0 &\bar f_I(S_T,I_T) &\bar f_A(S_T,A_T) \\ 0 &0 &0 \\ 0 &0 &0
\end{pmatrix}, \qquad
\bf{\Sigma} = 
\begin{pmatrix} 
	\bar a &0 &0 \\ \hat a &-\bar \gamma_I &0 \\ \tilde a &0 & -\bar \gamma_A 
\end{pmatrix}. \]
The basic reproduction number $R_0$ is the spectral radius of the \textit{next-generation operator}, $\bf{K_L} = -\bf T \bf \Sigma^{-1}$, and results composed by the sum of two components, deriving from the two compartments of the model contributing to the spread of the epidemic ($I$ and $A$)
\begin{equation}
R_0 = \rho \left( \bf K_L \right)  = R_0^I + R_0^A
\label{eq.R0_1}
\end{equation}
with
\begin{equation}
R_0^I  = \frac{\bar f_I(S_T,I_T) \, \hat a}{\bar\gamma_I \, \bar a}, \quad 
R_0^A= \frac{\bar f_A(S_T,A_T) \, \tilde a}{\bar\gamma_A \, \bar a} \,.
\label{eq.R0_IA}
\end{equation}
and therefore, recalling the previous definitions
\begin{equation}
\begin{aligned}
R_0 (t) &= \frac{\int_\Omega f_I(S_T,I_T) \,dx}{\int_\Omega \gamma_I(x) I_T(x,t) \,dx} \cdot \frac{\int_\Omega a(x)\sigma (x) E_T(x,t) \,dx}{\int_\Omega a(x) E_T(x,t) \,dx} \\&+ \frac{\int_\Omega f_A(S_T,A_T) \,dx}{\int_\Omega \gamma_A(x) A_T(x,t)\, dx} \cdot \frac{\int_\Omega a(x)(1-\sigma(x)) E_T(x,t) \,dx}{\int_\Omega a(x) E_T(x,t)\, dx} \,.
\end{aligned}
\label{eq:Rt}
\end{equation}
Note that, the dependence of additional stochastic parameters will not modify the above reasoning leading to the same definition \eqref{eq:Rt} including dependence from the uncertainty variable $\boldsymbol{z} = (z_1,\ldots,z_{d})^T \in \mathbb{R}^{d}$. 

\section{Stochastic Collocation Method}
\label{appendix:SCM}
Following a stochastic collocation approach \cite{bertaglia2021a,Xiu2010}, the solution of the stochastic problem \eqref{systcompactform} can be computed employing a generalized Polynomial Chaos (gPC) expansion \cite{pareschi2020,jin2018,jin2017}. Let us consider, for simplicity, a single source of uncertainty $z\in\RR$ and that the probability density function (PDF) of this random input, $\rho_z: \Gamma \rightarrow \mathbb{R}^+$, is known. The approximated solution of the problem, ${\bf{Q}}^h(x,t,z)=({\bf{U}}_c^h, {\bf{V}}^h,{\bf{U}}_0^h)$, where ${\bf{U}}_0^h$ is the state vector of non-commuters, referring to system \eqref{eq.SEIARkinetic_noncommuters}, can be expressed as a finite series of orthonormal polynomials in terms of the stochastic parameter
\begin{equation}\label{eq:expansion}
{\bf{Q}}^h(x,t,z)=\sum_{j=1}^{M} \hat{\bf{Q}}_j(x,t) \phi_j(z),
\end{equation}
where $M$ is the number of terms of the truncated series and $\phi_j(z)$ are orthonormal polynomials, with respect to the measure $\rho_z(z)\, \d z$. The expansion coefficients $\hat{\bf{Q}}_j(x,t)$ is obtained as
\begin{equation}
\label{eq:exp_coeff_int}
\hat{\bf{Q}}_j(x,t) = \int_{\Gamma} {\bf{Q}}  (x,t,z)\, \phi_j(z)\, \rho_z(z)\, \d z, \qquad j=1,\ldots,M.
\end{equation}
Following the stochastic collocation method \cite{xiu2005,jin2017}, the integrals for the expansion coefficients in Eq.~\eqref{eq:exp_coeff_int} are replaced by a suitable quadrature $\mathcal{U}^{M_p}$ characterized by the set $\{z_n, w_n \}_{n=1}^{M_p}$, where $z_n$ is the $n$-th collocation point, $w_n$ is the corresponding weight and $M_p$ represents the number of quadrature points. For instance, for a stochastic parameter with a uniform distribution the quadrature is defined by the Gauss-Legendre weights and nodes; while for a random variable associated to a Gaussian PDF, we will rely on a Gauss-Hermite quadrature, which reads
\begin{equation}
\label{eq:exp_coeff_quad}
\hat{\bf{Q}}_j(x,t) \approx \mathcal{U}^{M_p}\left[{\bf{Q}}^d(x,t,z)\, \phi_j(z)\right] = \sum_{n=1}^{M_p} {\bf{Q}}^d(x,t,z_n)\, \phi_j(z_n)\, w_n, \qquad j=1,\ldots,M
\end{equation}
where ${\bf{Q}}^d(x,t,z_n)$, with $n=1,\ldots,M_p$, is the deterministic solution of the problem evaluated at the $n$-th collocation point. 
After the computation of the expansion coefficients, substituting eq.~\eqref{eq:exp_coeff_quad} in eq.~\eqref{eq:expansion}, an approximated solution is available. In particular, it is possible to continue with the post-processing step, evaluating the expectation of the variables of interest
\begin{equation}
\label{eq:mean_an}
\mathbb{E}\left[{\bf{Q}}\right] =\int_{\Gamma} {\bf{Q}}(x,t,z)\, \rho_z(z)\, \d z ,
\end{equation}
which are approximated as
\begin{equation}
\label{eq:mean_apx}
\mathbb{E}\left[{\bf{Q}}\right] \approx \mathbb{E}\left[{\bf{Q}}^h\right] =\int_{\Gamma} {\bf{Q}}(x,t,z)\, \rho_z(z)\, \d z \approx \sum_{n=1}^{M_p}  {\bf{Q}}^d(x,t,z_n)\, w_n .
\end{equation}
Similarly, other statistical quantities of interest can be computed \cite{Xiu2010}, like the variance
\begin{equation}
\label{eq:variance_apx}
\mathbb{V}\left[{\bf{Q}}\right] = \mathbb{E}\left[ \left( {\bf{Q}} - \mathbb{E}\left[{\bf{Q}}\right] \right)^2 \right] \approx \mathbb{E}\left[\left({\bf{Q}}^h \right)^2\right] - \mathbb{E}\left[{\bf{Q}}^h \right]^2 .
\end{equation}

When more than one stochastic parameter is considered in the problem, $\boldsymbol{z}\in \RR^{d}$, the joint PDF $\rho_z\left(\boldsymbol{z}\right)$ of the random vector composed by the random parameters (assuming independence of the variables) is given by \begin{equation}
\rho_z\left(\boldsymbol{z}\right) = \prod_{k=1}^{d} \rho_{z,k}\left(z_k\right),
\end{equation} 
where $\rho_{z,k}$ is the PDF of the $k$-th variable.
The extension of the collocation method then follows in a similar way \cite{xiu2005}.

\section{AP-IMEX Runge-Kutta Finite Volume scheme}
\label{appendix:IMEX}
To compute the solution at each collocation point, an IMEX Runge-Kutta Finite Volume method for hyperbolic systems with multiscale relaxation is adopted \cite{boscarino2017,bertaglia2021}. 
The IMEX discretization for the commuters' dynamics \eqref{systcompactform}, coupled with the equations for the non commuter population \eqref{eq.SEIARkinetic_noncommuters} through identities \eqref{eq:seiart}, results with internal Runge-Kutta stages
\begin{equation}
\begin{split}
	&\U_c^{(k)} = \U_c^n -  \Delta t \sum_{j=1}^{k} a_{kj} \partial_x \V^{(j)}  + \Delta t \sum_{j=1}^{k-1} \tilde{a}_{kj} \F_c\left(\U_T^{(j)},\U_c^{(j)}\right)
	\\
	&\V^{(k)} = \V^n -  \Delta t \sum_{j=1}^{k-1} \tilde{a}_{kj} \left( \boldsymbol{\Lambda}^2 \partial_x \U_c^{(j)} - \G\left(\U_T^{(j)},\V^{(j)}\right)\right)  + \Delta t \sum_{j=1}^{k} a_{kj} \H\left(\V^{(j)}\right)
	\\
	&\U_0^{(k)} = \U_0^n + \Delta t \sum_{j=1}^{k-1} \tilde{a}_{kj} \F_0\left(\U_T^{(j)},\U_0^{(j)}\right),
	\end{split}
\label{eq.iterIMEX}
\end{equation}
and final numerical solution
\begin{equation}
\begin{split}		
	& \U_c^{n+1} = \U_c^n - \Delta t \sum_{k=1}^{s} b_{k} \partial_x \V^{(k)} + \Delta t \sum_{k=1}^{s} \tilde{b}_{k} \F_c\left(\U_T^{(k)},\U_c^{(k)}\right)
	\\
	& \V^{n+1} = \V^n - \Delta t \sum_{k=1}^{s} \tilde{b}_{k} \left( \boldsymbol{\Lambda}^2 \partial_x \U_c^{(k)} - \G\left(\U_T^{(k)},\V^{(k)}\right) \right) + \Delta t \sum_{k=1}^{s} b_{k} \H\left(\V^{(k)}\right)
	\\
	& \U_0^{n+1} = \U_0^n + \Delta t \sum_{k=1}^{s} \tilde{b}_{k} \F_0\left(\U_T^{(k)},\U_0^{(k)}\right).
\end{split}
\label{eq.finalIMEX}
\end{equation}
Here $\Delta t = t^{n+1} - t^n$ is the time step size that follows the less restrictive between the standard hyperbolic Courant-Friedrichs-Levy condition, $\Delta t \leq \CFL \frac{\Delta x}{max\left\{\lambda_i\right\}}$, and the parabolic stability restriction, $\Delta t \leq \nu \frac{\Delta x^2}{max\left\{D_i\right\}}$, given by the diffusive components of the system, with $\CFL$ and $\nu$ suitable stability constants \cite{boscarino2017} and $\Delta x = x_{i+\frac{1}{2}} - x_{i-\frac{1}{2}}$ space grid size. In this work, we fix $\CFL = 0.9$ and $\nu = \frac{max\left\{D_i\right\}}{2}$.

The numerical scheme is characterized by two $s \times s$ matrices, $\tilde A = (\tilde a_{kj})$, with $\tilde a_{kj} = 0$ for $ j\geq k$, and $A = (a_{kj})$, with $a_{kj} = 0$ for $j > k$, and by the weights vectors $\tilde b = (\tilde b_1, ...,\tilde b_s)^T$, $b = (b_1, ...,b_s)^T$ (with $s$ identifying the number of the Runge-Kutta stages), which can be represented in the following Butcher notation \cite{pareschi2005}
\begin{center}
\begin{tabular}{c | c}
$\tilde c$ & $\tilde A$ \\  
\hline \\[-1.0em] 
 &  $\tilde b^T$ \\ 
\end{tabular}
\hspace{2.0cm}
\begin{tabular}{c | c}
$c$ & $A$ \\  
\hline \\[-1.0em] 
 &  $b^T$ \\ 
\end{tabular}
\end{center}
where $\tilde c$ and $c$ are the time coefficient vectors
\begin{equation*}
\tilde c_k = \sum^{k-1}_{j=1} \tilde a_{kj} , \qquad \qquad c_k = \sum^{k}_{j=1} a_{kj}.
\end{equation*}
The distribution of matrices $\tilde A$ and $A$ permits to treat implicitly the stiff terms (hence those depending on the scaling parameters, $\tau_i$ and $\lambda_i$) and explicitly all the rest. Moreover, if a proper set of matrices $\tilde A, A$ and vectors $\tilde b, b$ are chosen, the AP property is satisfied, which means that the scheme maintains a consistent discretization of the asymptotic system in the diffusive (parabolic) regime (see \cite{boscarino2017, bertaglia2021}). For example, the second order GSA BPR(4,4,2) scheme proposed in \cite{boscarino2017} satisfies the AP property and is defined by the following double Butcher tableau (explicit on the left and implicit on the right)
\begin{equation}
\begin{tabular}{c | c c c c c}
0 & 0 & 0 & 0 & 0 & 0 \\
1/4 & 1/4 & 0 & 0 & 0 & 0 \\
1/4 & 13/4 & -3 & 0 & 0 & 0 \\
3/4 & 1/4 & 0 & 1/2 & 0 & 0 \\
1 & 0 & 1/3 & 1/6 & 1/2 & 0 \\ \hline
  & 0 & 1/3 & 1/6 & 1/2 & 0  \\
\end{tabular}
\hspace{1.0cm}
\begin{tabular}{c | c c c c c}
0 & 0 & 0 & 0 & 0 & 0 \\
1/4 & 0 & 1/4 & 0 & 0 & 0 \\
1/4 & 0 & 0 & 1/4 & 0 & 0 \\
3/4 & 0 & 1/24 & 11/24 & 1/4 & 0 \\
1 & 0 & 11/24 & 1/6 & 1/8 & 1/4 \\ \hline
  & 0 & 11/24 & 1/6 & 1/8 & 1/4   \\
\end{tabular}
\label{eq:tables}
\end{equation}

At each internal stage of the IMEX Runge-Kutta scheme \eqref{eq.iterIMEX}, we apply a TVD Finite Volume discretization \cite{bertaglia2021a,dumbser2011}. 
To achieve second order accuracy also in space, while avoiding the occurrence of spurious oscillations, a classical minmod slope limiter has been adopted.

Finally, let us remark that, given the non-intrusive nature of the stochastic collocation method which only requires the evaluation of the solutions of the corresponding deterministic problem at each collocation point, the AP property of the deterministic IMEX scheme is preserved, leading to a stochastic asymptotic-preserving scheme \cite{jin2015}. 

\bibliographystyle{abbrv}
\bibliography{SEIARnetwork}

\end{document}